\documentclass[aps,prl,reprint,nofootinbib]{revtex4-2}

\usepackage{graphicx}
\usepackage{amsmath,amssymb}
\usepackage{bm}
\usepackage{xcolor}
\usepackage[colorlinks=true,linkcolor=blue!60!black,citecolor=blue!60!black,urlcolor=blue!60!black]{hyperref}

\usepackage{float}

\newcommand{\smfloats}{\renewcommand{\topfraction}{.9}\renewcommand{\bottomfraction}{.9}%
  \renewcommand{\textfraction}{.08}\renewcommand{\floatpagefraction}{.75}\raggedbottom}
\newcommand{\sub}[3]{%
  \begin{minipage}[t]{#1\linewidth}\raggedright\textbf{(#2)}\par\vspace{1pt}%
  \includegraphics[width=\linewidth]{#3}\end{minipage}}
\newcommand{\subc}[3]{%
  \par\noindent\hfil\begin{minipage}[t]{#1\linewidth}\raggedright\textbf{(#2)}\par\vspace{1pt}%
  \includegraphics[width=\linewidth]{#3}\end{minipage}\hfil\par\vspace{5pt}}
\newcommand{\subnt}[2]{%
  \begin{minipage}[t]{#1\linewidth}\centering\includegraphics[width=\linewidth]{#2}\end{minipage}}

\begin{document}

\title{Pair density waves in altermagnets: thermal stability and spectroscopic fingerprints}

\author{Amrutha N Madhusuthanan}
\author{Madhuparna Karmakar$^{*}$}
\affiliation{Department of Physics and Nanotechnology, SRM Institute of Science and Technology, Kattankulathur, Chennai 603203, India}

\date{\today}

\begin{abstract}
We show that altermagnetism stabilizes finite-momentum superconductivity at finite temperature in two dimensions without an applied field. Using 
static path approximation Monte Carlo, we find that a d-wave altermagnet hosts unidirectional and bidirectional pair density waves (PDWs) that survive 
strong thermal fluctuations over a finite temperature window. A Zeeman field induces a Fulde-Ferrell-Larkin-Ovchinnikov-type state. We identify the 
thermal scales for phase coherence, gap closing, and pseudogap formation, and predict fingerprints for ARPES and quasiparticle interference.
\end{abstract}

\maketitle

\textit{Introduction:} A finite momentum paired superconducting (SC) state is the one in which the superconducting correlations vary periodically 
in space such that the spatial average vanishes. Finite momentum superconductivity dates back to Fulde-Ferrell-Larkin-Ovchinnikov (FFLO) \cite{FF1964,LO1964} and Sarma phases \cite{Sarma1963} in Pauli limited superconductors, wherein an applied Zeeman field induces imbalance 
in the fermionic population and promotes finite-momentum pairing instabilities \cite{Casalbuoni2004,Bianchi2003,Matsuda2007,Lortz2007,Kenzelmann2008,Mayaffre2014,Shin2008}. A pair density wave (PDW) state is thermodynamically distinct 
from FFLO in that it doesn't require an explicit time-reversal symmetry breaking and is stabilized instead by electronic interactions and Fermi 
surface anisotropy \cite{Agterberg2020,Fradkin2015,Loder2010,Berg2009NJP,Soto2014,Agterberg2008,Berg2009}.  Despite extensive theoretical work \cite{Agterberg2020,Fradkin2015,Loder2010,Berg2009NJP,Soto2014} and growing experimental evidence in systems such as Kagome metals \cite{Jiang2021,Chen2021}, UTe$_{2}$ \cite{Gu2023,Aishwarya2023}, EuRbFe$_{4}$As$_{4}$ \cite{Zhao2023}, stabilizing PDW order at finite temperatures in two dimensions (2D) remains an open challenge due to its sensitivity to thermal fluctuations.

Of late altermagnet (ALM) with its non-trivial magnetic space-group properties \cite{Smejkal2022beyond,Smejkal2022emerging,Song2025,Jungwirth2025} has proved to be a natural set up to stabilize the PDW state. With their momentum-dependent spin-splitting with zero net magnetization \cite{Krempasky2024,Lee2024,Osumi2024,Ding2024,Jiang2025} ALM naturally mimics a $k$-space Zeeman field that favors finite-${\bf q}$ pairing channels while avoiding orbital depairing effects \cite{Sun2023,Ouassou2023,Lu2024,Maeda2025,Fukaya2025}. PDW is inherently a strong coupling phenomenon that energetically competes with the uniform $d$-wave pairing and is expected to stabilize in materials such as, the high-temperature cuprate whose normal state is a non-Fermi liquid \cite{Agterberg2020}. Evidently, its 
microscopic understanding requires non-perturbative numerical approaches, particularly at finite temperatures where thermal fluctuations are often known 
to be detrimental towards ${\bf q} \neq 0$ phases. Existing theoretical studies, largely based on mean-field theory (MFT) and phenomenological approaches have largely established that ALM can stabilize PDW as the ground state in the putative phase diagram  \cite{Sumita2025,Zhang2024,Hu2026,Sim2025,Chakraborty2024,Chakraborty2025,Bose2024,Zou2026,Jasiewicz2025,Hong2025,Wei2024,Ouassou2023,Papaj2023}. Its fate at $T \neq 0$ however, remains an open question owing largely to the lack of a suitable numerical approach that can handle spatially modulated SC phases in reasonably large system sizes and down to low temperatures while accounting for spatial fluctuations at all orders.   
\begin{figure}
  \centering
  \includegraphics[width=8cm]{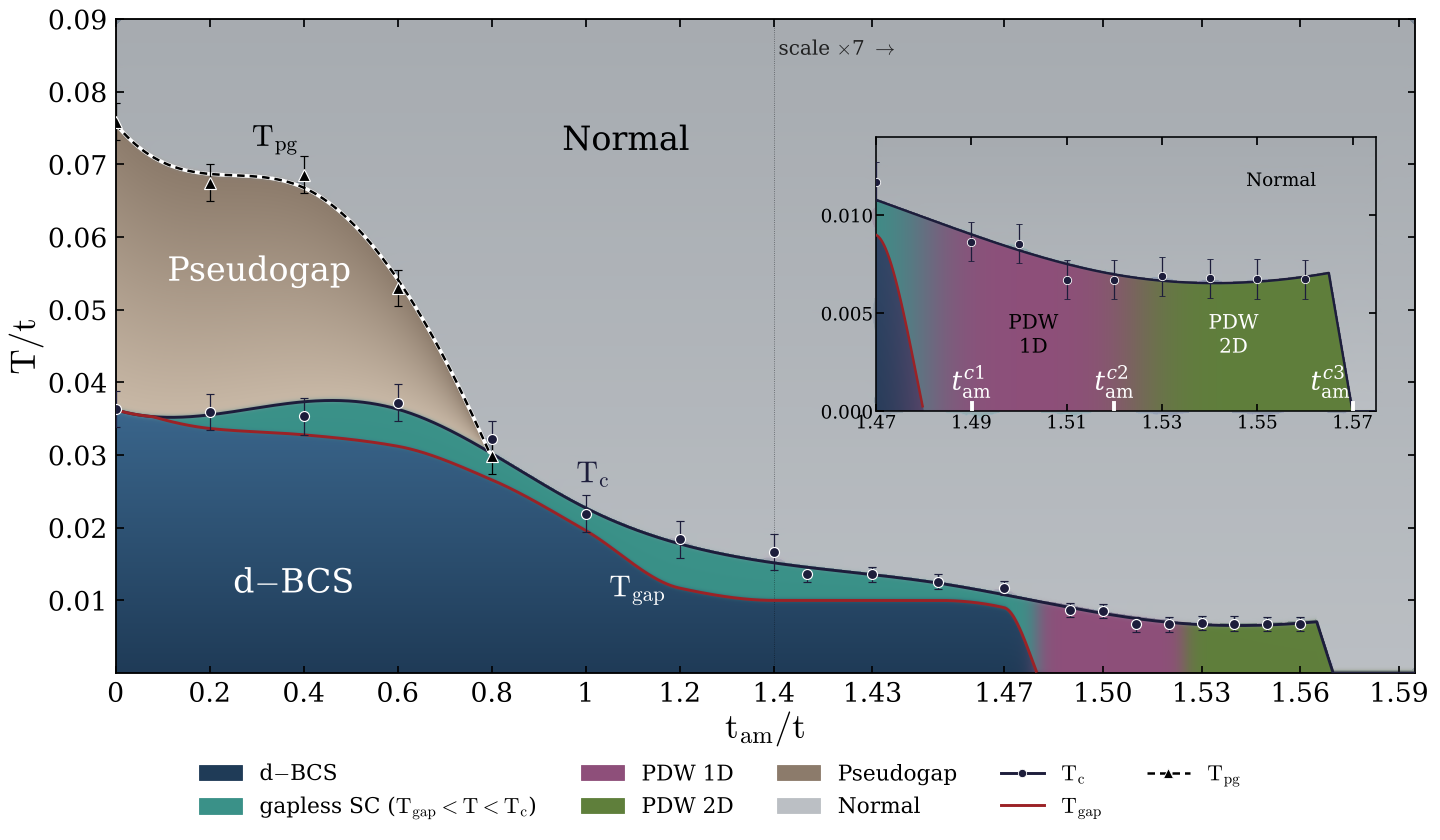}
  \caption{Thermal phase diagrams of the $d_{x^{2}-y^{2}}$-wave ALM superconductor in the $t_{\mathrm{am}}$--$T$ plane at $h=0$, $\vert U\vert = 4t$, $\mu = -1.5t$ and $L=24$,  showing the thermodynamic phases $d$-BCS,  $\text{PDW 1D}$, $\text{PDW 2D}$, normal, and the thermal scales $T_{c}$, $T_{gap}$ and $T_{pg}$ (see text). Note the $\times 7$ horizontal scale expansion above $t_{\mathrm{am}}=1.4t$; the inset magnifies the finite-momentum PDW window.}
  \label{fig:phase}
\end{figure}  

In this work we use static path approximation (SPA) Monte Carlo simulations to access the finite temperature physics of the ALM tuned PDW phase in 
a $d$-wave superconductor \cite{Mayr2005,Kumar2006,Karmakar2016,Karmakar2018,Tarat2015}. Taking into account the spatial fluctuations of the superconducting pairing field at all orders we base our analysis on the thermodynamic and transport signatures of the system, which should be accessible in candidate materials to standard probes such as, angle resolved photoemission spectroscopy (ARPES), quasiparticle interference (QPI) and neutron scattering measurements. The principal result from this work is distilled in Figure~\ref{fig:phase} as the ALM-temperature phase diagram which maps  out the regime 
of thermal stability of the PDW phase in a $d_{x^{2}-y^{2}}$ ALM material with $d_{x^{2}-y^{2}}$ SC pairing.  

\textit{Model and Methods:}
We model the system on a square lattice with $d_{x^{2}-y^{2}}$ ALM spin splitting \cite{Giuli2025,Kaushal2025,Durrnagel2025} and an effective inter-site attractive interaction that stabilizes a 
$d_{x^{2}-y^{2}}$ pairing state symmetry \cite{Micnas1990,Mayr2005,dagotto_prb2005} (see SM for details), 

\begin{equation}
\begin{split}
\hat{H} ={}& \sum_{\langle ij\rangle,\sigma}\left(-t_{ij}+\sigma\, t_{\mathrm{am}}\,\eta_{ij}\right)\left(\hat{c}^{\dagger}_{i\sigma}\hat{c}_{j\sigma}+\mathrm{h.c.}\right)\\
&-\sum_{i,\sigma}\left(\mu+\sigma h\right)\hat{n}_{i\sigma}-|U|\sum_{\langle ij\rangle}\hat{n}_{i}\hat{n}_{j},
\end{split}
\label{eq:ham}
\end{equation}
where, $t_{ij}=t=1$ is the nearest neighbor hopping amplitude and sets the reference energy scale of the model. The ALM splitting of the electronic 
band is quantified in terms of $t_{am}$ with $t_x = t - \sigma t_{\mathrm{am}}/4$ and $t_y = t + \sigma t_{\mathrm{am}}/4$ being the effective 
spin-momentum dependent hopping amplitudes. We work in the grand canonical ensemble with chemical potential $\mu = -1.5t$ corresponding 
to the fermionic filling of $n \sim 0.5$. The uniform Zeeman field $h$ accounts for the FFLO physics and the finite-${\bf q}$ paired states in the FFLO 
and PDW phases are quantified in terms of the net magnetization ($m$): PDW ($m = 0$) and FFLO ($m \neq 0$). The model is made numerically 
tractable via Hubbard-Stratonovich decomposition of the quartic interaction \cite{Hubbard1959,Schulz1990}. The resulting quadratic Hamiltonian 
contains complex bosonic pairing field $\Delta_{ij}(\tau)=|\Delta_{ij}(\tau)|e^{i\phi_{ij}(\tau)}$ coupled to the singlet ($\hat{c}_{i\uparrow}\hat{c}_{j\downarrow}+\hat{c}_{j\uparrow}\hat{c}_{i\downarrow}$) pairing channel. Within SPA we retain the spatial fluctuations of $\Delta_{ij}$ to all orders while keeping 
only its zero mode Matsubara-frequency ($\Delta_{ij}(\tau)\rightarrow\Delta_{ij}$) \cite{Mayr2005,dagotto_prb2005,Karmakar2016}. 
With this adiabatic approximation the Hamiltonian maps onto an effective spin-fermion model wherein the fermions are subjected to a random, fluctuating 
background of the bosonic auxiliary field \cite{Fratini2021,Ciuchi2023,Murthy2023} (see SM). To complement our finite-temperature simulations at low $T$, we also perform a variational mean-field theory (MFT) optimization at $T=0.005t$ across system sizes $L=24$--$100$, comparing uniform $d$-wave, unidirectional $\Delta_{ij}\sim|\Delta|\cos(q_x x_i+q_y y_i)$, and bidirectional $\Delta_{ij}\sim|\Delta|\left[\cos(q_x x_i)+\cos(q_y y_i)\right]$ ansatz for the pairing field with fixed relative $x$-$y$ bond phase $\phi_{ij}^{\mathrm{rel}}=\pi$ \cite{Karmakar2020,Karmakar2024} (see SM for the $L=80$ benchmarks).
\begin{figure}
  \centering
  \includegraphics[height=6cm,width=8cm,angle=0]{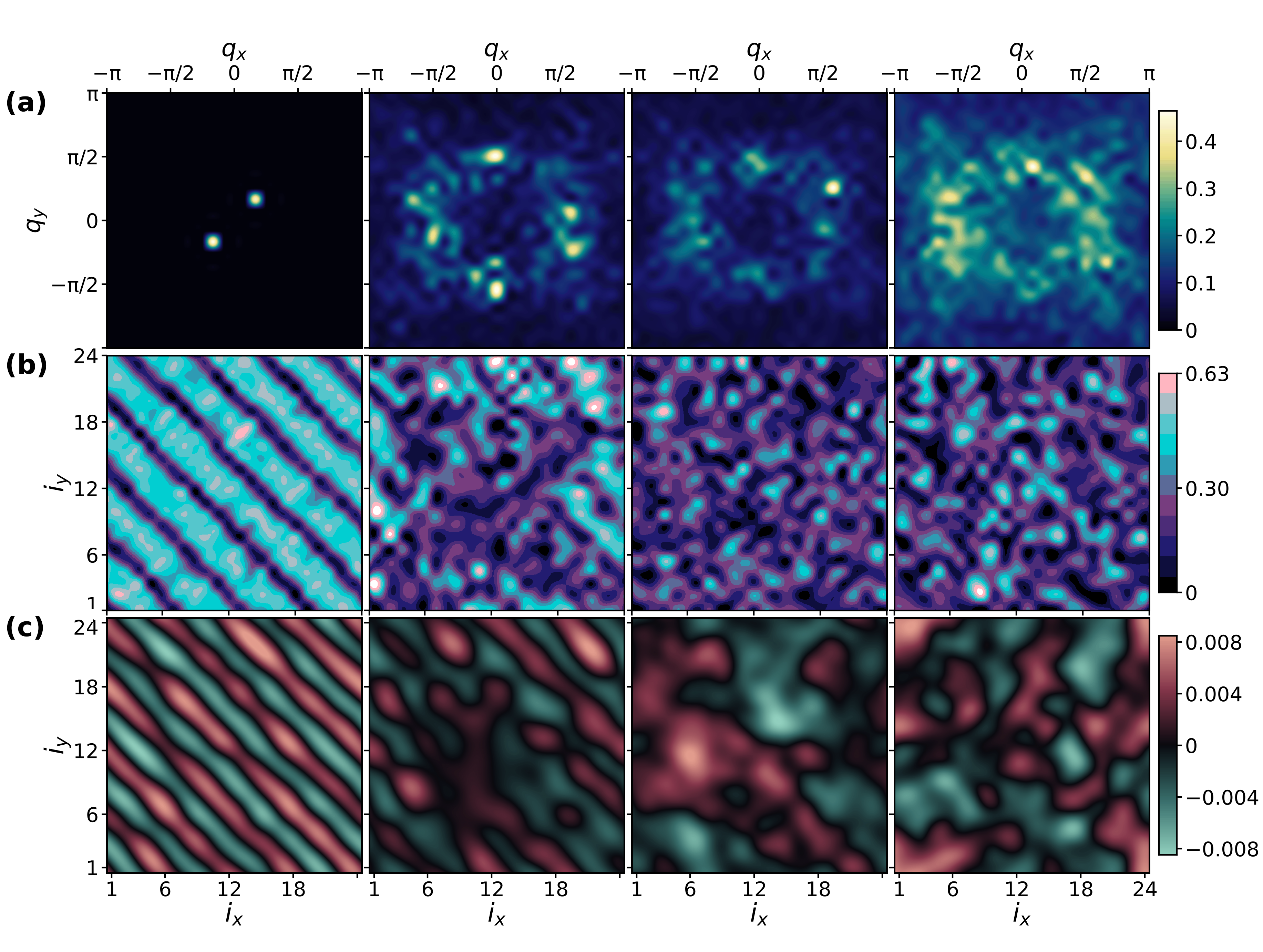}
  \caption{Thermal disordering of the $\text{PDW 1D}$ state at $t_{\mathrm{am}}=1.49t$ and $h=0$ for $T=0.002t$, $0.006t$, $0.008t$, and $0.012t$ (from left to right). (a) Structure factor $S_x(\mathbf{q})$ of the pairing field, (b) single-snapshot real-space maps of the pairing amplitude $\vert \Delta_i \vert$, 
  (c) single-snapshot real-space maps of the local magnetization $m_i = n_{i\uparrow} - n_{i\downarrow}$.}
  \label{fig:pdw1d}
\end{figure}

Our key indicators for analyzing the thermodynamic phases and transitions include:  (i) pairing-field structure factor $S_x(\mathbf{q})$ quantifying the superconducting thermal transition scale ($T_{c}$; see SM for its extraction), (ii) real-space pairing field amplitude $\vert \Delta_{ij} \vert$ and local spin polarization $m_i = n_{i\uparrow} - n_{i\downarrow}$, (iii) the single-particle density of states $N(\omega)$, with the corresponding energy gap $E_{gap}$ quantifying the 
gap to gapless transition in the superconducting phase and (iv) spin-resolved spectral function $A_\sigma(\mathbf{k},\omega)$, whose zero-energy weight $A_\sigma(\mathbf{k},0)$ and spin difference $A_\uparrow(\mathbf{k},\omega) - A_\downarrow(\mathbf{k},\omega)$ map out Bogoliubov Fermi-surface and ALM spin splitting across the phase diagram, respectively. The Monte Carlo calculations are carried out on a system size of $L=24$ and a larger system 
size of $L=40$ is used at selected parameters to circumvent finite system size effects (see SM).
  
\textit{Thermodynamic phases and thermal scales:}
Figure~\ref{fig:phase} encodes the primary result from this work as the thermal phase diagram in the $t_{am}-T$ plane at $h=0$, with six thermodynamic phases and three thermal scales: (a) the global superconducting ordering temperature $T_c$, (b) the spectral gap-closing scale $T_{\mathrm{gap}}$, 
and (c) the pseudogap crossover temperature $T_{\mathrm{pg}}$. 

At $t_{\mathrm{am}}=0$, the system is a uniform $d_{x^2-y^2}$ BCS superconductor for $T < T_c \approx 0.036t$, where the single-particle spectral gap closes simultaneously with the loss of (quasi) long range phase coherence ($T_{\mathrm{gap}} \approx T_c$). At $T > T_c$, strong classical phase fluctuations of the pairing field sustain a pseudogap up to $T_{\mathrm{pg}} \approx 0.075t$ \cite{Mayr2005,Karmakar2020,Kadge2018}. ALM progressively suppresses $T_{\mathrm{pg}}$, narrowing the pseudogap window until $T_{\mathrm{pg}}$ merges with $T_c$ near $t_{\mathrm{am}} \approx 0.8t$; for stronger altermagnetic splitting, the system lacks a pseudogap phase. $t_{am}^{c1} \sim 1.49t$ marks the first critical point of this system, at which the 
BCS ($\mathbf{Q}=(0,0)$) superconducting state breaks the translation symmetry and stabilizes a finite-${\bf Q}$ paired PDW state. In the regime 
$0 < t_{am} \lesssim t_{am}^{c1}$ the $T_{c}$ undergoes monotonic suppression from $0.036t$ down to $\sim 0.012t$ at $t_{\mathrm{am}} \approx 1.47t$. 
The BCS phase is characterized by a gapless temperature regime $T_{gap} < T \lesssim T_{c}$ that corresponds to a globally phase coherent state with weak spatial fragmentation, a virtue of the intermediate electronic interaction ($\vert U\vert = 4t$) with intertwined bosonic and fermionic degrees of freedom 
\cite{Tarat2015}.

ALM progressively distorts the $d_{x^2-y^2}$ Fermi surface, particularly along the anti-nodal directions, such that for $t_{am} > t_{am}^{c1}$ a $\mathbf{Q}=(0,0)$ state is no longer stabilized and the system undergoes a weakly first order transition to the gapless finite-momentum paired PDW phase ($\text{PDW 1D}$) with unidirectional modulation (see inset of Fig.~\ref{fig:phase}). The second critical point $t_{am}^{c2} \sim 1.52t$ marks the cross-over to a bi-directional PDW phase ($\text{PDW 2D}$) with checkerboard modulations. Within the limits of our temperature discretization the $T_c$ is a nearly flat 
plateau around $0.007t$, a strong suppression of $\sim 5$ times from the $t_{am} = 0.0t$ limit. Superconducting correlations are destroyed across the 
third critical point $t_{am}^{c3} \sim 1.57t$ as the system undergoes a second order phase transition to the normal state (see SM for the thermal evolution of the Fermi surface in each regime).  

\begin{figure}
\centering
\includegraphics[height=6cm,width=8cm,angle=0]{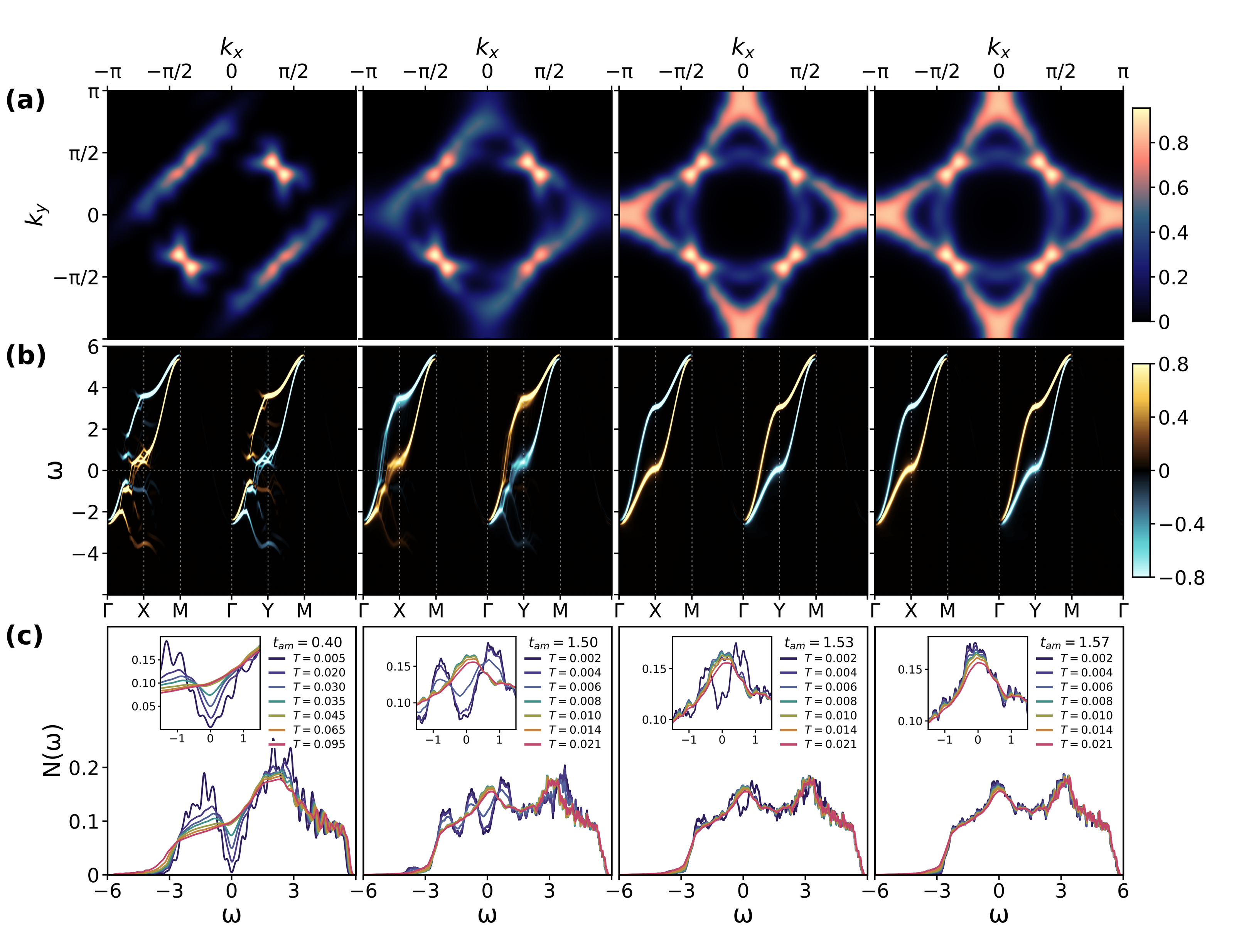}
\caption{Spectroscopic signatures of the $\text{PDW 1D}$ phase at $t_{\mathrm{am}}=1.49t$ for $T=0.002t$, $0.006t$, $0.008t$, and $0.012t$ 
(left to right). (a) Low-energy spectral weight distribution $A(\mathbf{k},0)$ mapping the Fermi surface, (b) Spectral function $\delta A({\bf k}, \omega) = A_\uparrow(\mathbf{k},\omega)-A_\downarrow(\mathbf{k},\omega)$ along the high-symmetry path $\Gamma\text{--}X\text{--}M\text{--}\Gamma\text{--}Y\text{--}M\text{--}\Gamma$ (gold and cyan denote spin-$\uparrow$ and spin-$\downarrow$ excess, respectively). (c) Thermal evolution of the single-particle DOS $N(\omega)$ across four representative regimes: $t_{\mathrm{am}}=0.40t$ ($d$-BCS), $t_{\mathrm{am}}=1.50t$ ($\text{PDW 1D}$), $t_{\mathrm{am}}=1.53t$ ($\text{PDW 2D}$), and $t_{\mathrm{am}}=1.57t$ (normal metal). Insets zoom in on the low-energy window $\vert \omega \vert \le 1.5t$ around the Fermi level.}
\label{fig:spectra_h0}
\end{figure}

\textit{The PDW phase:}
Figure~\ref{fig:pdw1d} and Figure~\ref{fig:spectra_h0} quantify the $\text{PDW 1D}$ phase (see SM for $\text{PDW 2D}$) based on the thermal evolution of 
its thermodynamic and spectroscopic signatures.  At the representative $t_{am} = 1.49t$ the $S_x(\mathbf{q})$ of the $\text{PDW 1D}$ shows two sharp symmetry-related Bragg peaks at $\pm\mathbf{Q}$ along the Brillouin-zone diagonal with a peak weight of $\approx 0.48$ at $T \ll T_{c}$ (Figure~\ref{fig:pdw1d}(a)). The peaks undergo fluctuation induced thermal broadening, restore the translation symmetry before undergoing the order-disorder transition to the normal state at $T > T_{c} \approx 0.0085t$. The corresponding pairing field amplitude $\vert \Delta_i \vert$ organizes into a regular diagonal stripe pattern in real space with a peak $\vert \Delta_i\vert \approx 0.63$ (Figure~\ref{fig:pdw1d}(b)). Because $|\Delta_i|$ is the modulus of an order parameter that changes sign across each nodal line, its stripes repeat with half the PDW wavelength, i.e.\ at wavevector $2\mathbf{Q}$. Thermal fluctuations break the diagonal ridges into meandering, orientationally disordered domains. The $T > T_{c}$ phase is short range correlated with local large $\vert \Delta_{i} \vert$, phase uncorrelated regimes. The modulated pairing field induces a weak spin modulation ($\vert m_{i}\vert \sim 0.008$) of period \ $2\mathbf{Q}$ without a net magnetization as shown in Figure~\ref{fig:pdw1d}(c). Temperature leads to progressive fragmentation of the spin modulation 
and their eventual destruction at $T \gtrsim T_{c}$.  

The spectroscopic signatures for the $\text{PDW 1D}$ phase at $t_{am} = 1.49t$ are presented in Figure~\ref{fig:spectra_h0} in terms of the thermal 
evolution of the Bogoliubov Fermi surface $A({\bf k}, 0)$ (Figure~\ref{fig:spectra_h0}(a)) and spectral function $\delta A({\bf k}, \omega) = A_{\uparrow}({\bf k}, \omega) - A_{\downarrow}({\bf k}, \omega)$ (Figure~\ref{fig:spectra_h0}(b)). The $C_4$-to-$C_2$ rotational symmetry breaking of the $\text{PDW 1D}$ at low temperatures is observed in terms of the reconstruction of the Bogoliubov Fermi surface \cite{Baruch2008,Agterberg2020}. The pairing wave vector is directed along the nodal (diagonal) direction. For the pair $({\bf k}_{\uparrow}, {\bf -k + Q}_{\downarrow})$, the energy mismatch is $\approx 2\delta_{\bf k} + v_{k}.{\bf Q}$ to leading order in 
${\bf Q}$, where $\delta_{\bf k} \propto \cos k_{x} - \cos k_{y}$ is the altermagnetic spin splitting and $v_{\bf k} . {\bf Q}$ is the Doppler shift \cite{Lee2014,YuanFu2021}. The splitting vanishes on the nodal diagonals and is largest along the antinodal axes. A unidirectional ${\bf Q}$ along one axis could compensate the mismatch only on that axis, whereas ${\bf Q} = (q ,q)$ has equal projections on both, so the condition $2\delta_{\bf k} = -v_{\bf k}.{\bf Q}$ is satisfied at points on both the 
$x$- and $y$-axes (at opposite signs of $k$).

The hallmark of the momentum dependent spin-split ALM state is observed through the $\delta A({\bf k}, \omega)$ maps (Figure~\ref{fig:spectra_h0}(b)) 
that exhibits opposite spin polarizations along the orthogonal $\Gamma\text{--}X\text{--}M$ and $\Gamma\text{--}Y\text{--}M$ directions. The finite-momentum pairing scatters quasiparticles between $|\mathbf{k},\uparrow\rangle$ and $|-\mathbf{k}\pm\mathbf{Q},\downarrow\rangle$, generating multiple folded Bogoliubov shadow branches, replica bands, and hybridized anti-crossings near $\omega=0$. Temperature broadens and smears the pair correlation induced shadow branches and at $T > T_{c}$ only two sharp, oppositely spin-split normal-state bands cross the Fermi level. The momentum dependent spin splitting 
survives for $t_{am} \neq 0$ independent of the superconducting correlations. 

We compare and contrast the thermal evolution of the single particle DOS across the $t_{am}$ regimes in Figure~\ref{fig:spectra_h0}(c). At $t_{\mathrm{am}}=0.40t$ (uniform $d$-BCS), the low-temperature spectrum at $T=0.005t$ displays a classic V-shaped $d$-wave nodal gap with $N(0)\rightarrow 0$ flanked by prominent coherence peaks.  Temperature accumulates spectral weight at $\omega  \approx 0$ leading to the nodal gap closure at $T \approx 0.020t$ 
and suppresses the coherence peaks via large transfer of spectral weight away from the Fermi level. The pseudogap phase typified by the depletion in 
the single particle DOS at $\omega \approx 0$ survives over the regime $0.020t < T_{pg} \lesssim 0.068t$ and gives way to the featureless normal state single particle DOS at $T > T_{pg}$. The $\text{PDW 1D}$ state is gapless ($N(0)\approx 0.08$) as evidenced by the low temperature single particle DOS 
at the representative $t_{am} = 1.50t$. The multiple subgap shoulders and secondary van Hove singularities are produced by the folded Bogoliubov branches akin to the Zeeman field induced standard FFLO state. Thermal fluctuation broadens the spectra, smears out the sub-gap features at $T \sim T_c\approx 0.0085t$. The sub-gap features multiply in the $\text{PDW 2D}$ regime ($N(0)\approx 0.11$) as the electrons undergo ${\bf q}$-dependent scattering.   
The PDW-2D to normal state transition takes place at $T_{c} \approx 0.007t$ and the single particle DOS becomes featureless. $t_{\mathrm{am}}=1.57t$ 
represents the normal state devoid of any superconducting correlations and characterized by the temperature independent single particle DOS with ALM 
van Hove singularities. 
\begin{figure}
\centering
\includegraphics[width=7cm]{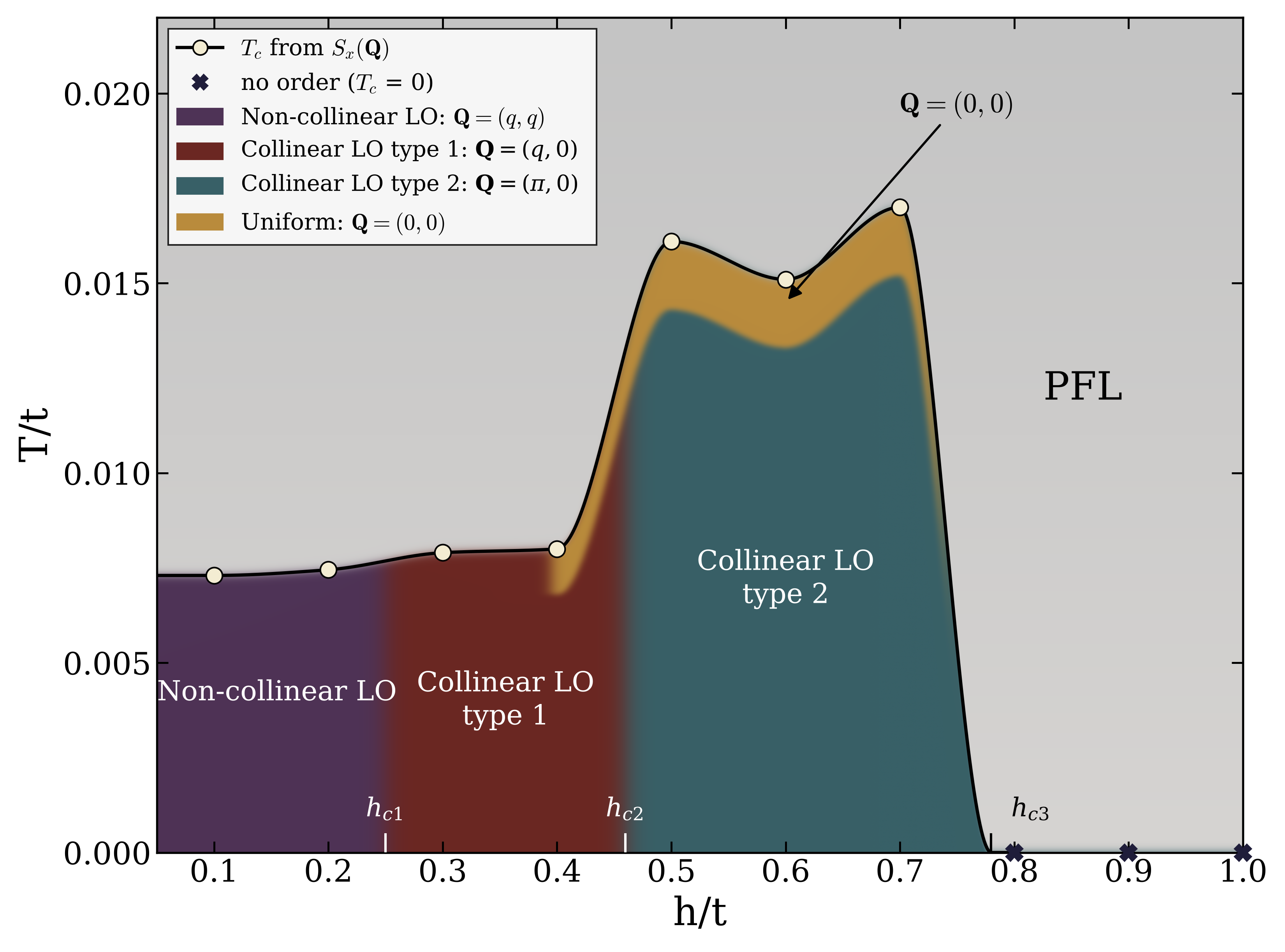}
\caption{Thermal phase diagrams of the $d_{x^{2}-y^{2}}$-wave ALM superconductor in the $h$--$T$ plane at $t_{am}=1.49t$, showing the 
thermodynamic phases non-collinear LO,  collinear Type-I LO, collinear Type-II LO, QBP and normal (see text).}
\label{fig:phase_lo}
\end{figure}  
 
\textit{The FFLO phase:} An applied Zeeman field $h \hat z$ is known to stabilize a range of ${\bf Q} \neq 0$ FFLO along with a re-entrant ${\bf Q} = 0$ phase at the ground state of the $d_{x^{2}-y^{2}}$-wave altermagnetic superconductor \cite{Chakraborty2024,Zhang2024,Wei2024}. The ground state phase diagram obtained by our variational 
MFT (at $L = 24$, $40$, and $80$) and SPA (at $L = 24, 40$) simulations show that with $h \neq 0$ the weak $t_{am}$ regime stabilizes a quantum breached pair (QBP) state with ${\bf Q} = 0$ pairing and non-zero magnetization, up to a critical Zeeman field $h_{QBP}$ \cite{Karmakar2020,Karmakar2024}. The strong $t_{am}$ and $h$ regimes host a plethora of 1D (unidirectional) and 2D (bidirectional) FFLO phases typified as non-collinear 1D LO ($\mathbf{Q}=(q,q)$), incommensurate Type-I collinear LO ($\mathbf{Q}=(q,0)$), commensurate Type-II LO ($\mathbf{Q}_x=(\pi,0)$), 2D LO ($\mathbf{Q}=(\pm q, \pm q)$), based on the pairing wave vector (see SM). Further, strong $t_{am}, h$ promotes the re-entrance of the gapless QBP phase which differ from the standard $d$-wave BCS state in terms of finite magnetization that spatially co-exists with ${\bf Q} = 0$ superconductivity as local, isolated islands (see SM). The first order translation symmetry breaking transition between the QBP to the FFLO phase is dictated by the Clogston criterion $h \gtrsim \Delta_{0}$, where $\Delta_{0}$ is the pairing field amplitude at $h = 0$. The first order transition between the LO phases is dictated by energetics and is tantamount to solving a two-component Ginzburg-Landau energy functional $F(\Delta_{x}, \Delta_{y}) = \alpha(t_{am}, h)(\Delta_{x}^{2} + \Delta_{y}^{2}) + \beta_{1}(\Delta_{x}^{4} + \Delta_{y}^{4}) + \beta_{2}\Delta_{x}^{2}\Delta_{y}^{2} + \cdot$. Uni-directional wins if $\beta_{2} > \beta_{1}$, biaxial wins if $\beta_{1} > \beta_{2}$. Beyond a $t_{am}$ dependent upper critical 
field the FFLO phase undergoes Pauli paramagnetic pair breaking and gives way to the polarized Fermi liquid (PFL) via second order phase transition (see SM) \cite{Hu2025}.

At $T \neq 0$ the effect of an applied Zeeman field $h \hat z$ on the PDW state is encoded in Figure~\ref{fig:phase_lo} as the thermal phase diagram at a representative $t_{am} = 1.49t$. The pairing wave vectors are tracked via $S_x(\mathbf{q})$ and demarcates three FFLO regimes with the corresponding critical fields as follows: the weak field $h \lesssim h_{c1} \approx 0.25t$ regime is the cross-over from the $h = 0$ $\text{PDW 1D}$ phase to the noncollinear  LO phase with the pairing wave vector $\mathbf{Q}=(q,q)$ remaining locked to the diagonal modulation. Increasing $h$ rotates the ordering wavevector onto the crystal axes, such that the $h_{c1} < h \lesssim h_{c2} \approx 0.46t$ regime represents an incommensurate Type-I collinear LO phase with a $\mathbf{Q}=(q,0)$ pairing wave vector and $h_{c2} < h \lesssim h_{c3} \approx 0.78t$ corresponds to a commensurate Type-II collinear LO phase with 2${\bf Q}$ pairing wave vector $\mathbf{Q}_x=(\pi,0)$. While the $T_c$ stays around $0.007$--$0.008t$ at low fields, it roughly doubles to $0.015$-$0.017t$ in the $0.5 \lesssim h/t \lesssim 0.7$ window. A narrow sliver of thermal counterpart of the QBP phase is stabilized just below the $T_{c}$ in this regime \cite{Karmakar2020}. A plausible mechanism behind the enhancement of the $T_{c}$ in the collinear LO window is that the Zeeman shift partly compensates the $d$-wave altermagnetic splitting along the nodal direction leading to a gain in the low-energy pairing phase space.  At $h \gtrsim 0.8t$ the system undergoes transition to the PFL phase. 
\begin{figure}
\centering
\includegraphics[height=10cm,width=8cm,angle=0]{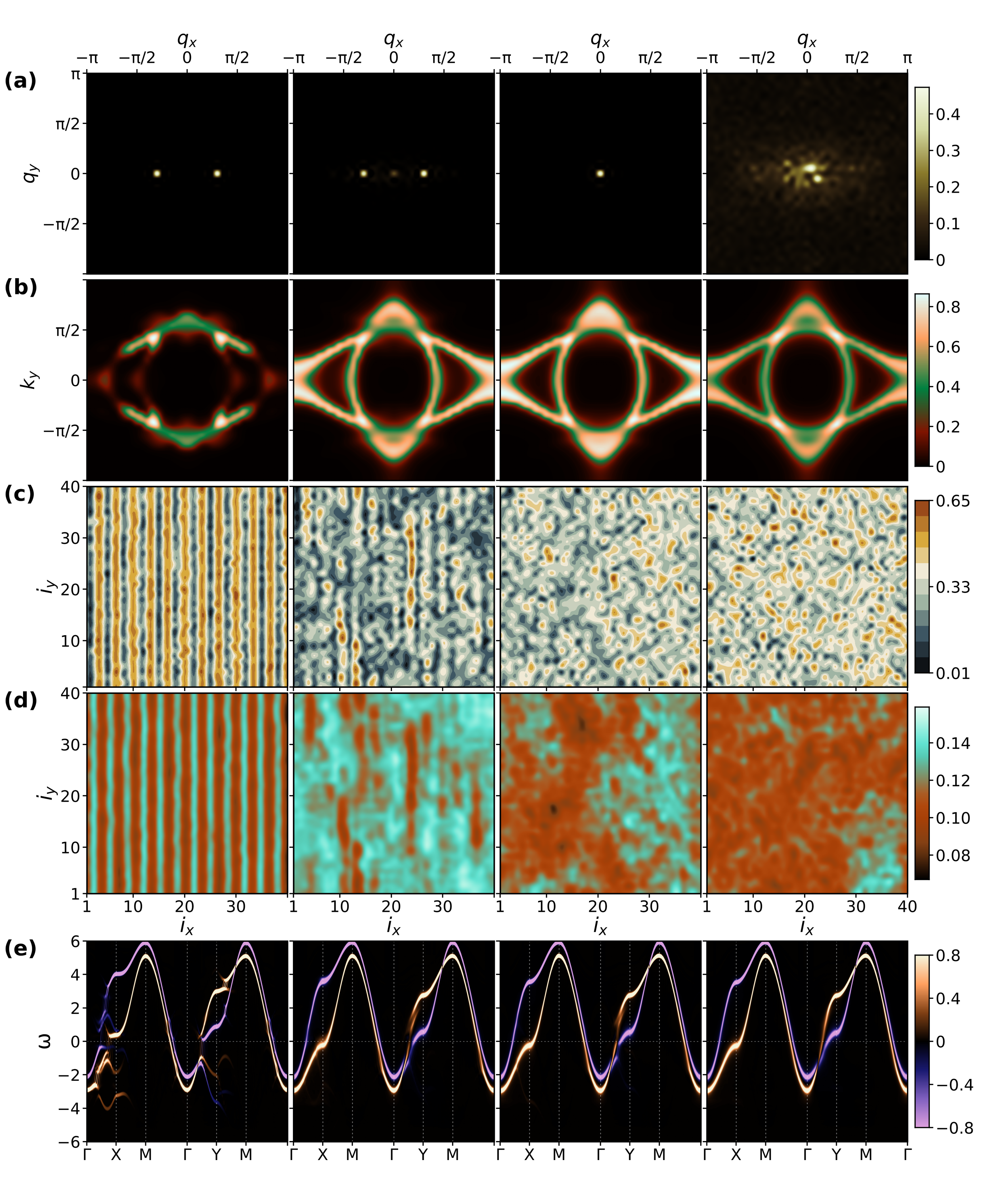}
\caption{Thermal evolution of the collinear Type-I LO phase on an $L=40$ lattice at $t_{\mathrm{am}}=1.49t$ and $h=0.40t$ for $T=0.002t$, $0.006t$, $0.008t$, and $0.012t$ (from left to right). (a) Pairing field structure factor $S_x(\mathbf{q})$, (b) zero-energy spectral weight distribution $A(\mathbf{k},0)$ mapping out the Fermi-surface reconstruction. Single-snapshot real-space maps of (c) the pairing amplitude $\vert \Delta_i\vert$ and (d) the local magnetization $m_i$. (e) Spectral function $\delta A({\bf k}, \omega) = A_\uparrow(\mathbf{k},\omega)-A_\downarrow(\mathbf{k},\omega)$ along the high-symmetry trajectory $\Gamma\text{--}X\text{--}M\text{--}\Gamma\text{--}Y\text{--}M\text{--}\Gamma$.}
\label{fig:lo}
\end{figure}

We analyze the thermodynamic and spectroscopic signatures at $t_{am} = 1.49t$, $h = 0.40t$ representative of the Type-I collinear LO phase in Figure~\ref{fig:lo}. Deep in the LO regime ($T = 0.002t$). The $S_x(\mathbf{q})$ map (Figure~\ref{fig:lo}(a)) exhibits two sharp collinear Bragg peaks along 
the $q_x$ axis at $\mathbf{Q}=(\pm q,0)$ with peak intensity $\approx 0.48$ with the corresponding real space signature being a periodic array of unidirectional stripes along $i_y$ with $\vert \Delta_i \vert \approx 0.65$ (Figure~\ref{fig:lo}(c)). The magnetization ($m_{i}$) is robust and complementary to $\vert \Delta_i \vert$ in agreement with the LO physics \cite{LO1964,Buzdin2005} (Figure~\ref{fig:lo}(d)). The spectroscopic signatures are depicted in Figure~\ref{fig:lo}(b) and Figure~\ref{fig:lo}(e) in terms of $A({\bf k}, \omega)$ and $\delta A({\bf k}, \omega)$, respectively. The pairing gaps out the antinodal segments near $(\pm\pi,0)$, leaving behind segmented Bogoliubov arcs near $(0,\pm\pi/2)$ that break $C_4$ rotational symmetry (Figure~\ref{fig:lo}(b)). 
$\delta A({\bf k}, \omega)$ undergoes a global Zeeman shift between the spin-$\uparrow$ (gold) and spin-$\downarrow$ (violet) bands superimposed on the $d$-wave altermagnetic anisotropy, together with multiple folded LO shadow replicas and anticrossings near $\omega=0$ induced by finite-$\mathbf{Q}$ quasiparticle scattering (Fig.~\ref{fig:lo}(e)). 

At $T=0.006t$, the $\mathbf{q} \neq 0$ structure-factor peaks weaken ($S_x\approx 0.13$) while a central peak begins to emerge at $\mathbf{Q}=(0,0)$, accompanied by fragmentation of the vertical stripes in $\vert \Delta_i\vert$ and $m_i$ into short-range filaments, reminiscent of the thermal QBP phase. 
Robust signatures of the thermal QBP phase is observed at $T = 0.008t$ in terms of the $\mathbf{Q}=(0,0)$ peak in $S_{x}(\bf q)$, inhomogeneous $\vert \Delta_i\vert$ complementary to the $m_i$. At $T > T_{c}$, the LO shadow bands vanish from $\delta A({\bf k}, \omega)$, and $A(\mathbf{k},0)$ shows two nested, spin-split Fermi-surface contours, as expected for a PFL. Thermodynamic and spectroscopic signatures for the 2D-LO phases are analyzed in the SM. 

\textit{Discussion and Conclusion:} ALM provides a field-free mechanism for stabilizing PDW superconductivity at finite temperatures in 2D, overcoming the conventional fragility of finite-momentum pairing in low dimensions. The intrinsic momentum dependent spin splitting in altermagnets acts as an effective $k$-space Zeeman field, selectively enhancing pairing at finite center of mass momentum while suppressing uniform superconductivity. In contrast to Zeeman field induced FFLO states, which are rapidly suppressed by thermal fluctuations, the ALM route yields a qualitatively more stable finite-momentum superconducting state.

PDW has been extensively discussed in the context of the pseudogap phase and competing orders in high-$T_{c}$ cuprates \cite{Baruch2008,nishida_jpsj2007,nishida_jpsj2013,tamegai_natcom2016,Edkins2019,Kohsaka2007,eisaki_natphys2016,Hamidian2016,Fujita2014}. For ALM materials, experimental evidence of PDW superconductivity continues to be lacking, though there are prospective candidates such as, Kagome superconductors AV$_{3}$Sb$_{5}$ ($A=K, Cs$) \cite{Jiang2021,Chen2021,Zhang2023}. Recent transport measurements on the ALM material, alkali vanadium oxychalcogenide CsV$_{2}$Se$_{2}$O brought forth a pressure induced $d$-wave SC state, originating from a density-wave parent state \cite{Li2026}. It was further shown that the high temperature phase of this system comprises a strange metal (pseudogap) phase with suppressed SC correlations. For a model Hamiltonian, the electronic band structure changes arising due to an applied pressure can be modeled in terms of effective electronic interaction. As a function of ALM interaction we define a parameter as, $U_{eff}=U/t_{am}$ such that, the effective fermionic interaction is dictated by the ALM coupling. 

PDW correlations with $d$-wave symmetry are highly susceptible to thermal fluctuations, making their stability in 2D challenging. Standard non-perturbative methods cannot access sufficiently low temperatures, while MFT overestimates PDW stability at finite $T$ (see SM). Our approach overcomes these limitations and provides quantitative estimates of thermal transition and crossover scales in a $d$-wave ALM-superconductor, establishing finite temperature PDW stability. The method has been widely applied to many-body systems, including unconventional superconductors \cite{Karmakar2016,Karmakar2018}, frustrated lattices and Mott transitions \cite{Swain2022,Karmakar2022}, flat and multiband systems \cite{Swain2020,Kunwar2024,Kunwar2026}, and altermagnetic metals \cite{Kannan2026}. It captures thermal fluctuations but neglects quantum fluctuations, reducing to MFT as $T \rightarrow 0$. The dominant low-energy fluctuations arise from the $U(1)$ phase, with lattice effects gapping translational and rotational modes, leaving $XY$-type excitations.

\textit{Acknowledgment:} M.K. would like to acknowledge the use of the high performance computing facility (AQUA) at the Indian Institute of Technology, Madras, India. MK and ANM acknowledges the support from Anusandhan National Research Foundation (ANRF), Govt. of India through the grant ANRF/ARG/2025/001620.  

\vspace{4pt}
\noindent$^{*}$\,Corresponding author: \href{mailto:madhuparna.k@gmail.com}{madhuparna.k@gmail.com}

\bibliography{references}

\begin{thebibliography}{87}%
\makeatletter
\providecommand \@ifxundefined [1]{%
 \@ifx{#1\undefined}
}%
\providecommand \@ifnum [1]{%
 \ifnum #1\expandafter \@firstoftwo
 \else \expandafter \@secondoftwo
 \fi
}%
\providecommand \@ifx [1]{%
 \ifx #1\expandafter \@firstoftwo
 \else \expandafter \@secondoftwo
 \fi
}%
\providecommand \natexlab [1]{#1}%
\providecommand \enquote  [1]{``#1''}%
\providecommand \bibnamefont  [1]{#1}%
\providecommand \bibfnamefont [1]{#1}%
\providecommand \citenamefont [1]{#1}%
\providecommand \href@noop [0]{\@secondoftwo}%
\providecommand \href [0]{\begingroup \@sanitize@url \@href}%
\providecommand \@href[1]{\@@startlink{#1}\@@href}%
\providecommand \@@href[1]{\endgroup#1\@@endlink}%
\providecommand \@sanitize@url [0]{\catcode `\\12\catcode `\$12\catcode
  `\&12\catcode `\#12\catcode `\^12\catcode `\_12\catcode `\%12\relax}%
\providecommand \@@startlink[1]{}%
\providecommand \@@endlink[0]{}%
\providecommand \url  [0]{\begingroup\@sanitize@url \@url }%
\providecommand \@url [1]{\endgroup\@href {#1}{\urlprefix }}%
\providecommand \urlprefix  [0]{URL }%
\providecommand \Eprint [0]{\href }%
\providecommand \doibase [0]{https://doi.org/}%
\providecommand \selectlanguage [0]{\@gobble}%
\providecommand \bibinfo  [0]{\@secondoftwo}%
\providecommand \bibfield  [0]{\@secondoftwo}%
\providecommand \translation [1]{[#1]}%
\providecommand \BibitemOpen [0]{}%
\providecommand \bibitemStop [0]{}%
\providecommand \bibitemNoStop [0]{.\EOS\space}%
\providecommand \EOS [0]{\spacefactor3000\relax}%
\providecommand \BibitemShut  [1]{\csname bibitem#1\endcsname}%
\let\auto@bib@innerbib\@empty
\bibitem [{\citenamefont {Fulde}\ and\ \citenamefont {Ferrell}(1964)}]{FF1964}%
  \BibitemOpen
  \bibfield  {author} {\bibinfo {author} {\bibfnamefont {P.}~\bibnamefont
  {Fulde}}\ and\ \bibinfo {author} {\bibfnamefont {R.~A.}\ \bibnamefont
  {Ferrell}},\ }\bibfield  {title} {\bibinfo {title} {Superconductivity in a
  strong spin-exchange field},\ }\href
  {https://doi.org/10.1103/PhysRev.135.A550} {\bibfield  {journal} {\bibinfo
  {journal} {Phys. Rev.}\ }\textbf {\bibinfo {volume} {135}},\ \bibinfo {pages}
  {A550} (\bibinfo {year} {1964})}\BibitemShut {NoStop}%
\bibitem [{\citenamefont {Larkin}\ and\ \citenamefont
  {Ovchinnikov}(1964)}]{LO1964}%
  \BibitemOpen
  \bibfield  {author} {\bibinfo {author} {\bibfnamefont {A.~I.}\ \bibnamefont
  {Larkin}}\ and\ \bibinfo {author} {\bibfnamefont {Y.~N.}\ \bibnamefont
  {Ovchinnikov}},\ }\bibfield  {title} {\bibinfo {title} {Nonuniform state of
  superconductors},\ }\href@noop {} {\bibfield  {journal} {\bibinfo  {journal}
  {Zh. Eksp. Teor. Fiz.}\ }\textbf {\bibinfo {volume} {47}},\ \bibinfo {pages}
  {1136} (\bibinfo {year} {1964})},\ \bibinfo {note} {[Sov. Phys. JETP
  \textbf{20}, 762 (1965)]}\BibitemShut {NoStop}%
\bibitem [{\citenamefont {Sarma}(1963)}]{Sarma1963}%
  \BibitemOpen
  \bibfield  {author} {\bibinfo {author} {\bibfnamefont {G.}~\bibnamefont
  {Sarma}},\ }\bibfield  {title} {\bibinfo {title} {On the influence of a
  uniform exchange field acting on the spins of the conduction electrons in a
  superconductor},\ }\href {https://doi.org/10.1016/0022-3697(63)90007-6}
  {\bibfield  {journal} {\bibinfo  {journal} {J. Phys. Chem. Solids}\ }\textbf
  {\bibinfo {volume} {24}},\ \bibinfo {pages} {1029} (\bibinfo {year}
  {1963})}\BibitemShut {NoStop}%
\bibitem [{\citenamefont {Casalbuoni}\ and\ \citenamefont
  {Nardulli}(2004)}]{Casalbuoni2004}%
  \BibitemOpen
  \bibfield  {author} {\bibinfo {author} {\bibfnamefont {R.}~\bibnamefont
  {Casalbuoni}}\ and\ \bibinfo {author} {\bibfnamefont {G.}~\bibnamefont
  {Nardulli}},\ }\bibfield  {title} {\bibinfo {title} {Inhomogeneous
  superconductivity in condensed matter and {QCD}},\ }\href
  {https://doi.org/10.1103/RevModPhys.76.263} {\bibfield  {journal} {\bibinfo
  {journal} {Rev. Mod. Phys.}\ }\textbf {\bibinfo {volume} {76}},\ \bibinfo
  {pages} {263} (\bibinfo {year} {2004})}\BibitemShut {NoStop}%
\bibitem [{\citenamefont {Bianchi}\ \emph {et~al.}(2003)\citenamefont
  {Bianchi}, \citenamefont {Movshovich}, \citenamefont {Capan}, \citenamefont
  {Pagliuso},\ and\ \citenamefont {Sarrao}}]{Bianchi2003}%
  \BibitemOpen
  \bibfield  {author} {\bibinfo {author} {\bibfnamefont {A.}~\bibnamefont
  {Bianchi}}, \bibinfo {author} {\bibfnamefont {R.}~\bibnamefont {Movshovich}},
  \bibinfo {author} {\bibfnamefont {C.}~\bibnamefont {Capan}}, \bibinfo
  {author} {\bibfnamefont {P.~G.}\ \bibnamefont {Pagliuso}},\ and\ \bibinfo
  {author} {\bibfnamefont {J.~L.}\ \bibnamefont {Sarrao}},\ }\bibfield  {title}
  {\bibinfo {title} {Possible {Fulde-Ferrell-Larkin-Ovchinnikov}
  superconducting state in {CeCoIn$_5$}},\ }\href
  {https://doi.org/10.1103/PhysRevLett.91.187004} {\bibfield  {journal}
  {\bibinfo  {journal} {Phys. Rev. Lett.}\ }\textbf {\bibinfo {volume} {91}},\
  \bibinfo {pages} {187004} (\bibinfo {year} {2003})}\BibitemShut {NoStop}%
\bibitem [{\citenamefont {Matsuda}\ and\ \citenamefont
  {Shimahara}(2007)}]{Matsuda2007}%
  \BibitemOpen
  \bibfield  {author} {\bibinfo {author} {\bibfnamefont {Y.}~\bibnamefont
  {Matsuda}}\ and\ \bibinfo {author} {\bibfnamefont {H.}~\bibnamefont
  {Shimahara}},\ }\bibfield  {title} {\bibinfo {title}
  {{Fulde--Ferrell--Larkin--Ovchinnikov} state in heavy fermion
  superconductors},\ }\href {https://doi.org/10.1143/JPSJ.76.051005} {\bibfield
   {journal} {\bibinfo  {journal} {J. Phys. Soc. Jpn.}\ }\textbf {\bibinfo
  {volume} {76}},\ \bibinfo {pages} {051005} (\bibinfo {year}
  {2007})}\BibitemShut {NoStop}%
\bibitem [{\citenamefont {Lortz}\ \emph {et~al.}(2007)\citenamefont {Lortz},
  \citenamefont {Wang}, \citenamefont {Demuer}, \citenamefont {B{\"o}ttger},
  \citenamefont {Bergk}, \citenamefont {Zwicknagl}, \citenamefont {Nakazawa},\
  and\ \citenamefont {Wosnitza}}]{Lortz2007}%
  \BibitemOpen
  \bibfield  {author} {\bibinfo {author} {\bibfnamefont {R.}~\bibnamefont
  {Lortz}}, \bibinfo {author} {\bibfnamefont {Y.}~\bibnamefont {Wang}},
  \bibinfo {author} {\bibfnamefont {A.}~\bibnamefont {Demuer}}, \bibinfo
  {author} {\bibfnamefont {P.~H.~M.}\ \bibnamefont {B{\"o}ttger}}, \bibinfo
  {author} {\bibfnamefont {B.}~\bibnamefont {Bergk}}, \bibinfo {author}
  {\bibfnamefont {G.}~\bibnamefont {Zwicknagl}}, \bibinfo {author}
  {\bibfnamefont {Y.}~\bibnamefont {Nakazawa}},\ and\ \bibinfo {author}
  {\bibfnamefont {J.}~\bibnamefont {Wosnitza}},\ }\bibfield  {title} {\bibinfo
  {title} {Calorimetric evidence for a {Fulde-Ferrell-Larkin-Ovchinnikov}
  superconducting state in the layered organic superconductor
  {$\kappa$-(BEDT-TTF)$_2$Cu(NCS)$_2$}},\ }\href
  {https://doi.org/10.1103/PhysRevLett.99.187002} {\bibfield  {journal}
  {\bibinfo  {journal} {Phys. Rev. Lett.}\ }\textbf {\bibinfo {volume} {99}},\
  \bibinfo {pages} {187002} (\bibinfo {year} {2007})}\BibitemShut {NoStop}%
\bibitem [{\citenamefont {Kenzelmann}\ \emph {et~al.}(2008)\citenamefont
  {Kenzelmann}, \citenamefont {Strassle}, \citenamefont {Niedermayer},
  \citenamefont {Sigrist}, \citenamefont {Padmanabhan}, \citenamefont
  {Zolliker}, \citenamefont {Bianchi}, \citenamefont {Movshovich},
  \citenamefont {Bauer}, \citenamefont {Sarrao},\ and\ \citenamefont
  {Thompson}}]{Kenzelmann2008}%
  \BibitemOpen
  \bibfield  {author} {\bibinfo {author} {\bibfnamefont {M.}~\bibnamefont
  {Kenzelmann}}, \bibinfo {author} {\bibfnamefont {T.}~\bibnamefont
  {Strassle}}, \bibinfo {author} {\bibfnamefont {C.}~\bibnamefont
  {Niedermayer}}, \bibinfo {author} {\bibfnamefont {M.}~\bibnamefont
  {Sigrist}}, \bibinfo {author} {\bibfnamefont {B.}~\bibnamefont
  {Padmanabhan}}, \bibinfo {author} {\bibfnamefont {M.}~\bibnamefont
  {Zolliker}}, \bibinfo {author} {\bibfnamefont {A.~D.}\ \bibnamefont
  {Bianchi}}, \bibinfo {author} {\bibfnamefont {R.}~\bibnamefont {Movshovich}},
  \bibinfo {author} {\bibfnamefont {E.~D.}\ \bibnamefont {Bauer}}, \bibinfo
  {author} {\bibfnamefont {J.~L.}\ \bibnamefont {Sarrao}},\ and\ \bibinfo
  {author} {\bibfnamefont {J.~D.}\ \bibnamefont {Thompson}},\ }\bibfield
  {title} {\bibinfo {title} {Coupled superconducting and magnetic order in
  {CeCoIn$_5$}},\ }\href {https://doi.org/10.1126/science.1161818} {\bibfield
  {journal} {\bibinfo  {journal} {Science}\ }\textbf {\bibinfo {volume}
  {321}},\ \bibinfo {pages} {1652} (\bibinfo {year} {2008})}\BibitemShut
  {NoStop}%
\bibitem [{\citenamefont {Mayaffre}\ \emph {et~al.}(2014)\citenamefont
  {Mayaffre}, \citenamefont {Kr{\"a}mer}, \citenamefont {Horvati{\'c}},
  \citenamefont {Berthier}, \citenamefont {Miyagawa}, \citenamefont {Kanoda},\
  and\ \citenamefont {Mitrovi{\'c}}}]{Mayaffre2014}%
  \BibitemOpen
  \bibfield  {author} {\bibinfo {author} {\bibfnamefont {H.}~\bibnamefont
  {Mayaffre}}, \bibinfo {author} {\bibfnamefont {S.}~\bibnamefont
  {Kr{\"a}mer}}, \bibinfo {author} {\bibfnamefont {M.}~\bibnamefont
  {Horvati{\'c}}}, \bibinfo {author} {\bibfnamefont {C.}~\bibnamefont
  {Berthier}}, \bibinfo {author} {\bibfnamefont {K.}~\bibnamefont {Miyagawa}},
  \bibinfo {author} {\bibfnamefont {K.}~\bibnamefont {Kanoda}},\ and\ \bibinfo
  {author} {\bibfnamefont {V.~F.}\ \bibnamefont {Mitrovi{\'c}}},\ }\bibfield
  {title} {\bibinfo {title} {Evidence of {Andreev} bound states as a hallmark
  of the {FFLO} phase in {$\kappa$-(BEDT-TTF)$_2$Cu(NCS)$_2$}},\ }\href
  {https://doi.org/10.1038/nphys3126} {\bibfield  {journal} {\bibinfo
  {journal} {Nat. Phys.}\ }\textbf {\bibinfo {volume} {10}},\ \bibinfo {pages}
  {928} (\bibinfo {year} {2014})}\BibitemShut {NoStop}%
\bibitem [{\citenamefont {Shin}\ \emph {et~al.}(2008)\citenamefont {Shin},
  \citenamefont {Schunck}, \citenamefont {Schirotzek},\ and\ \citenamefont
  {Ketterle}}]{Shin2008}%
  \BibitemOpen
  \bibfield  {author} {\bibinfo {author} {\bibfnamefont {Y.-i.}\ \bibnamefont
  {Shin}}, \bibinfo {author} {\bibfnamefont {C.~H.}\ \bibnamefont {Schunck}},
  \bibinfo {author} {\bibfnamefont {A.}~\bibnamefont {Schirotzek}},\ and\
  \bibinfo {author} {\bibfnamefont {W.}~\bibnamefont {Ketterle}},\ }\bibfield
  {title} {\bibinfo {title} {Phase diagram of a two-component {Fermi} gas with
  resonant interactions},\ }\href {https://doi.org/10.1038/nature06473}
  {\bibfield  {journal} {\bibinfo  {journal} {Nature}\ }\textbf {\bibinfo
  {volume} {451}},\ \bibinfo {pages} {689} (\bibinfo {year}
  {2008})}\BibitemShut {NoStop}%
\bibitem [{\citenamefont {Agterberg}\ \emph {et~al.}(2020)\citenamefont
  {Agterberg}, \citenamefont {Davis}, \citenamefont {Edkins}, \citenamefont
  {Fradkin}, \citenamefont {Van~Harlingen}, \citenamefont {Kivelson},
  \citenamefont {Lee}, \citenamefont {Radzihovsky}, \citenamefont {Tranquada},\
  and\ \citenamefont {Wang}}]{Agterberg2020}%
  \BibitemOpen
  \bibfield  {author} {\bibinfo {author} {\bibfnamefont {D.~F.}\ \bibnamefont
  {Agterberg}}, \bibinfo {author} {\bibfnamefont {J.~C.~S.}\ \bibnamefont
  {Davis}}, \bibinfo {author} {\bibfnamefont {S.~D.}\ \bibnamefont {Edkins}},
  \bibinfo {author} {\bibfnamefont {E.}~\bibnamefont {Fradkin}}, \bibinfo
  {author} {\bibfnamefont {D.~J.}\ \bibnamefont {Van~Harlingen}}, \bibinfo
  {author} {\bibfnamefont {S.~A.}\ \bibnamefont {Kivelson}}, \bibinfo {author}
  {\bibfnamefont {P.~A.}\ \bibnamefont {Lee}}, \bibinfo {author} {\bibfnamefont
  {L.}~\bibnamefont {Radzihovsky}}, \bibinfo {author} {\bibfnamefont {J.~M.}\
  \bibnamefont {Tranquada}},\ and\ \bibinfo {author} {\bibfnamefont
  {Y.}~\bibnamefont {Wang}},\ }\bibfield  {title} {\bibinfo {title} {The
  physics of pair-density waves: Cuprate superconductors and beyond},\ }\href
  {https://doi.org/10.1146/annurev-conmatphys-031119-050711} {\bibfield
  {journal} {\bibinfo  {journal} {Annu. Rev. Condens. Matter Phys.}\ }\textbf
  {\bibinfo {volume} {11}},\ \bibinfo {pages} {231} (\bibinfo {year}
  {2020})}\BibitemShut {NoStop}%
\bibitem [{\citenamefont {Fradkin}\ \emph {et~al.}(2015)\citenamefont
  {Fradkin}, \citenamefont {Kivelson},\ and\ \citenamefont
  {Tranquada}}]{Fradkin2015}%
  \BibitemOpen
  \bibfield  {author} {\bibinfo {author} {\bibfnamefont {E.}~\bibnamefont
  {Fradkin}}, \bibinfo {author} {\bibfnamefont {S.~A.}\ \bibnamefont
  {Kivelson}},\ and\ \bibinfo {author} {\bibfnamefont {J.~M.}\ \bibnamefont
  {Tranquada}},\ }\bibfield  {title} {\bibinfo {title} {Colloquium: Theory of
  intertwined orders in high temperature superconductors},\ }\href
  {https://doi.org/10.1103/RevModPhys.87.457} {\bibfield  {journal} {\bibinfo
  {journal} {Rev. Mod. Phys.}\ }\textbf {\bibinfo {volume} {87}},\ \bibinfo
  {pages} {457} (\bibinfo {year} {2015})}\BibitemShut {NoStop}%
\bibitem [{\citenamefont {Loder}\ \emph {et~al.}(2010)\citenamefont {Loder},
  \citenamefont {Kampf},\ and\ \citenamefont {Kopp}}]{Loder2010}%
  \BibitemOpen
  \bibfield  {author} {\bibinfo {author} {\bibfnamefont {F.}~\bibnamefont
  {Loder}}, \bibinfo {author} {\bibfnamefont {A.~P.}\ \bibnamefont {Kampf}},\
  and\ \bibinfo {author} {\bibfnamefont {T.}~\bibnamefont {Kopp}},\ }\bibfield
  {title} {\bibinfo {title} {Superconducting state with a finite-momentum
  pairing mechanism in zero external magnetic field},\ }\href
  {https://doi.org/10.1103/PhysRevB.81.020511} {\bibfield  {journal} {\bibinfo
  {journal} {Phys. Rev. B}\ }\textbf {\bibinfo {volume} {81}},\ \bibinfo
  {pages} {020511(R)} (\bibinfo {year} {2010})}\BibitemShut {NoStop}%
\bibitem [{\citenamefont {Berg}\ \emph
  {et~al.}(2009{\natexlab{a}})\citenamefont {Berg}, \citenamefont {Fradkin},
  \citenamefont {Kivelson},\ and\ \citenamefont {Tranquada}}]{Berg2009NJP}%
  \BibitemOpen
  \bibfield  {author} {\bibinfo {author} {\bibfnamefont {E.}~\bibnamefont
  {Berg}}, \bibinfo {author} {\bibfnamefont {E.}~\bibnamefont {Fradkin}},
  \bibinfo {author} {\bibfnamefont {S.~A.}\ \bibnamefont {Kivelson}},\ and\
  \bibinfo {author} {\bibfnamefont {J.~M.}\ \bibnamefont {Tranquada}},\
  }\bibfield  {title} {\bibinfo {title} {Striped superconductors: how spin,
  charge and superconducting orders intertwine in the cuprates},\ }\href
  {https://doi.org/10.1088/1367-2630/11/11/115004} {\bibfield  {journal}
  {\bibinfo  {journal} {New J. Phys.}\ }\textbf {\bibinfo {volume} {11}},\
  \bibinfo {pages} {115004} (\bibinfo {year} {2009}{\natexlab{a}})}\BibitemShut
  {NoStop}%
\bibitem [{\citenamefont {Soto-Garrido}\ and\ \citenamefont
  {Fradkin}(2014)}]{Soto2014}%
  \BibitemOpen
  \bibfield  {author} {\bibinfo {author} {\bibfnamefont {R.}~\bibnamefont
  {Soto-Garrido}}\ and\ \bibinfo {author} {\bibfnamefont {E.}~\bibnamefont
  {Fradkin}},\ }\bibfield  {title} {\bibinfo {title} {Pair-density-wave
  superconducting states and electronic liquid-crystal phases},\ }\href@noop {}
  {\bibfield  {journal} {\bibinfo  {journal} {Phys. Rev. B}\ }\textbf {\bibinfo
  {volume} {89}},\ \bibinfo {pages} {165126} (\bibinfo {year}
  {2014})}\BibitemShut {NoStop}%
\bibitem [{\citenamefont {Agterberg}\ and\ \citenamefont
  {Tsunetsugu}(2008)}]{Agterberg2008}%
  \BibitemOpen
  \bibfield  {author} {\bibinfo {author} {\bibfnamefont {D.~F.}\ \bibnamefont
  {Agterberg}}\ and\ \bibinfo {author} {\bibfnamefont {H.}~\bibnamefont
  {Tsunetsugu}},\ }\bibfield  {title} {\bibinfo {title} {Dislocations and
  vortices in pair-density-wave superconductors},\ }\href
  {https://doi.org/10.1038/nphys999} {\bibfield  {journal} {\bibinfo  {journal}
  {Nat. Phys.}\ }\textbf {\bibinfo {volume} {4}},\ \bibinfo {pages} {639}
  (\bibinfo {year} {2008})}\BibitemShut {NoStop}%
\bibitem [{\citenamefont {Berg}\ \emph
  {et~al.}(2009{\natexlab{b}})\citenamefont {Berg}, \citenamefont {Fradkin},\
  and\ \citenamefont {Kivelson}}]{Berg2009}%
  \BibitemOpen
  \bibfield  {author} {\bibinfo {author} {\bibfnamefont {E.}~\bibnamefont
  {Berg}}, \bibinfo {author} {\bibfnamefont {E.}~\bibnamefont {Fradkin}},\ and\
  \bibinfo {author} {\bibfnamefont {S.~A.}\ \bibnamefont {Kivelson}},\
  }\bibfield  {title} {\bibinfo {title} {Charge-4e superconductivity from
  pair-density-wave order in certain high-temperature superconductors},\ }\href
  {https://doi.org/10.1038/nphys1389} {\bibfield  {journal} {\bibinfo
  {journal} {Nat. Phys.}\ }\textbf {\bibinfo {volume} {5}},\ \bibinfo {pages}
  {830} (\bibinfo {year} {2009}{\natexlab{b}})}\BibitemShut {NoStop}%
\bibitem [{\citenamefont {Jiang}\ \emph {et~al.}(2021)\citenamefont {Jiang},
  \citenamefont {Yin}, \citenamefont {Denner}, \citenamefont {Shumiya},
  \citenamefont {Ortiz}, \citenamefont {Xu}, \citenamefont {Guguchia},
  \citenamefont {He}, \citenamefont {Hossain}, \citenamefont {Liu},
  \citenamefont {Ruff}, \citenamefont {Kautzsch}, \citenamefont {Zhang},
  \citenamefont {Chang}, \citenamefont {Belopolski}, \citenamefont {Zhang},
  \citenamefont {Cochran}, \citenamefont {Multer}, \citenamefont {Litskevich},
  \citenamefont {Cheng}, \citenamefont {Yang}, \citenamefont {Wang},
  \citenamefont {Thomale}, \citenamefont {Neupert}, \citenamefont {Wilson},\
  and\ \citenamefont {Hasan}}]{Jiang2021}%
  \BibitemOpen
  \bibfield  {author} {\bibinfo {author} {\bibfnamefont {Y.-X.}\ \bibnamefont
  {Jiang}}, \bibinfo {author} {\bibfnamefont {J.-X.}\ \bibnamefont {Yin}},
  \bibinfo {author} {\bibfnamefont {M.~M.}\ \bibnamefont {Denner}}, \bibinfo
  {author} {\bibfnamefont {N.}~\bibnamefont {Shumiya}}, \bibinfo {author}
  {\bibfnamefont {B.~R.}\ \bibnamefont {Ortiz}}, \bibinfo {author}
  {\bibfnamefont {G.}~\bibnamefont {Xu}}, \bibinfo {author} {\bibfnamefont
  {Z.}~\bibnamefont {Guguchia}}, \bibinfo {author} {\bibfnamefont
  {J.}~\bibnamefont {He}}, \bibinfo {author} {\bibfnamefont {M.~S.}\
  \bibnamefont {Hossain}}, \bibinfo {author} {\bibfnamefont {X.}~\bibnamefont
  {Liu}}, \bibinfo {author} {\bibfnamefont {J.}~\bibnamefont {Ruff}}, \bibinfo
  {author} {\bibfnamefont {L.}~\bibnamefont {Kautzsch}}, \bibinfo {author}
  {\bibfnamefont {S.~S.}\ \bibnamefont {Zhang}}, \bibinfo {author}
  {\bibfnamefont {G.}~\bibnamefont {Chang}}, \bibinfo {author} {\bibfnamefont
  {I.}~\bibnamefont {Belopolski}}, \bibinfo {author} {\bibfnamefont
  {Q.}~\bibnamefont {Zhang}}, \bibinfo {author} {\bibfnamefont {T.~A.}\
  \bibnamefont {Cochran}}, \bibinfo {author} {\bibfnamefont {D.}~\bibnamefont
  {Multer}}, \bibinfo {author} {\bibfnamefont {M.}~\bibnamefont {Litskevich}},
  \bibinfo {author} {\bibfnamefont {Z.-J.}\ \bibnamefont {Cheng}}, \bibinfo
  {author} {\bibfnamefont {X.~P.}\ \bibnamefont {Yang}}, \bibinfo {author}
  {\bibfnamefont {Z.}~\bibnamefont {Wang}}, \bibinfo {author} {\bibfnamefont
  {R.}~\bibnamefont {Thomale}}, \bibinfo {author} {\bibfnamefont
  {T.}~\bibnamefont {Neupert}}, \bibinfo {author} {\bibfnamefont {S.~D.}\
  \bibnamefont {Wilson}},\ and\ \bibinfo {author} {\bibfnamefont {M.~Z.}\
  \bibnamefont {Hasan}},\ }\bibfield  {title} {\bibinfo {title} {Unconventional
  chiral charge order in kagome superconductor {KV$_3$Sb$_5$}},\ }\href
  {https://doi.org/10.1038/s41563-021-01034-y} {\bibfield  {journal} {\bibinfo
  {journal} {Nat. Mater.}\ }\textbf {\bibinfo {volume} {20}},\ \bibinfo {pages}
  {1353} (\bibinfo {year} {2021})}\BibitemShut {NoStop}%
\bibitem [{\citenamefont {Chen}\ \emph {et~al.}(2021)\citenamefont {Chen},
  \citenamefont {Yang}, \citenamefont {Hu}, \citenamefont {Zhao}, \citenamefont
  {Yuan}, \citenamefont {Xing}, \citenamefont {Qian}, \citenamefont {Huang},
  \citenamefont {Li}, \citenamefont {Ye}, \citenamefont {Ma}, \citenamefont
  {Ni}, \citenamefont {Zhang}, \citenamefont {Yin}, \citenamefont {Gong},
  \citenamefont {Tu}, \citenamefont {Lei}, \citenamefont {Tan}, \citenamefont
  {Zhou}, \citenamefont {Shen}, \citenamefont {Dong}, \citenamefont {Yan},
  \citenamefont {Wang},\ and\ \citenamefont {Gao}}]{Chen2021}%
  \BibitemOpen
  \bibfield  {author} {\bibinfo {author} {\bibfnamefont {H.}~\bibnamefont
  {Chen}}, \bibinfo {author} {\bibfnamefont {H.}~\bibnamefont {Yang}}, \bibinfo
  {author} {\bibfnamefont {B.}~\bibnamefont {Hu}}, \bibinfo {author}
  {\bibfnamefont {Z.}~\bibnamefont {Zhao}}, \bibinfo {author} {\bibfnamefont
  {J.}~\bibnamefont {Yuan}}, \bibinfo {author} {\bibfnamefont {Y.}~\bibnamefont
  {Xing}}, \bibinfo {author} {\bibfnamefont {G.}~\bibnamefont {Qian}}, \bibinfo
  {author} {\bibfnamefont {Z.}~\bibnamefont {Huang}}, \bibinfo {author}
  {\bibfnamefont {G.}~\bibnamefont {Li}}, \bibinfo {author} {\bibfnamefont
  {Y.}~\bibnamefont {Ye}}, \bibinfo {author} {\bibfnamefont {S.}~\bibnamefont
  {Ma}}, \bibinfo {author} {\bibfnamefont {S.}~\bibnamefont {Ni}}, \bibinfo
  {author} {\bibfnamefont {H.}~\bibnamefont {Zhang}}, \bibinfo {author}
  {\bibfnamefont {Q.}~\bibnamefont {Yin}}, \bibinfo {author} {\bibfnamefont
  {C.}~\bibnamefont {Gong}}, \bibinfo {author} {\bibfnamefont {Z.}~\bibnamefont
  {Tu}}, \bibinfo {author} {\bibfnamefont {H.}~\bibnamefont {Lei}}, \bibinfo
  {author} {\bibfnamefont {H.}~\bibnamefont {Tan}}, \bibinfo {author}
  {\bibfnamefont {S.}~\bibnamefont {Zhou}}, \bibinfo {author} {\bibfnamefont
  {C.}~\bibnamefont {Shen}}, \bibinfo {author} {\bibfnamefont {X.}~\bibnamefont
  {Dong}}, \bibinfo {author} {\bibfnamefont {B.}~\bibnamefont {Yan}}, \bibinfo
  {author} {\bibfnamefont {Z.}~\bibnamefont {Wang}},\ and\ \bibinfo {author}
  {\bibfnamefont {H.-J.}\ \bibnamefont {Gao}},\ }\bibfield  {title} {\bibinfo
  {title} {Roton pair density wave in a strong-coupling kagome
  superconductor},\ }\href {https://doi.org/10.1038/s41586-021-03983-5}
  {\bibfield  {journal} {\bibinfo  {journal} {Nature}\ }\textbf {\bibinfo
  {volume} {599}},\ \bibinfo {pages} {222} (\bibinfo {year}
  {2021})}\BibitemShut {NoStop}%
\bibitem [{\citenamefont {Gu}\ \emph {et~al.}(2023)\citenamefont {Gu},
  \citenamefont {Carroll}, \citenamefont {Wang}, \citenamefont {Ran},
  \citenamefont {Broyles}, \citenamefont {Siddiquee}, \citenamefont {Butch},
  \citenamefont {Saha}, \citenamefont {Paglione}, \citenamefont {Davis},\ and\
  \citenamefont {Liu}}]{Gu2023}%
  \BibitemOpen
  \bibfield  {author} {\bibinfo {author} {\bibfnamefont {Q.}~\bibnamefont
  {Gu}}, \bibinfo {author} {\bibfnamefont {J.~P.}\ \bibnamefont {Carroll}},
  \bibinfo {author} {\bibfnamefont {S.}~\bibnamefont {Wang}}, \bibinfo {author}
  {\bibfnamefont {S.}~\bibnamefont {Ran}}, \bibinfo {author} {\bibfnamefont
  {C.}~\bibnamefont {Broyles}}, \bibinfo {author} {\bibfnamefont
  {H.}~\bibnamefont {Siddiquee}}, \bibinfo {author} {\bibfnamefont {N.~P.}\
  \bibnamefont {Butch}}, \bibinfo {author} {\bibfnamefont {S.~R.}\ \bibnamefont
  {Saha}}, \bibinfo {author} {\bibfnamefont {J.}~\bibnamefont {Paglione}},
  \bibinfo {author} {\bibfnamefont {J.~C.~S.}\ \bibnamefont {Davis}},\ and\
  \bibinfo {author} {\bibfnamefont {X.}~\bibnamefont {Liu}},\ }\bibfield
  {title} {\bibinfo {title} {Detection of a pair density wave state in
  {UTe$_2$}},\ }\href {https://doi.org/10.1038/s41586-023-05919-7} {\bibfield
  {journal} {\bibinfo  {journal} {Nature}\ }\textbf {\bibinfo {volume} {618}},\
  \bibinfo {pages} {921} (\bibinfo {year} {2023})}\BibitemShut {NoStop}%
\bibitem [{\citenamefont {Aishwarya}\ \emph {et~al.}(2023)\citenamefont
  {Aishwarya}, \citenamefont {May-Mann}, \citenamefont {Raghavan},
  \citenamefont {Nie}, \citenamefont {Romanelli}, \citenamefont {Ran},
  \citenamefont {Saha}, \citenamefont {Paglione}, \citenamefont {Butch},
  \citenamefont {Fradkin},\ and\ \citenamefont {Madhavan}}]{Aishwarya2023}%
  \BibitemOpen
  \bibfield  {author} {\bibinfo {author} {\bibfnamefont {A.}~\bibnamefont
  {Aishwarya}}, \bibinfo {author} {\bibfnamefont {J.}~\bibnamefont {May-Mann}},
  \bibinfo {author} {\bibfnamefont {A.}~\bibnamefont {Raghavan}}, \bibinfo
  {author} {\bibfnamefont {L.}~\bibnamefont {Nie}}, \bibinfo {author}
  {\bibfnamefont {M.}~\bibnamefont {Romanelli}}, \bibinfo {author}
  {\bibfnamefont {S.}~\bibnamefont {Ran}}, \bibinfo {author} {\bibfnamefont
  {S.~R.}\ \bibnamefont {Saha}}, \bibinfo {author} {\bibfnamefont
  {J.}~\bibnamefont {Paglione}}, \bibinfo {author} {\bibfnamefont {N.~P.}\
  \bibnamefont {Butch}}, \bibinfo {author} {\bibfnamefont {E.}~\bibnamefont
  {Fradkin}},\ and\ \bibinfo {author} {\bibfnamefont {V.}~\bibnamefont
  {Madhavan}},\ }\bibfield  {title} {\bibinfo {title} {Magnetic-field-sensitive
  charge density waves in the superconductor {UTe$_2$}},\ }\href
  {https://doi.org/10.1038/s41586-023-06005-8} {\bibfield  {journal} {\bibinfo
  {journal} {Nature}\ }\textbf {\bibinfo {volume} {618}},\ \bibinfo {pages}
  {928} (\bibinfo {year} {2023})}\BibitemShut {NoStop}%
\bibitem [{\citenamefont {Zhao}\ \emph {et~al.}(2023)\citenamefont {Zhao},
  \citenamefont {Blackwell}, \citenamefont {Thinel}, \citenamefont {Handa},
  \citenamefont {Ishida}, \citenamefont {Zhu}, \citenamefont {Iyo},
  \citenamefont {Eisaki}, \citenamefont {Pasupathy},\ and\ \citenamefont
  {Fujita}}]{Zhao2023}%
  \BibitemOpen
  \bibfield  {author} {\bibinfo {author} {\bibfnamefont {H.}~\bibnamefont
  {Zhao}}, \bibinfo {author} {\bibfnamefont {R.}~\bibnamefont {Blackwell}},
  \bibinfo {author} {\bibfnamefont {M.}~\bibnamefont {Thinel}}, \bibinfo
  {author} {\bibfnamefont {T.}~\bibnamefont {Handa}}, \bibinfo {author}
  {\bibfnamefont {S.}~\bibnamefont {Ishida}}, \bibinfo {author} {\bibfnamefont
  {X.}~\bibnamefont {Zhu}}, \bibinfo {author} {\bibfnamefont {A.}~\bibnamefont
  {Iyo}}, \bibinfo {author} {\bibfnamefont {H.}~\bibnamefont {Eisaki}},
  \bibinfo {author} {\bibfnamefont {A.~N.}\ \bibnamefont {Pasupathy}},\ and\
  \bibinfo {author} {\bibfnamefont {K.}~\bibnamefont {Fujita}},\ }\bibfield
  {title} {\bibinfo {title} {Smectic pair-density-wave order in
  {EuRbFe$_4$As$_4$}},\ }\href {https://doi.org/10.1038/s41586-023-06103-7}
  {\bibfield  {journal} {\bibinfo  {journal} {Nature}\ }\textbf {\bibinfo
  {volume} {618}},\ \bibinfo {pages} {940} (\bibinfo {year}
  {2023})}\BibitemShut {NoStop}%
\bibitem [{\citenamefont {{\v{S}}mejkal}\ \emph
  {et~al.}(2022{\natexlab{a}})\citenamefont {{\v{S}}mejkal}, \citenamefont
  {Sinova},\ and\ \citenamefont {Jungwirth}}]{Smejkal2022beyond}%
  \BibitemOpen
  \bibfield  {author} {\bibinfo {author} {\bibfnamefont {L.}~\bibnamefont
  {{\v{S}}mejkal}}, \bibinfo {author} {\bibfnamefont {J.}~\bibnamefont
  {Sinova}},\ and\ \bibinfo {author} {\bibfnamefont {T.}~\bibnamefont
  {Jungwirth}},\ }\bibfield  {title} {\bibinfo {title} {Beyond conventional
  ferromagnetism and antiferromagnetism: A phase with nonrelativistic spin and
  crystal rotation symmetry},\ }\href
  {https://doi.org/10.1103/PhysRevX.12.031042} {\bibfield  {journal} {\bibinfo
  {journal} {Phys. Rev. X}\ }\textbf {\bibinfo {volume} {12}},\ \bibinfo
  {pages} {031042} (\bibinfo {year} {2022}{\natexlab{a}})}\BibitemShut
  {NoStop}%
\bibitem [{\citenamefont {{\v{S}}mejkal}\ \emph
  {et~al.}(2022{\natexlab{b}})\citenamefont {{\v{S}}mejkal}, \citenamefont
  {Sinova},\ and\ \citenamefont {Jungwirth}}]{Smejkal2022emerging}%
  \BibitemOpen
  \bibfield  {author} {\bibinfo {author} {\bibfnamefont {L.}~\bibnamefont
  {{\v{S}}mejkal}}, \bibinfo {author} {\bibfnamefont {J.}~\bibnamefont
  {Sinova}},\ and\ \bibinfo {author} {\bibfnamefont {T.}~\bibnamefont
  {Jungwirth}},\ }\bibfield  {title} {\bibinfo {title} {Emerging research
  landscape of altermagnetism},\ }\href
  {https://doi.org/10.1103/PhysRevX.12.040501} {\bibfield  {journal} {\bibinfo
  {journal} {Phys. Rev. X}\ }\textbf {\bibinfo {volume} {12}},\ \bibinfo
  {pages} {040501} (\bibinfo {year} {2022}{\natexlab{b}})}\BibitemShut
  {NoStop}%
\bibitem [{\citenamefont {Song}\ \emph {et~al.}(2025)\citenamefont {Song},
  \citenamefont {Bai}, \citenamefont {Zhou}, \citenamefont {Han}, \citenamefont
  {Reichlova}, \citenamefont {Dil}, \citenamefont {Liu}, \citenamefont {Chen},\
  and\ \citenamefont {Pan}}]{Song2025}%
  \BibitemOpen
  \bibfield  {author} {\bibinfo {author} {\bibfnamefont {C.}~\bibnamefont
  {Song}}, \bibinfo {author} {\bibfnamefont {H.}~\bibnamefont {Bai}}, \bibinfo
  {author} {\bibfnamefont {Z.}~\bibnamefont {Zhou}}, \bibinfo {author}
  {\bibfnamefont {L.}~\bibnamefont {Han}}, \bibinfo {author} {\bibfnamefont
  {H.}~\bibnamefont {Reichlova}}, \bibinfo {author} {\bibfnamefont {J.~H.}\
  \bibnamefont {Dil}}, \bibinfo {author} {\bibfnamefont {J.}~\bibnamefont
  {Liu}}, \bibinfo {author} {\bibfnamefont {X.}~\bibnamefont {Chen}},\ and\
  \bibinfo {author} {\bibfnamefont {F.}~\bibnamefont {Pan}},\ }\bibfield
  {title} {\bibinfo {title} {Altermagnets as a new class of functional
  materials},\ }\href {https://doi.org/10.1038/s41578-025-00779-1} {\bibfield
  {journal} {\bibinfo  {journal} {Nat. Rev. Mater.}\ }\textbf {\bibinfo
  {volume} {10}},\ \bibinfo {pages} {473} (\bibinfo {year} {2025})}\BibitemShut
  {NoStop}%
\bibitem [{\citenamefont {Jungwirth}\ \emph {et~al.}(2025)\citenamefont
  {Jungwirth}, \citenamefont {Sinova}, \citenamefont {Fernandes}, \citenamefont
  {Liu}, \citenamefont {Watanabe}, \citenamefont {Murakami}, \citenamefont
  {Nakatsuji},\ and\ \citenamefont {{\v{S}}mejkal}}]{Jungwirth2025}%
  \BibitemOpen
  \bibfield  {author} {\bibinfo {author} {\bibfnamefont {T.}~\bibnamefont
  {Jungwirth}}, \bibinfo {author} {\bibfnamefont {J.}~\bibnamefont {Sinova}},
  \bibinfo {author} {\bibfnamefont {R.~M.}\ \bibnamefont {Fernandes}}, \bibinfo
  {author} {\bibfnamefont {Q.}~\bibnamefont {Liu}}, \bibinfo {author}
  {\bibfnamefont {H.}~\bibnamefont {Watanabe}}, \bibinfo {author}
  {\bibfnamefont {S.}~\bibnamefont {Murakami}}, \bibinfo {author}
  {\bibfnamefont {S.}~\bibnamefont {Nakatsuji}},\ and\ \bibinfo {author}
  {\bibfnamefont {L.}~\bibnamefont {{\v{S}}mejkal}},\ }\href@noop {} {\bibinfo
  {title} {Symmetry, microscopy and spectroscopy signatures of altermagnetism}}
  (\bibinfo {year} {2025}),\ \Eprint {https://arxiv.org/abs/2506.22860}
  {arXiv:2506.22860 [cond-mat.mtrl-sci]} \BibitemShut {NoStop}%
\bibitem [{\citenamefont {Krempask{\'y}}\ \emph {et~al.}(2024)\citenamefont
  {Krempask{\'y}}, \citenamefont {{\v{S}}mejkal}, \citenamefont {D'Souza},
  \citenamefont {Hajlaoui}, \citenamefont {Springholz}, \citenamefont
  {Uhl{\'\i}{\v{r}}ov{\'a}}, \citenamefont {Alarab}, \citenamefont
  {Constantinou}, \citenamefont {Strocov}, \citenamefont {Usanov},
  \citenamefont {Pudelko}, \citenamefont {Gonz{\'a}lez-Hern{\'a}ndez},
  \citenamefont {Birk~Hellenes}, \citenamefont {Jansa}, \citenamefont
  {Reichlov{\'a}}, \citenamefont {{\v{S}}ob{\'a}{\v{n}}}, \citenamefont
  {Gonzalez~Betancourt}, \citenamefont {Wadley}, \citenamefont {Sinova},
  \citenamefont {Kriegner}, \citenamefont {Min{\'a}r}, \citenamefont {Dil},\
  and\ \citenamefont {Jungwirth}}]{Krempasky2024}%
  \BibitemOpen
  \bibfield  {author} {\bibinfo {author} {\bibfnamefont {J.}~\bibnamefont
  {Krempask{\'y}}}, \bibinfo {author} {\bibfnamefont {L.}~\bibnamefont
  {{\v{S}}mejkal}}, \bibinfo {author} {\bibfnamefont {S.~W.}\ \bibnamefont
  {D'Souza}}, \bibinfo {author} {\bibfnamefont {M.}~\bibnamefont {Hajlaoui}},
  \bibinfo {author} {\bibfnamefont {G.}~\bibnamefont {Springholz}}, \bibinfo
  {author} {\bibfnamefont {K.}~\bibnamefont {Uhl{\'\i}{\v{r}}ov{\'a}}},
  \bibinfo {author} {\bibfnamefont {F.}~\bibnamefont {Alarab}}, \bibinfo
  {author} {\bibfnamefont {P.~C.}\ \bibnamefont {Constantinou}}, \bibinfo
  {author} {\bibfnamefont {V.}~\bibnamefont {Strocov}}, \bibinfo {author}
  {\bibfnamefont {D.}~\bibnamefont {Usanov}}, \bibinfo {author} {\bibfnamefont
  {W.~R.}\ \bibnamefont {Pudelko}}, \bibinfo {author} {\bibfnamefont
  {R.}~\bibnamefont {Gonz{\'a}lez-Hern{\'a}ndez}}, \bibinfo {author}
  {\bibfnamefont {A.}~\bibnamefont {Birk~Hellenes}}, \bibinfo {author}
  {\bibfnamefont {Z.}~\bibnamefont {Jansa}}, \bibinfo {author} {\bibfnamefont
  {H.}~\bibnamefont {Reichlov{\'a}}}, \bibinfo {author} {\bibfnamefont
  {Z.}~\bibnamefont {{\v{S}}ob{\'a}{\v{n}}}}, \bibinfo {author} {\bibfnamefont
  {R.~D.}\ \bibnamefont {Gonzalez~Betancourt}}, \bibinfo {author}
  {\bibfnamefont {P.}~\bibnamefont {Wadley}}, \bibinfo {author} {\bibfnamefont
  {J.}~\bibnamefont {Sinova}}, \bibinfo {author} {\bibfnamefont
  {D.}~\bibnamefont {Kriegner}}, \bibinfo {author} {\bibfnamefont
  {J.}~\bibnamefont {Min{\'a}r}}, \bibinfo {author} {\bibfnamefont {J.~H.}\
  \bibnamefont {Dil}},\ and\ \bibinfo {author} {\bibfnamefont {T.}~\bibnamefont
  {Jungwirth}},\ }\bibfield  {title} {\bibinfo {title} {Altermagnetic lifting
  of {Kramers} spin degeneracy},\ }\href
  {https://doi.org/10.1038/s41586-023-06907-7} {\bibfield  {journal} {\bibinfo
  {journal} {Nature}\ }\textbf {\bibinfo {volume} {626}},\ \bibinfo {pages}
  {517} (\bibinfo {year} {2024})}\BibitemShut {NoStop}%
\bibitem [{\citenamefont {Lee}\ \emph {et~al.}(2024)\citenamefont {Lee},
  \citenamefont {Lee}, \citenamefont {Jung}, \citenamefont {Jung},
  \citenamefont {Kim}, \citenamefont {Lee}, \citenamefont {Seok}, \citenamefont
  {Kim}, \citenamefont {Park}, \citenamefont {{\v{S}}mejkal}, \citenamefont
  {Kang},\ and\ \citenamefont {Kim}}]{Lee2024}%
  \BibitemOpen
  \bibfield  {author} {\bibinfo {author} {\bibfnamefont {S.}~\bibnamefont
  {Lee}}, \bibinfo {author} {\bibfnamefont {S.}~\bibnamefont {Lee}}, \bibinfo
  {author} {\bibfnamefont {S.}~\bibnamefont {Jung}}, \bibinfo {author}
  {\bibfnamefont {J.}~\bibnamefont {Jung}}, \bibinfo {author} {\bibfnamefont
  {D.}~\bibnamefont {Kim}}, \bibinfo {author} {\bibfnamefont {Y.}~\bibnamefont
  {Lee}}, \bibinfo {author} {\bibfnamefont {B.}~\bibnamefont {Seok}}, \bibinfo
  {author} {\bibfnamefont {J.}~\bibnamefont {Kim}}, \bibinfo {author}
  {\bibfnamefont {B.~G.}\ \bibnamefont {Park}}, \bibinfo {author}
  {\bibfnamefont {L.}~\bibnamefont {{\v{S}}mejkal}}, \bibinfo {author}
  {\bibfnamefont {C.-J.}\ \bibnamefont {Kang}},\ and\ \bibinfo {author}
  {\bibfnamefont {C.}~\bibnamefont {Kim}},\ }\bibfield  {title} {\bibinfo
  {title} {Broken {Kramers} degeneracy in altermagnetic {MnTe}},\ }\href
  {https://doi.org/10.1103/PhysRevLett.132.036702} {\bibfield  {journal}
  {\bibinfo  {journal} {Phys. Rev. Lett.}\ }\textbf {\bibinfo {volume} {132}},\
  \bibinfo {pages} {036702} (\bibinfo {year} {2024})}\BibitemShut {NoStop}%
\bibitem [{\citenamefont {Osumi}\ \emph {et~al.}(2024)\citenamefont {Osumi},
  \citenamefont {Souma}, \citenamefont {Aoyama}, \citenamefont {Yamauchi},
  \citenamefont {Honma}, \citenamefont {Nakayama}, \citenamefont {Takahashi},
  \citenamefont {Ohgushi},\ and\ \citenamefont {Sato}}]{Osumi2024}%
  \BibitemOpen
  \bibfield  {author} {\bibinfo {author} {\bibfnamefont {T.}~\bibnamefont
  {Osumi}}, \bibinfo {author} {\bibfnamefont {S.}~\bibnamefont {Souma}},
  \bibinfo {author} {\bibfnamefont {T.}~\bibnamefont {Aoyama}}, \bibinfo
  {author} {\bibfnamefont {K.}~\bibnamefont {Yamauchi}}, \bibinfo {author}
  {\bibfnamefont {A.}~\bibnamefont {Honma}}, \bibinfo {author} {\bibfnamefont
  {K.}~\bibnamefont {Nakayama}}, \bibinfo {author} {\bibfnamefont
  {T.}~\bibnamefont {Takahashi}}, \bibinfo {author} {\bibfnamefont
  {K.}~\bibnamefont {Ohgushi}},\ and\ \bibinfo {author} {\bibfnamefont
  {T.}~\bibnamefont {Sato}},\ }\bibfield  {title} {\bibinfo {title}
  {Observation of a giant band splitting in altermagnetic {MnTe}},\ }\href
  {https://doi.org/10.1103/PhysRevB.109.115102} {\bibfield  {journal} {\bibinfo
   {journal} {Phys. Rev. B}\ }\textbf {\bibinfo {volume} {109}},\ \bibinfo
  {pages} {115102} (\bibinfo {year} {2024})}\BibitemShut {NoStop}%
\bibitem [{\citenamefont {Ding}\ \emph {et~al.}(2024)\citenamefont {Ding},
  \citenamefont {Jiang}, \citenamefont {Chen}, \citenamefont {Tao},
  \citenamefont {Liu}, \citenamefont {Li}, \citenamefont {Liu}, \citenamefont
  {Sun}, \citenamefont {Cheng}, \citenamefont {Liu}, \citenamefont {Yang},
  \citenamefont {Zhang}, \citenamefont {Deng}, \citenamefont {Jing},
  \citenamefont {Huang}, \citenamefont {Shi}, \citenamefont {Ye}, \citenamefont
  {Qiao}, \citenamefont {Wang}, \citenamefont {Guo}, \citenamefont {Feng},\
  and\ \citenamefont {Shen}}]{Ding2024}%
  \BibitemOpen
  \bibfield  {author} {\bibinfo {author} {\bibfnamefont {J.}~\bibnamefont
  {Ding}}, \bibinfo {author} {\bibfnamefont {Z.}~\bibnamefont {Jiang}},
  \bibinfo {author} {\bibfnamefont {X.}~\bibnamefont {Chen}}, \bibinfo {author}
  {\bibfnamefont {Z.}~\bibnamefont {Tao}}, \bibinfo {author} {\bibfnamefont
  {Z.}~\bibnamefont {Liu}}, \bibinfo {author} {\bibfnamefont {T.}~\bibnamefont
  {Li}}, \bibinfo {author} {\bibfnamefont {J.}~\bibnamefont {Liu}}, \bibinfo
  {author} {\bibfnamefont {J.}~\bibnamefont {Sun}}, \bibinfo {author}
  {\bibfnamefont {J.}~\bibnamefont {Cheng}}, \bibinfo {author} {\bibfnamefont
  {J.}~\bibnamefont {Liu}}, \bibinfo {author} {\bibfnamefont {Y.}~\bibnamefont
  {Yang}}, \bibinfo {author} {\bibfnamefont {R.}~\bibnamefont {Zhang}},
  \bibinfo {author} {\bibfnamefont {L.}~\bibnamefont {Deng}}, \bibinfo {author}
  {\bibfnamefont {W.}~\bibnamefont {Jing}}, \bibinfo {author} {\bibfnamefont
  {Y.}~\bibnamefont {Huang}}, \bibinfo {author} {\bibfnamefont
  {Y.}~\bibnamefont {Shi}}, \bibinfo {author} {\bibfnamefont {M.}~\bibnamefont
  {Ye}}, \bibinfo {author} {\bibfnamefont {S.}~\bibnamefont {Qiao}}, \bibinfo
  {author} {\bibfnamefont {Y.}~\bibnamefont {Wang}}, \bibinfo {author}
  {\bibfnamefont {Y.}~\bibnamefont {Guo}}, \bibinfo {author} {\bibfnamefont
  {D.}~\bibnamefont {Feng}},\ and\ \bibinfo {author} {\bibfnamefont
  {D.}~\bibnamefont {Shen}},\ }\bibfield  {title} {\bibinfo {title} {Large band
  splitting in $g$-wave altermagnet {CrSb}},\ }\href
  {https://doi.org/10.1103/PhysRevLett.133.206401} {\bibfield  {journal}
  {\bibinfo  {journal} {Phys. Rev. Lett.}\ }\textbf {\bibinfo {volume} {133}},\
  \bibinfo {pages} {206401} (\bibinfo {year} {2024})}\BibitemShut {NoStop}%
\bibitem [{\citenamefont {Jiang}\ \emph {et~al.}(2025)\citenamefont {Jiang},
  \citenamefont {Hu}, \citenamefont {Bai}, \citenamefont {Song}, \citenamefont
  {Mu}, \citenamefont {Qu}, \citenamefont {Li}, \citenamefont {Zhu},
  \citenamefont {Pi}, \citenamefont {Wei}, \citenamefont {Sun}, \citenamefont
  {Huang}, \citenamefont {Zheng}, \citenamefont {Peng}, \citenamefont {He},
  \citenamefont {Li}, \citenamefont {Luo}, \citenamefont {Li}, \citenamefont
  {Chen}, \citenamefont {Li}, \citenamefont {Weng},\ and\ \citenamefont
  {Qian}}]{Jiang2025}%
  \BibitemOpen
  \bibfield  {author} {\bibinfo {author} {\bibfnamefont {B.}~\bibnamefont
  {Jiang}}, \bibinfo {author} {\bibfnamefont {M.}~\bibnamefont {Hu}}, \bibinfo
  {author} {\bibfnamefont {J.}~\bibnamefont {Bai}}, \bibinfo {author}
  {\bibfnamefont {Z.}~\bibnamefont {Song}}, \bibinfo {author} {\bibfnamefont
  {C.}~\bibnamefont {Mu}}, \bibinfo {author} {\bibfnamefont {G.}~\bibnamefont
  {Qu}}, \bibinfo {author} {\bibfnamefont {W.}~\bibnamefont {Li}}, \bibinfo
  {author} {\bibfnamefont {W.}~\bibnamefont {Zhu}}, \bibinfo {author}
  {\bibfnamefont {H.}~\bibnamefont {Pi}}, \bibinfo {author} {\bibfnamefont
  {Z.}~\bibnamefont {Wei}}, \bibinfo {author} {\bibfnamefont {Y.}~\bibnamefont
  {Sun}}, \bibinfo {author} {\bibfnamefont {Y.}~\bibnamefont {Huang}}, \bibinfo
  {author} {\bibfnamefont {X.}~\bibnamefont {Zheng}}, \bibinfo {author}
  {\bibfnamefont {Y.}~\bibnamefont {Peng}}, \bibinfo {author} {\bibfnamefont
  {L.}~\bibnamefont {He}}, \bibinfo {author} {\bibfnamefont {S.}~\bibnamefont
  {Li}}, \bibinfo {author} {\bibfnamefont {J.}~\bibnamefont {Luo}}, \bibinfo
  {author} {\bibfnamefont {Z.}~\bibnamefont {Li}}, \bibinfo {author}
  {\bibfnamefont {G.}~\bibnamefont {Chen}}, \bibinfo {author} {\bibfnamefont
  {H.}~\bibnamefont {Li}}, \bibinfo {author} {\bibfnamefont {H.}~\bibnamefont
  {Weng}},\ and\ \bibinfo {author} {\bibfnamefont {T.}~\bibnamefont {Qian}},\
  }\bibfield  {title} {\bibinfo {title} {A metallic room-temperature $d$-wave
  altermagnet},\ }\href {https://doi.org/10.1038/s41567-025-02822-y} {\bibfield
   {journal} {\bibinfo  {journal} {Nat. Phys.}\ }\textbf {\bibinfo {volume}
  {21}},\ \bibinfo {pages} {754} (\bibinfo {year} {2025})}\BibitemShut
  {NoStop}%
\bibitem [{\citenamefont {Sun}\ \emph {et~al.}(2023)\citenamefont {Sun},
  \citenamefont {Brataas},\ and\ \citenamefont {Linder}}]{Sun2023}%
  \BibitemOpen
  \bibfield  {author} {\bibinfo {author} {\bibfnamefont {C.}~\bibnamefont
  {Sun}}, \bibinfo {author} {\bibfnamefont {A.}~\bibnamefont {Brataas}},\ and\
  \bibinfo {author} {\bibfnamefont {J.}~\bibnamefont {Linder}},\ }\bibfield
  {title} {\bibinfo {title} {Andreev reflection in altermagnets},\ }\href
  {https://doi.org/10.1103/PhysRevB.108.054511} {\bibfield  {journal} {\bibinfo
   {journal} {Phys. Rev. B}\ }\textbf {\bibinfo {volume} {108}},\ \bibinfo
  {pages} {054511} (\bibinfo {year} {2023})}\BibitemShut {NoStop}%
\bibitem [{\citenamefont {Ouassou}\ \emph {et~al.}(2023)\citenamefont
  {Ouassou}, \citenamefont {Brataas},\ and\ \citenamefont
  {Linder}}]{Ouassou2023}%
  \BibitemOpen
  \bibfield  {author} {\bibinfo {author} {\bibfnamefont {J.~A.}\ \bibnamefont
  {Ouassou}}, \bibinfo {author} {\bibfnamefont {A.}~\bibnamefont {Brataas}},\
  and\ \bibinfo {author} {\bibfnamefont {J.}~\bibnamefont {Linder}},\
  }\bibfield  {title} {\bibinfo {title} {{dc} {Josephson} effect in
  altermagnets},\ }\href {https://doi.org/10.1103/PhysRevLett.131.076003}
  {\bibfield  {journal} {\bibinfo  {journal} {Phys. Rev. Lett.}\ }\textbf
  {\bibinfo {volume} {131}},\ \bibinfo {pages} {076003} (\bibinfo {year}
  {2023})}\BibitemShut {NoStop}%
\bibitem [{\citenamefont {Lu}\ \emph {et~al.}(2024)\citenamefont {Lu},
  \citenamefont {Maeda}, \citenamefont {Ito}, \citenamefont {Yada},\ and\
  \citenamefont {Tanaka}}]{Lu2024}%
  \BibitemOpen
  \bibfield  {author} {\bibinfo {author} {\bibfnamefont {B.}~\bibnamefont
  {Lu}}, \bibinfo {author} {\bibfnamefont {K.}~\bibnamefont {Maeda}}, \bibinfo
  {author} {\bibfnamefont {H.}~\bibnamefont {Ito}}, \bibinfo {author}
  {\bibfnamefont {K.}~\bibnamefont {Yada}},\ and\ \bibinfo {author}
  {\bibfnamefont {Y.}~\bibnamefont {Tanaka}},\ }\bibfield  {title} {\bibinfo
  {title} {$\varphi$ {Josephson} junction induced by altermagnetism},\
  }\href@noop {} {\bibfield  {journal} {\bibinfo  {journal} {Phys. Rev. Lett.}\
  }\textbf {\bibinfo {volume} {133}},\ \bibinfo {pages} {226002} (\bibinfo
  {year} {2024})}\BibitemShut {NoStop}%
\bibitem [{\citenamefont {Maeda}\ \emph {et~al.}(2025)\citenamefont {Maeda},
  \citenamefont {Fukaya}, \citenamefont {Yada}, \citenamefont {Lu},
  \citenamefont {Tanaka},\ and\ \citenamefont {Cayao}}]{Maeda2025}%
  \BibitemOpen
  \bibfield  {author} {\bibinfo {author} {\bibfnamefont {K.}~\bibnamefont
  {Maeda}}, \bibinfo {author} {\bibfnamefont {Y.}~\bibnamefont {Fukaya}},
  \bibinfo {author} {\bibfnamefont {K.}~\bibnamefont {Yada}}, \bibinfo {author}
  {\bibfnamefont {B.}~\bibnamefont {Lu}}, \bibinfo {author} {\bibfnamefont
  {Y.}~\bibnamefont {Tanaka}},\ and\ \bibinfo {author} {\bibfnamefont
  {J.}~\bibnamefont {Cayao}},\ }\bibfield  {title} {\bibinfo {title}
  {Classification of pair symmetries in superconductors with unconventional
  magnetism},\ }\href@noop {} {\bibfield  {journal} {\bibinfo  {journal} {Phys.
  Rev. B}\ }\textbf {\bibinfo {volume} {111}},\ \bibinfo {pages} {144508}
  (\bibinfo {year} {2025})}\BibitemShut {NoStop}%
\bibitem [{\citenamefont {Fukaya}\ \emph {et~al.}(2025)\citenamefont {Fukaya},
  \citenamefont {Lu}, \citenamefont {Yada}, \citenamefont {Tanaka},\ and\
  \citenamefont {Cayao}}]{Fukaya2025}%
  \BibitemOpen
  \bibfield  {author} {\bibinfo {author} {\bibfnamefont {Y.}~\bibnamefont
  {Fukaya}}, \bibinfo {author} {\bibfnamefont {B.}~\bibnamefont {Lu}}, \bibinfo
  {author} {\bibfnamefont {K.}~\bibnamefont {Yada}}, \bibinfo {author}
  {\bibfnamefont {Y.}~\bibnamefont {Tanaka}},\ and\ \bibinfo {author}
  {\bibfnamefont {J.}~\bibnamefont {Cayao}},\ }\bibfield  {title} {\bibinfo
  {title} {Superconducting phenomena in systems with unconventional magnets},\
  }\href@noop {} {\bibfield  {journal} {\bibinfo  {journal} {J. Phys.: Condens.
  Matter}\ }\textbf {\bibinfo {volume} {37}},\ \bibinfo {pages} {313003}
  (\bibinfo {year} {2025})},\ \bibinfo {note} {arXiv:2502.15400}\BibitemShut
  {NoStop}%
\bibitem [{\citenamefont {Sumita}\ \emph {et~al.}(2025)\citenamefont {Sumita},
  \citenamefont {Naka},\ and\ \citenamefont {Seo}}]{Sumita2025}%
  \BibitemOpen
  \bibfield  {author} {\bibinfo {author} {\bibfnamefont {S.}~\bibnamefont
  {Sumita}}, \bibinfo {author} {\bibfnamefont {M.}~\bibnamefont {Naka}},\ and\
  \bibinfo {author} {\bibfnamefont {H.}~\bibnamefont {Seo}},\ }\bibfield
  {title} {\bibinfo {title} {Phase-modulated superconductivity via
  altermagnetism},\ }\href {https://doi.org/10.1103/PhysRevB.112.144510}
  {\bibfield  {journal} {\bibinfo  {journal} {Phys. Rev. B}\ }\textbf {\bibinfo
  {volume} {112}},\ \bibinfo {pages} {144510} (\bibinfo {year}
  {2025})}\BibitemShut {NoStop}%
\bibitem [{\citenamefont {Zhang}\ \emph {et~al.}(2024)\citenamefont {Zhang},
  \citenamefont {Hu},\ and\ \citenamefont {Neupert}}]{Zhang2024}%
  \BibitemOpen
  \bibfield  {author} {\bibinfo {author} {\bibfnamefont {S.-B.}\ \bibnamefont
  {Zhang}}, \bibinfo {author} {\bibfnamefont {L.-H.}\ \bibnamefont {Hu}},\ and\
  \bibinfo {author} {\bibfnamefont {T.}~\bibnamefont {Neupert}},\ }\bibfield
  {title} {\bibinfo {title} {Finite-momentum {Cooper} pairing in proximitized
  altermagnets},\ }\href {https://doi.org/10.1038/s41467-024-45951-3}
  {\bibfield  {journal} {\bibinfo  {journal} {Nat. Commun.}\ }\textbf {\bibinfo
  {volume} {15}},\ \bibinfo {pages} {1801} (\bibinfo {year}
  {2024})}\BibitemShut {NoStop}%
\bibitem [{\citenamefont {Hu}\ \emph {et~al.}(2026)\citenamefont {Hu},
  \citenamefont {Liu}, \citenamefont {Wang}, \citenamefont {Liu},\ and\
  \citenamefont {Ohashi}}]{Hu2026}%
  \BibitemOpen
  \bibfield  {author} {\bibinfo {author} {\bibfnamefont {H.}~\bibnamefont
  {Hu}}, \bibinfo {author} {\bibfnamefont {Z.}~\bibnamefont {Liu}}, \bibinfo
  {author} {\bibfnamefont {J.}~\bibnamefont {Wang}}, \bibinfo {author}
  {\bibfnamefont {X.-J.}\ \bibnamefont {Liu}},\ and\ \bibinfo {author}
  {\bibfnamefont {Y.}~\bibnamefont {Ohashi}},\ }\href@noop {} {\bibinfo {title}
  {Finite-momentum superconductivity with singlet-triplet mixing in an
  altermagnetic metal: A pairing instability analysis}} (\bibinfo {year}
  {2026}),\ \Eprint {https://arxiv.org/abs/2603.12897} {arXiv:2603.12897
  [cond-mat.supr-con]} \BibitemShut {NoStop}%
\bibitem [{\citenamefont {Sim}\ and\ \citenamefont {Knolle}(2025)}]{Sim2025}%
  \BibitemOpen
  \bibfield  {author} {\bibinfo {author} {\bibfnamefont {G.}~\bibnamefont
  {Sim}}\ and\ \bibinfo {author} {\bibfnamefont {J.}~\bibnamefont {Knolle}},\
  }\bibfield  {title} {\bibinfo {title} {Pair density waves and supercurrent
  diode effect in altermagnets},\ }\href
  {https://doi.org/10.1103/PhysRevB.112.L020502} {\bibfield  {journal}
  {\bibinfo  {journal} {Phys. Rev. B}\ }\textbf {\bibinfo {volume} {112}},\
  \bibinfo {pages} {L020502} (\bibinfo {year} {2025})}\BibitemShut {NoStop}%
\bibitem [{\citenamefont {Chakraborty}\ and\ \citenamefont
  {Black-Schaffer}(2024)}]{Chakraborty2024}%
  \BibitemOpen
  \bibfield  {author} {\bibinfo {author} {\bibfnamefont {D.}~\bibnamefont
  {Chakraborty}}\ and\ \bibinfo {author} {\bibfnamefont {A.~M.}\ \bibnamefont
  {Black-Schaffer}},\ }\bibfield  {title} {\bibinfo {title} {Zero-field
  finite-momentum and field-induced superconductivity in altermagnets},\ }\href
  {https://doi.org/10.1103/PhysRevB.110.L060508} {\bibfield  {journal}
  {\bibinfo  {journal} {Phys. Rev. B}\ }\textbf {\bibinfo {volume} {110}},\
  \bibinfo {pages} {L060508} (\bibinfo {year} {2024})}\BibitemShut {NoStop}%
\bibitem [{\citenamefont {Chakraborty}\ and\ \citenamefont
  {Black-Schaffer}(2025)}]{Chakraborty2025}%
  \BibitemOpen
  \bibfield  {author} {\bibinfo {author} {\bibfnamefont {D.}~\bibnamefont
  {Chakraborty}}\ and\ \bibinfo {author} {\bibfnamefont {A.~M.}\ \bibnamefont
  {Black-Schaffer}},\ }\bibfield  {title} {\bibinfo {title} {Perfect
  superconducting diode effect in altermagnets},\ }\href
  {https://doi.org/10.1103/cv8s-tk4c} {\bibfield  {journal} {\bibinfo
  {journal} {Phys. Rev. Lett.}\ }\textbf {\bibinfo {volume} {135}},\ \bibinfo
  {pages} {026001} (\bibinfo {year} {2025})}\BibitemShut {NoStop}%
\bibitem [{\citenamefont {Bose}\ \emph {et~al.}(2024)\citenamefont {Bose},
  \citenamefont {Vadnais},\ and\ \citenamefont {Paramekanti}}]{Bose2024}%
  \BibitemOpen
  \bibfield  {author} {\bibinfo {author} {\bibfnamefont {A.}~\bibnamefont
  {Bose}}, \bibinfo {author} {\bibfnamefont {S.}~\bibnamefont {Vadnais}},\ and\
  \bibinfo {author} {\bibfnamefont {A.}~\bibnamefont {Paramekanti}},\
  }\bibfield  {title} {\bibinfo {title} {Altermagnetism and superconductivity
  in a multiorbital $t$-$j$ model},\ }\href
  {https://doi.org/10.1103/PhysRevB.110.205120} {\bibfield  {journal} {\bibinfo
   {journal} {Phys. Rev. B}\ }\textbf {\bibinfo {volume} {110}},\ \bibinfo
  {pages} {205120} (\bibinfo {year} {2024})}\BibitemShut {NoStop}%
\bibitem [{\citenamefont {Zou}\ \emph {et~al.}(2026)\citenamefont {Zou},
  \citenamefont {Fernandes},\ and\ \citenamefont {Fradkin}}]{Zou2026}%
  \BibitemOpen
  \bibfield  {author} {\bibinfo {author} {\bibfnamefont {X.}~\bibnamefont
  {Zou}}, \bibinfo {author} {\bibfnamefont {R.~M.}\ \bibnamefont {Fernandes}},\
  and\ \bibinfo {author} {\bibfnamefont {E.}~\bibnamefont {Fradkin}},\
  }\bibfield  {title} {\bibinfo {title} {Superconducting states and intertwined
  orders in metallic altermagnets},\ }\bibfield  {journal} {\bibinfo  {journal}
  {Phys. Rev. B}\ }\href {https://doi.org/10.1103/v8yz-2vp5}
  {10.1103/v8yz-2vp5} (\bibinfo {year} {2026}),\ \bibinfo {note}
  {arXiv:2603.04503}\BibitemShut {NoStop}%
\bibitem [{\citenamefont {Jasiewicz}\ \emph {et~al.}(2025)\citenamefont
  {Jasiewicz}, \citenamefont {W{\'o}jcik}, \citenamefont {Nowak},\ and\
  \citenamefont {Zegrodnik}}]{Jasiewicz2025}%
  \BibitemOpen
  \bibfield  {author} {\bibinfo {author} {\bibfnamefont {K.}~\bibnamefont
  {Jasiewicz}}, \bibinfo {author} {\bibfnamefont {P.}~\bibnamefont
  {W{\'o}jcik}}, \bibinfo {author} {\bibfnamefont {M.~P.}\ \bibnamefont
  {Nowak}},\ and\ \bibinfo {author} {\bibfnamefont {M.}~\bibnamefont
  {Zegrodnik}},\ }\bibfield  {title} {\bibinfo {title} {Interplay between
  altermagnetism and superconductivity in two dimensions: intertwined
  symmetries and singlet-triplet mixing},\ }\bibfield  {journal} {\bibinfo
  {journal} {npj Quantum Mater.}\ }\href
  {https://doi.org/10.1038/s41535-025-00840-w} {10.1038/s41535-025-00840-w}
  (\bibinfo {year} {2025})\BibitemShut {NoStop}%
\bibitem [{\citenamefont {Hong}\ \emph {et~al.}(2025)\citenamefont {Hong},
  \citenamefont {Park},\ and\ \citenamefont {Kim}}]{Hong2025}%
  \BibitemOpen
  \bibfield  {author} {\bibinfo {author} {\bibfnamefont {S.}~\bibnamefont
  {Hong}}, \bibinfo {author} {\bibfnamefont {M.~J.}\ \bibnamefont {Park}},\
  and\ \bibinfo {author} {\bibfnamefont {K.-M.}\ \bibnamefont {Kim}},\
  }\bibfield  {title} {\bibinfo {title} {Unconventional $p$-wave and
  finite-momentum superconductivity induced by altermagnetism through the
  formation of {Bogoliubov Fermi} surface},\ }\href
  {https://doi.org/10.1103/PhysRevB.111.054501} {\bibfield  {journal} {\bibinfo
   {journal} {Phys. Rev. B}\ }\textbf {\bibinfo {volume} {111}},\ \bibinfo
  {pages} {054501} (\bibinfo {year} {2025})}\BibitemShut {NoStop}%
\bibitem [{\citenamefont {Wei}\ \emph {et~al.}(2024)\citenamefont {Wei},
  \citenamefont {Xiang}, \citenamefont {Xu}, \citenamefont {Zhang},
  \citenamefont {Tang},\ and\ \citenamefont {Wang}}]{Wei2024}%
  \BibitemOpen
  \bibfield  {author} {\bibinfo {author} {\bibfnamefont {M.}~\bibnamefont
  {Wei}}, \bibinfo {author} {\bibfnamefont {L.}~\bibnamefont {Xiang}}, \bibinfo
  {author} {\bibfnamefont {F.}~\bibnamefont {Xu}}, \bibinfo {author}
  {\bibfnamefont {L.}~\bibnamefont {Zhang}}, \bibinfo {author} {\bibfnamefont
  {G.}~\bibnamefont {Tang}},\ and\ \bibinfo {author} {\bibfnamefont
  {J.}~\bibnamefont {Wang}},\ }\bibfield  {title} {\bibinfo {title} {Gapless
  superconducting state and mirage gap in altermagnets},\ }\href@noop {}
  {\bibfield  {journal} {\bibinfo  {journal} {Phys. Rev. B}\ }\textbf {\bibinfo
  {volume} {109}},\ \bibinfo {pages} {L201404} (\bibinfo {year}
  {2024})}\BibitemShut {NoStop}%
\bibitem [{\citenamefont {Papaj}(2023)}]{Papaj2023}%
  \BibitemOpen
  \bibfield  {author} {\bibinfo {author} {\bibfnamefont {M.}~\bibnamefont
  {Papaj}},\ }\bibfield  {title} {\bibinfo {title} {Andreev reflection at the
  altermagnet-superconductor interface},\ }\href
  {https://doi.org/10.1103/PhysRevB.108.L060508} {\bibfield  {journal}
  {\bibinfo  {journal} {Phys. Rev. B}\ }\textbf {\bibinfo {volume} {108}},\
  \bibinfo {pages} {L060508} (\bibinfo {year} {2023})}\BibitemShut {NoStop}%
\bibitem [{\citenamefont {Mayr}\ \emph {et~al.}(2005)\citenamefont {Mayr},
  \citenamefont {Alvarez}, \citenamefont {{\c{S}}en},\ and\ \citenamefont
  {Dagotto}}]{Mayr2005}%
  \BibitemOpen
  \bibfield  {author} {\bibinfo {author} {\bibfnamefont {M.}~\bibnamefont
  {Mayr}}, \bibinfo {author} {\bibfnamefont {G.}~\bibnamefont {Alvarez}},
  \bibinfo {author} {\bibfnamefont {C.}~\bibnamefont {{\c{S}}en}},\ and\
  \bibinfo {author} {\bibfnamefont {E.}~\bibnamefont {Dagotto}},\ }\bibfield
  {title} {\bibinfo {title} {Phase fluctuations in strongly coupled $d$-wave
  superconductors},\ }\href {https://doi.org/10.1103/PhysRevLett.94.217001}
  {\bibfield  {journal} {\bibinfo  {journal} {Phys. Rev. Lett.}\ }\textbf
  {\bibinfo {volume} {94}},\ \bibinfo {pages} {217001} (\bibinfo {year}
  {2005})}\BibitemShut {NoStop}%
\bibitem [{\citenamefont {Kumar}\ and\ \citenamefont
  {Majumdar}(2006)}]{Kumar2006}%
  \BibitemOpen
  \bibfield  {author} {\bibinfo {author} {\bibfnamefont {S.}~\bibnamefont
  {Kumar}}\ and\ \bibinfo {author} {\bibfnamefont {P.}~\bibnamefont
  {Majumdar}},\ }\bibfield  {title} {\bibinfo {title} {A travelling cluster
  approximation for lattice fermions strongly coupled to classical degrees of
  freedom},\ }\href {https://doi.org/10.1140/epjb/e2006-00173-2} {\bibfield
  {journal} {\bibinfo  {journal} {Eur. Phys. J. B}\ }\textbf {\bibinfo {volume}
  {50}},\ \bibinfo {pages} {571} (\bibinfo {year} {2006})}\BibitemShut
  {NoStop}%
\bibitem [{\citenamefont {Karmakar}\ and\ \citenamefont
  {Majumdar}(2016)}]{Karmakar2016}%
  \BibitemOpen
  \bibfield  {author} {\bibinfo {author} {\bibfnamefont {M.}~\bibnamefont
  {Karmakar}}\ and\ \bibinfo {author} {\bibfnamefont {P.}~\bibnamefont
  {Majumdar}},\ }\bibfield  {title} {\bibinfo {title} {Population-imbalanced
  lattice fermions near the {BCS-BEC} crossover: Thermal physics of the
  breached pair and {Fulde-Ferrell-Larkin-Ovchinnikov} phases},\ }\href
  {https://doi.org/10.1103/PhysRevA.93.053609} {\bibfield  {journal} {\bibinfo
  {journal} {Phys. Rev. A}\ }\textbf {\bibinfo {volume} {93}},\ \bibinfo
  {pages} {053609} (\bibinfo {year} {2016})}\BibitemShut {NoStop}%
\bibitem [{\citenamefont {Karmakar}(2018)}]{Karmakar2018}%
  \BibitemOpen
  \bibfield  {author} {\bibinfo {author} {\bibfnamefont {M.}~\bibnamefont
  {Karmakar}},\ }\bibfield  {title} {\bibinfo {title} {Thermal transitions,
  pseudogap behavior, and {BCS-BEC} crossover in {Fermi-Fermi} mixtures},\
  }\href {https://doi.org/10.1103/PhysRevA.97.033617} {\bibfield  {journal}
  {\bibinfo  {journal} {Phys. Rev. A}\ }\textbf {\bibinfo {volume} {97}},\
  \bibinfo {pages} {033617} (\bibinfo {year} {2018})}\BibitemShut {NoStop}%
\bibitem [{\citenamefont {Tarat}\ and\ \citenamefont
  {Majumdar}(2015)}]{Tarat2015}%
  \BibitemOpen
  \bibfield  {author} {\bibinfo {author} {\bibfnamefont {S.}~\bibnamefont
  {Tarat}}\ and\ \bibinfo {author} {\bibfnamefont {P.}~\bibnamefont
  {Majumdar}},\ }\bibfield  {title} {\bibinfo {title} {A real space auxiliary
  field approach to the {BCS-BEC} crossover},\ }\href
  {https://doi.org/10.1140/epjb/e2015-50794-3} {\bibfield  {journal} {\bibinfo
  {journal} {Eur. Phys. J. B}\ }\textbf {\bibinfo {volume} {88}},\ \bibinfo
  {pages} {68} (\bibinfo {year} {2015})}\BibitemShut {NoStop}%
\bibitem [{\citenamefont {Giuli}\ \emph {et~al.}(2025)\citenamefont {Giuli},
  \citenamefont {Mejuto-Zaera},\ and\ \citenamefont {Capone}}]{Giuli2025}%
  \BibitemOpen
  \bibfield  {author} {\bibinfo {author} {\bibfnamefont {S.}~\bibnamefont
  {Giuli}}, \bibinfo {author} {\bibfnamefont {C.}~\bibnamefont
  {Mejuto-Zaera}},\ and\ \bibinfo {author} {\bibfnamefont {M.}~\bibnamefont
  {Capone}},\ }\bibfield  {title} {\bibinfo {title} {Altermagnetism from
  interaction-driven itinerant magnetism},\ }\href
  {https://doi.org/10.1103/PhysRevB.111.L020401} {\bibfield  {journal}
  {\bibinfo  {journal} {Phys. Rev. B}\ }\textbf {\bibinfo {volume} {111}},\
  \bibinfo {pages} {L020401} (\bibinfo {year} {2025})}\BibitemShut {NoStop}%
\bibitem [{\citenamefont {Kaushal}\ and\ \citenamefont
  {Franz}(2025)}]{Kaushal2025}%
  \BibitemOpen
  \bibfield  {author} {\bibinfo {author} {\bibfnamefont {N.}~\bibnamefont
  {Kaushal}}\ and\ \bibinfo {author} {\bibfnamefont {M.}~\bibnamefont
  {Franz}},\ }\bibfield  {title} {\bibinfo {title} {Altermagnetism in the
  modified {Lieb} lattice {Hubbard} model},\ }\href
  {https://doi.org/10.1103/s31h-hk2v} {\bibfield  {journal} {\bibinfo
  {journal} {Phys. Rev. Lett.}\ }\textbf {\bibinfo {volume} {135}},\ \bibinfo
  {pages} {156502} (\bibinfo {year} {2025})}\BibitemShut {NoStop}%
\bibitem [{\citenamefont {D{\"u}rrnagel}\ \emph {et~al.}(2025)\citenamefont
  {D{\"u}rrnagel}, \citenamefont {Hohmann}, \citenamefont {Maity},
  \citenamefont {Seufert}, \citenamefont {Klett}, \citenamefont {Klebl},\ and\
  \citenamefont {Thomale}}]{Durrnagel2025}%
  \BibitemOpen
  \bibfield  {author} {\bibinfo {author} {\bibfnamefont {M.}~\bibnamefont
  {D{\"u}rrnagel}}, \bibinfo {author} {\bibfnamefont {H.}~\bibnamefont
  {Hohmann}}, \bibinfo {author} {\bibfnamefont {A.}~\bibnamefont {Maity}},
  \bibinfo {author} {\bibfnamefont {J.}~\bibnamefont {Seufert}}, \bibinfo
  {author} {\bibfnamefont {M.}~\bibnamefont {Klett}}, \bibinfo {author}
  {\bibfnamefont {L.}~\bibnamefont {Klebl}},\ and\ \bibinfo {author}
  {\bibfnamefont {R.}~\bibnamefont {Thomale}},\ }\bibfield  {title} {\bibinfo
  {title} {Altermagnetic phase transition in a {Lieb} metal},\ }\href@noop {}
  {\bibfield  {journal} {\bibinfo  {journal} {Phys. Rev. Lett.}\ }\textbf
  {\bibinfo {volume} {135}},\ \bibinfo {pages} {036502} (\bibinfo {year}
  {2025})}\BibitemShut {NoStop}%
\bibitem [{\citenamefont {Micnas}\ \emph {et~al.}(1990)\citenamefont {Micnas},
  \citenamefont {Ranninger},\ and\ \citenamefont {Robaszkiewicz}}]{Micnas1990}%
  \BibitemOpen
  \bibfield  {author} {\bibinfo {author} {\bibfnamefont {R.}~\bibnamefont
  {Micnas}}, \bibinfo {author} {\bibfnamefont {J.}~\bibnamefont {Ranninger}},\
  and\ \bibinfo {author} {\bibfnamefont {S.}~\bibnamefont {Robaszkiewicz}},\
  }\bibfield  {title} {\bibinfo {title} {Superconductivity in narrow-band
  systems with local nonretarded attractive interactions},\ }\href
  {https://doi.org/10.1103/RevModPhys.62.113} {\bibfield  {journal} {\bibinfo
  {journal} {Rev. Mod. Phys.}\ }\textbf {\bibinfo {volume} {62}},\ \bibinfo
  {pages} {113} (\bibinfo {year} {1990})}\BibitemShut {NoStop}%
\bibitem [{\citenamefont {Alvarez}\ \emph {et~al.}(2005)\citenamefont
  {Alvarez}, \citenamefont {Mayr}, \citenamefont {Moreo},\ and\ \citenamefont
  {Dagotto}}]{dagotto_prb2005}%
  \BibitemOpen
  \bibfield  {author} {\bibinfo {author} {\bibfnamefont {G.}~\bibnamefont
  {Alvarez}}, \bibinfo {author} {\bibfnamefont {M.}~\bibnamefont {Mayr}},
  \bibinfo {author} {\bibfnamefont {A.}~\bibnamefont {Moreo}},\ and\ \bibinfo
  {author} {\bibfnamefont {E.}~\bibnamefont {Dagotto}},\ }\bibfield  {title}
  {\bibinfo {title} {Areas of superconductivity and giant proximity effects in
  underdoped cuprates},\ }\href {https://doi.org/10.1103/PhysRevB.71.014514}
  {\bibfield  {journal} {\bibinfo  {journal} {Phys. Rev. B}\ }\textbf {\bibinfo
  {volume} {71}},\ \bibinfo {pages} {014514} (\bibinfo {year}
  {2005})}\BibitemShut {NoStop}%
\bibitem [{\citenamefont {Hubbard}(1959)}]{Hubbard1959}%
  \BibitemOpen
  \bibfield  {author} {\bibinfo {author} {\bibfnamefont {J.}~\bibnamefont
  {Hubbard}},\ }\bibfield  {title} {\bibinfo {title} {Calculation of partition
  functions},\ }\href {https://doi.org/10.1103/PhysRevLett.3.77} {\bibfield
  {journal} {\bibinfo  {journal} {Phys. Rev. Lett.}\ }\textbf {\bibinfo
  {volume} {3}},\ \bibinfo {pages} {77} (\bibinfo {year} {1959})}\BibitemShut
  {NoStop}%
\bibitem [{\citenamefont {Schulz}(1990)}]{Schulz1990}%
  \BibitemOpen
  \bibfield  {author} {\bibinfo {author} {\bibfnamefont {H.~J.}\ \bibnamefont
  {Schulz}},\ }\bibfield  {title} {\bibinfo {title} {Effective action for
  strongly correlated fermions from functional integrals},\ }\href
  {https://doi.org/10.1103/PhysRevLett.65.2462} {\bibfield  {journal} {\bibinfo
   {journal} {Phys. Rev. Lett.}\ }\textbf {\bibinfo {volume} {65}},\ \bibinfo
  {pages} {2462} (\bibinfo {year} {1990})}\BibitemShut {NoStop}%
\bibitem [{\citenamefont {Fratini}\ and\ \citenamefont
  {Ciuchi}(2021)}]{Fratini2021}%
  \BibitemOpen
  \bibfield  {author} {\bibinfo {author} {\bibfnamefont {S.}~\bibnamefont
  {Fratini}}\ and\ \bibinfo {author} {\bibfnamefont {S.}~\bibnamefont
  {Ciuchi}},\ }\bibfield  {title} {\bibinfo {title} {Displaced {Drude} peak and
  bad metal from the interaction with slow fluctuations},\ }\href
  {https://doi.org/10.21468/SciPostPhys.11.2.039} {\bibfield  {journal}
  {\bibinfo  {journal} {SciPost Phys.}\ }\textbf {\bibinfo {volume} {11}},\
  \bibinfo {pages} {039} (\bibinfo {year} {2021})}\BibitemShut {NoStop}%
\bibitem [{\citenamefont {Ciuchi}\ and\ \citenamefont
  {Fratini}(2023)}]{Ciuchi2023}%
  \BibitemOpen
  \bibfield  {author} {\bibinfo {author} {\bibfnamefont {S.}~\bibnamefont
  {Ciuchi}}\ and\ \bibinfo {author} {\bibfnamefont {S.}~\bibnamefont
  {Fratini}},\ }\bibfield  {title} {\bibinfo {title} {Strange metal behavior
  from incoherent carriers scattered by local moments},\ }\href
  {https://doi.org/10.1103/PhysRevB.108.235173} {\bibfield  {journal} {\bibinfo
   {journal} {Phys. Rev. B}\ }\textbf {\bibinfo {volume} {108}},\ \bibinfo
  {pages} {235173} (\bibinfo {year} {2023})}\BibitemShut {NoStop}%
\bibitem [{\citenamefont {Murthy}\ \emph {et~al.}(2023)\citenamefont {Murthy},
  \citenamefont {Pandey}, \citenamefont {Esterlis},\ and\ \citenamefont
  {Kivelson}}]{Murthy2023}%
  \BibitemOpen
  \bibfield  {author} {\bibinfo {author} {\bibfnamefont {C.}~\bibnamefont
  {Murthy}}, \bibinfo {author} {\bibfnamefont {A.}~\bibnamefont {Pandey}},
  \bibinfo {author} {\bibfnamefont {I.}~\bibnamefont {Esterlis}},\ and\
  \bibinfo {author} {\bibfnamefont {S.~A.}\ \bibnamefont {Kivelson}},\
  }\bibfield  {title} {\bibinfo {title} {A stability bound on the {$T$}-linear
  resistivity of conventional metals},\ }\href
  {https://doi.org/10.1073/pnas.2216241120} {\bibfield  {journal} {\bibinfo
  {journal} {Proc. Natl. Acad. Sci. U.S.A.}\ }\textbf {\bibinfo {volume}
  {120}},\ \bibinfo {pages} {e2216241120} (\bibinfo {year} {2023})}\BibitemShut
  {NoStop}%
\bibitem [{\citenamefont {Karmakar}(2020)}]{Karmakar2020}%
  \BibitemOpen
  \bibfield  {author} {\bibinfo {author} {\bibfnamefont {M.}~\bibnamefont
  {Karmakar}},\ }\bibfield  {title} {\bibinfo {title} {Pauli limited $d$-wave
  superconductors: quantum breached pair phase and thermal transitions},\
  }\href {https://doi.org/10.1088/1361-648X/ab99f7} {\bibfield  {journal}
  {\bibinfo  {journal} {J. Phys.: Condens. Matter}\ }\textbf {\bibinfo {volume}
  {32}},\ \bibinfo {pages} {405604} (\bibinfo {year} {2020})}\BibitemShut
  {NoStop}%
\bibitem [{\citenamefont {Karmakar}(2024)}]{Karmakar2024}%
  \BibitemOpen
  \bibfield  {author} {\bibinfo {author} {\bibfnamefont {M.}~\bibnamefont
  {Karmakar}},\ }\bibfield  {title} {\bibinfo {title} {Magnetotransport and
  {Fermi} surface segmentation in {Pauli} limited superconductors},\ }\href
  {https://doi.org/10.1088/1361-648X/ad1f2b} {\bibfield  {journal} {\bibinfo
  {journal} {J. Phys.: Condens. Matter}\ }\textbf {\bibinfo {volume} {36}},\
  \bibinfo {pages} {165601} (\bibinfo {year} {2024})}\BibitemShut {NoStop}%
\bibitem [{\citenamefont {Kadge}\ and\ \citenamefont
  {Majumdar}(2018)}]{Kadge2018}%
  \BibitemOpen
  \bibfield  {author} {\bibinfo {author} {\bibfnamefont {S.}~\bibnamefont
  {Kadge}}\ and\ \bibinfo {author} {\bibfnamefont {P.}~\bibnamefont
  {Majumdar}},\ }\bibfield  {title} {\bibinfo {title} {The competition and
  coexistence of antiferromagnetism and $d$-wave superconductivity: probing
  coupled thermal fluctuations in a two dimensional minimal model},\ }\bibfield
   {journal} {\bibinfo  {journal} {Eur. Phys. J. B}\ }\textbf {\bibinfo
  {volume} {91}},\ \href {https://doi.org/10.1140/epjb/e2018-90351-x}
  {10.1140/epjb/e2018-90351-x} (\bibinfo {year} {2018})\BibitemShut {NoStop}%
\bibitem [{\citenamefont {Baruch}\ and\ \citenamefont
  {Orgad}(2008)}]{Baruch2008}%
  \BibitemOpen
  \bibfield  {author} {\bibinfo {author} {\bibfnamefont {S.}~\bibnamefont
  {Baruch}}\ and\ \bibinfo {author} {\bibfnamefont {D.}~\bibnamefont {Orgad}},\
  }\bibfield  {title} {\bibinfo {title} {Spectral signatures of modulated
  $d$-wave superconducting phases},\ }\href
  {https://doi.org/10.1103/PhysRevB.77.174502} {\bibfield  {journal} {\bibinfo
  {journal} {Phys. Rev. B}\ }\textbf {\bibinfo {volume} {77}},\ \bibinfo
  {pages} {174502} (\bibinfo {year} {2008})}\BibitemShut {NoStop}%
\bibitem [{\citenamefont {Lee}(2014)}]{Lee2014}%
  \BibitemOpen
  \bibfield  {author} {\bibinfo {author} {\bibfnamefont {P.~A.}\ \bibnamefont
  {Lee}},\ }\bibfield  {title} {\bibinfo {title} {Amperean pairing and the
  pseudogap phase of cuprate superconductors},\ }\href
  {https://doi.org/10.1103/PhysRevX.4.031017} {\bibfield  {journal} {\bibinfo
  {journal} {Phys. Rev. X}\ }\textbf {\bibinfo {volume} {4}},\ \bibinfo {pages}
  {031017} (\bibinfo {year} {2014})}\BibitemShut {NoStop}%
\bibitem [{\citenamefont {Yuan}\ and\ \citenamefont {Fu}(2021)}]{YuanFu2021}%
  \BibitemOpen
  \bibfield  {author} {\bibinfo {author} {\bibfnamefont {N.~F.~Q.}\
  \bibnamefont {Yuan}}\ and\ \bibinfo {author} {\bibfnamefont {L.}~\bibnamefont
  {Fu}},\ }\bibfield  {title} {\bibinfo {title} {Topological metals and
  finite-momentum superconductors},\ }\href
  {https://doi.org/10.1073/pnas.2019063118} {\bibfield  {journal} {\bibinfo
  {journal} {Proc. Natl. Acad. Sci. U.S.A.}\ }\textbf {\bibinfo {volume}
  {118}},\ \bibinfo {pages} {e2019063118} (\bibinfo {year} {2021})}\BibitemShut
  {NoStop}%
\bibitem [{\citenamefont {Hu}\ \emph {et~al.}(2025)\citenamefont {Hu},
  \citenamefont {Liu},\ and\ \citenamefont {Liu}}]{Hu2025}%
  \BibitemOpen
  \bibfield  {author} {\bibinfo {author} {\bibfnamefont {H.}~\bibnamefont
  {Hu}}, \bibinfo {author} {\bibfnamefont {Z.}~\bibnamefont {Liu}},\ and\
  \bibinfo {author} {\bibfnamefont {X.-J.}\ \bibnamefont {Liu}},\ }\bibfield
  {title} {\bibinfo {title} {Unconventional superconductivity of an
  altermagnetic metal: Polarized {BCS} and inhomogeneous {FFLO} states},\
  }\href@noop {} {\bibfield  {journal} {\bibinfo  {journal} {Phys. Rev. B}\
  }\textbf {\bibinfo {volume} {112}},\ \bibinfo {pages} {184501} (\bibinfo
  {year} {2025})}\BibitemShut {NoStop}%
\bibitem [{\citenamefont {Buzdin}(2005)}]{Buzdin2005}%
  \BibitemOpen
  \bibfield  {author} {\bibinfo {author} {\bibfnamefont {A.~I.}\ \bibnamefont
  {Buzdin}},\ }\bibfield  {title} {\bibinfo {title} {Proximity effects in
  superconductor-ferromagnet heterostructures},\ }\href
  {https://doi.org/10.1103/RevModPhys.77.935} {\bibfield  {journal} {\bibinfo
  {journal} {Rev. Mod. Phys.}\ }\textbf {\bibinfo {volume} {77}},\ \bibinfo
  {pages} {935} (\bibinfo {year} {2005})}\BibitemShut {NoStop}%
\bibitem [{\citenamefont {Matsuba}\ \emph {et~al.}(2007)\citenamefont
  {Matsuba}, \citenamefont {Yoshizawa}, \citenamefont {Mochizuki},
  \citenamefont {Mochiku}, \citenamefont {Hirata},\ and\ \citenamefont
  {Nishida}}]{nishida_jpsj2007}%
  \BibitemOpen
  \bibfield  {author} {\bibinfo {author} {\bibfnamefont {K.}~\bibnamefont
  {Matsuba}}, \bibinfo {author} {\bibfnamefont {S.}~\bibnamefont {Yoshizawa}},
  \bibinfo {author} {\bibfnamefont {Y.}~\bibnamefont {Mochizuki}}, \bibinfo
  {author} {\bibfnamefont {T.}~\bibnamefont {Mochiku}}, \bibinfo {author}
  {\bibfnamefont {K.}~\bibnamefont {Hirata}},\ and\ \bibinfo {author}
  {\bibfnamefont {N.}~\bibnamefont {Nishida}},\ }\bibfield  {title} {\bibinfo
  {title} {Anti-phase modulation of electron- and hole-like states in vortex
  core of bi2sr2cacu2ox probed by scanning tunneling spectroscopy},\ }\href
  {https://doi.org/10.1143/JPSJ.76.063704} {\bibfield  {journal} {\bibinfo
  {journal} {Journal of the Physical Society of Japan}\ }\textbf {\bibinfo
  {volume} {76}},\ \bibinfo {pages} {063704} (\bibinfo {year}
  {2007})}\BibitemShut {NoStop}%
\bibitem [{\citenamefont {Yoshizawa}\ \emph {et~al.}(2013)\citenamefont
  {Yoshizawa}, \citenamefont {Koseki}, \citenamefont {Matsuba}, \citenamefont
  {Mochiku}, \citenamefont {Hirata},\ and\ \citenamefont
  {Nishida}}]{nishida_jpsj2013}%
  \BibitemOpen
  \bibfield  {author} {\bibinfo {author} {\bibfnamefont {S.}~\bibnamefont
  {Yoshizawa}}, \bibinfo {author} {\bibfnamefont {T.}~\bibnamefont {Koseki}},
  \bibinfo {author} {\bibfnamefont {K.}~\bibnamefont {Matsuba}}, \bibinfo
  {author} {\bibfnamefont {T.}~\bibnamefont {Mochiku}}, \bibinfo {author}
  {\bibfnamefont {K.}~\bibnamefont {Hirata}},\ and\ \bibinfo {author}
  {\bibfnamefont {N.}~\bibnamefont {Nishida}},\ }\bibfield  {title} {\bibinfo
  {title} {High-resolution scanning tunneling spectroscopy of vortex cores in
  inhomogeneous electronic states of bi2sr2cacu2ox},\ }\href
  {https://doi.org/10.7566/JPSJ.82.083706} {\bibfield  {journal} {\bibinfo
  {journal} {Journal of the Physical Society of Japan}\ }\textbf {\bibinfo
  {volume} {82}},\ \bibinfo {pages} {083706} (\bibinfo {year}
  {2013})}\BibitemShut {NoStop}%
\bibitem [{\citenamefont {Machida}\ \emph {et~al.}(2016)\citenamefont
  {Machida}, \citenamefont {Kohsaka}, \citenamefont {Matsuoka}, \citenamefont
  {Iwaya}, \citenamefont {Hanaguri},\ and\ \citenamefont
  {Tamegai}}]{tamegai_natcom2016}%
  \BibitemOpen
  \bibfield  {author} {\bibinfo {author} {\bibfnamefont {T.}~\bibnamefont
  {Machida}}, \bibinfo {author} {\bibfnamefont {Y.}~\bibnamefont {Kohsaka}},
  \bibinfo {author} {\bibfnamefont {K.}~\bibnamefont {Matsuoka}}, \bibinfo
  {author} {\bibfnamefont {K.}~\bibnamefont {Iwaya}}, \bibinfo {author}
  {\bibfnamefont {T.}~\bibnamefont {Hanaguri}},\ and\ \bibinfo {author}
  {\bibfnamefont {T.}~\bibnamefont {Tamegai}},\ }\bibfield  {title} {\bibinfo
  {title} {Bipartite electronic superstructures in the vortex core of
  bi$_{2}$sr$_{2}$cacu$_{2}$o$_{8+\delta}$},\ }\href
  {https://doi.org/10.1038/ncomms11747} {\bibfield  {journal} {\bibinfo
  {journal} {Nature Communications}\ }\textbf {\bibinfo {volume} {7}},\
  \bibinfo {pages} {2041} (\bibinfo {year} {2016})}\BibitemShut {NoStop}%
\bibitem [{\citenamefont {Edkins}\ \emph {et~al.}(2019)\citenamefont {Edkins},
  \citenamefont {Kostin}, \citenamefont {Fujita}, \citenamefont {Mackenzie},
  \citenamefont {Eisaki}, \citenamefont {Uchida}, \citenamefont {Sachdev},
  \citenamefont {Lawler}, \citenamefont {Kim}, \citenamefont {Davis},\ and\
  \citenamefont {Hamidian}}]{Edkins2019}%
  \BibitemOpen
  \bibfield  {author} {\bibinfo {author} {\bibfnamefont {S.~D.}\ \bibnamefont
  {Edkins}}, \bibinfo {author} {\bibfnamefont {A.}~\bibnamefont {Kostin}},
  \bibinfo {author} {\bibfnamefont {K.}~\bibnamefont {Fujita}}, \bibinfo
  {author} {\bibfnamefont {A.~P.}\ \bibnamefont {Mackenzie}}, \bibinfo {author}
  {\bibfnamefont {H.}~\bibnamefont {Eisaki}}, \bibinfo {author} {\bibfnamefont
  {S.}~\bibnamefont {Uchida}}, \bibinfo {author} {\bibfnamefont
  {S.}~\bibnamefont {Sachdev}}, \bibinfo {author} {\bibfnamefont {M.~J.}\
  \bibnamefont {Lawler}}, \bibinfo {author} {\bibfnamefont {E.-A.}\
  \bibnamefont {Kim}}, \bibinfo {author} {\bibfnamefont {J.~C.~S.}\
  \bibnamefont {Davis}},\ and\ \bibinfo {author} {\bibfnamefont {M.~H.}\
  \bibnamefont {Hamidian}},\ }\bibfield  {title} {\bibinfo {title} {Magnetic
  field-induced pair density wave state in the cuprate vortex halo},\ }\href
  {https://doi.org/10.1126/science.aat1773} {\bibfield  {journal} {\bibinfo
  {journal} {Science}\ }\textbf {\bibinfo {volume} {364}},\ \bibinfo {pages}
  {976} (\bibinfo {year} {2019})}\BibitemShut {NoStop}%
\bibitem [{\citenamefont {Kohsaka}\ \emph {et~al.}(2007)\citenamefont
  {Kohsaka}, \citenamefont {Taylor}, \citenamefont {Fujita}, \citenamefont
  {Schmidt}, \citenamefont {Lupien}, \citenamefont {Hanaguri}, \citenamefont
  {Azuma}, \citenamefont {Takano}, \citenamefont {Eisaki}, \citenamefont
  {Takagi}, \citenamefont {Uchida},\ and\ \citenamefont {Davis}}]{Kohsaka2007}%
  \BibitemOpen
  \bibfield  {author} {\bibinfo {author} {\bibfnamefont {Y.}~\bibnamefont
  {Kohsaka}}, \bibinfo {author} {\bibfnamefont {C.}~\bibnamefont {Taylor}},
  \bibinfo {author} {\bibfnamefont {K.}~\bibnamefont {Fujita}}, \bibinfo
  {author} {\bibfnamefont {A.}~\bibnamefont {Schmidt}}, \bibinfo {author}
  {\bibfnamefont {C.}~\bibnamefont {Lupien}}, \bibinfo {author} {\bibfnamefont
  {T.}~\bibnamefont {Hanaguri}}, \bibinfo {author} {\bibfnamefont
  {M.}~\bibnamefont {Azuma}}, \bibinfo {author} {\bibfnamefont
  {M.}~\bibnamefont {Takano}}, \bibinfo {author} {\bibfnamefont
  {H.}~\bibnamefont {Eisaki}}, \bibinfo {author} {\bibfnamefont
  {H.}~\bibnamefont {Takagi}}, \bibinfo {author} {\bibfnamefont
  {S.}~\bibnamefont {Uchida}},\ and\ \bibinfo {author} {\bibfnamefont {J.~C.}\
  \bibnamefont {Davis}},\ }\bibfield  {title} {\bibinfo {title} {An intrinsic
  bond-centered electronic glass with unidirectional domains in underdoped
  cuprates},\ }\href {https://doi.org/10.1126/science.1138584} {\bibfield
  {journal} {\bibinfo  {journal} {Science}\ }\textbf {\bibinfo {volume}
  {315}},\ \bibinfo {pages} {1380} (\bibinfo {year} {2007})}\BibitemShut
  {NoStop}%
\bibitem [{\citenamefont {Hamidian}\ \emph
  {et~al.}(2016{\natexlab{a}})\citenamefont {Hamidian}, \citenamefont {Edkins},
  \citenamefont {Kim}, \citenamefont {Davis}, \citenamefont {Mackenzie},
  \citenamefont {Eisaki}, \citenamefont {Uchida}, \citenamefont {Lawler},
  \citenamefont {Kim}, \citenamefont {Sachdev},\ and\ \citenamefont
  {Fujita}}]{eisaki_natphys2016}%
  \BibitemOpen
  \bibfield  {author} {\bibinfo {author} {\bibfnamefont {M.~H.}\ \bibnamefont
  {Hamidian}}, \bibinfo {author} {\bibfnamefont {S.~D.}\ \bibnamefont
  {Edkins}}, \bibinfo {author} {\bibfnamefont {C.~K.}\ \bibnamefont {Kim}},
  \bibinfo {author} {\bibfnamefont {J.~C.}\ \bibnamefont {Davis}}, \bibinfo
  {author} {\bibfnamefont {A.~P.}\ \bibnamefont {Mackenzie}}, \bibinfo {author}
  {\bibfnamefont {H.}~\bibnamefont {Eisaki}}, \bibinfo {author} {\bibfnamefont
  {S.}~\bibnamefont {Uchida}}, \bibinfo {author} {\bibfnamefont {M.~J.}\
  \bibnamefont {Lawler}}, \bibinfo {author} {\bibfnamefont {E.-A.}\
  \bibnamefont {Kim}}, \bibinfo {author} {\bibfnamefont {S.}~\bibnamefont
  {Sachdev}},\ and\ \bibinfo {author} {\bibfnamefont {K.}~\bibnamefont
  {Fujita}},\ }\bibfield  {title} {\bibinfo {title} {Atomic-scale electronic
  structure of the cuprate d-symmetry form factor density wave state},\ }\href
  {https://doi.org/10.1038/nphys3519} {\bibfield  {journal} {\bibinfo
  {journal} {Nature Physics}\ }\textbf {\bibinfo {volume} {12}},\ \bibinfo
  {pages} {150} (\bibinfo {year} {2016}{\natexlab{a}})}\BibitemShut {NoStop}%
\bibitem [{\citenamefont {Hamidian}\ \emph
  {et~al.}(2016{\natexlab{b}})\citenamefont {Hamidian}, \citenamefont {Edkins},
  \citenamefont {Joo}, \citenamefont {Kostin}, \citenamefont {Eisaki},
  \citenamefont {Uchida}, \citenamefont {Lawler}, \citenamefont {Kim},
  \citenamefont {Mackenzie}, \citenamefont {Fujita}, \citenamefont {Lee},\ and\
  \citenamefont {Davis}}]{Hamidian2016}%
  \BibitemOpen
  \bibfield  {author} {\bibinfo {author} {\bibfnamefont {M.~H.}\ \bibnamefont
  {Hamidian}}, \bibinfo {author} {\bibfnamefont {S.~D.}\ \bibnamefont
  {Edkins}}, \bibinfo {author} {\bibfnamefont {S.~H.}\ \bibnamefont {Joo}},
  \bibinfo {author} {\bibfnamefont {A.}~\bibnamefont {Kostin}}, \bibinfo
  {author} {\bibfnamefont {H.}~\bibnamefont {Eisaki}}, \bibinfo {author}
  {\bibfnamefont {S.}~\bibnamefont {Uchida}}, \bibinfo {author} {\bibfnamefont
  {M.~J.}\ \bibnamefont {Lawler}}, \bibinfo {author} {\bibfnamefont {E.-A.}\
  \bibnamefont {Kim}}, \bibinfo {author} {\bibfnamefont {A.~P.}\ \bibnamefont
  {Mackenzie}}, \bibinfo {author} {\bibfnamefont {K.}~\bibnamefont {Fujita}},
  \bibinfo {author} {\bibfnamefont {J.}~\bibnamefont {Lee}},\ and\ \bibinfo
  {author} {\bibfnamefont {J.~C.~S.}\ \bibnamefont {Davis}},\ }\bibfield
  {title} {\bibinfo {title} {Detection of a {Cooper}-pair density wave in
  {Bi$_2$Sr$_2$CaCu$_2$O$_{8+x}$}},\ }\href
  {https://doi.org/10.1038/nature17411} {\bibfield  {journal} {\bibinfo
  {journal} {Nature}\ }\textbf {\bibinfo {volume} {532}},\ \bibinfo {pages}
  {343} (\bibinfo {year} {2016}{\natexlab{b}})}\BibitemShut {NoStop}%
\bibitem [{\citenamefont {Fujita}\ \emph {et~al.}(2014)\citenamefont {Fujita},
  \citenamefont {Hamidian}, \citenamefont {Edkins}, \citenamefont {Kim},
  \citenamefont {Kohsaka}, \citenamefont {Azuma}, \citenamefont {Takano},
  \citenamefont {Takagi}, \citenamefont {Eisaki}, \citenamefont {Uchida},
  \citenamefont {Allais}, \citenamefont {Lawler}, \citenamefont {Kim},
  \citenamefont {Sachdev},\ and\ \citenamefont {Davis}}]{Fujita2014}%
  \BibitemOpen
  \bibfield  {author} {\bibinfo {author} {\bibfnamefont {K.}~\bibnamefont
  {Fujita}}, \bibinfo {author} {\bibfnamefont {M.~H.}\ \bibnamefont
  {Hamidian}}, \bibinfo {author} {\bibfnamefont {S.~D.}\ \bibnamefont
  {Edkins}}, \bibinfo {author} {\bibfnamefont {C.~K.}\ \bibnamefont {Kim}},
  \bibinfo {author} {\bibfnamefont {Y.}~\bibnamefont {Kohsaka}}, \bibinfo
  {author} {\bibfnamefont {M.}~\bibnamefont {Azuma}}, \bibinfo {author}
  {\bibfnamefont {M.}~\bibnamefont {Takano}}, \bibinfo {author} {\bibfnamefont
  {H.}~\bibnamefont {Takagi}}, \bibinfo {author} {\bibfnamefont
  {H.}~\bibnamefont {Eisaki}}, \bibinfo {author} {\bibfnamefont {S.-i.}\
  \bibnamefont {Uchida}}, \bibinfo {author} {\bibfnamefont {A.}~\bibnamefont
  {Allais}}, \bibinfo {author} {\bibfnamefont {M.~J.}\ \bibnamefont {Lawler}},
  \bibinfo {author} {\bibfnamefont {E.-A.}\ \bibnamefont {Kim}}, \bibinfo
  {author} {\bibfnamefont {S.}~\bibnamefont {Sachdev}},\ and\ \bibinfo {author}
  {\bibfnamefont {J.~C.~S.}\ \bibnamefont {Davis}},\ }\bibfield  {title}
  {\bibinfo {title} {Direct phase-sensitive identification of a $d$-form factor
  density wave in underdoped cuprates},\ }\href
  {https://doi.org/10.1073/pnas.1406297111} {\bibfield  {journal} {\bibinfo
  {journal} {Proc. Natl. Acad. Sci. U.S.A.}\ }\textbf {\bibinfo {volume}
  {111}},\ \bibinfo {pages} {E3026} (\bibinfo {year} {2014})}\BibitemShut
  {NoStop}%
\bibitem [{\citenamefont {Zhang}\ \emph {et~al.}(2023)\citenamefont {Zhang},
  \citenamefont {Liu}, \citenamefont {Wang}, \citenamefont {Tsang},
  \citenamefont {Wang}, \citenamefont {Lam}, \citenamefont {Wang},
  \citenamefont {Xie}, \citenamefont {Zhou}, \citenamefont {Zhao},
  \citenamefont {Wang}, \citenamefont {Tallon}, \citenamefont {Lai},\ and\
  \citenamefont {Goh}}]{Zhang2023}%
  \BibitemOpen
  \bibfield  {author} {\bibinfo {author} {\bibfnamefont {W.}~\bibnamefont
  {Zhang}}, \bibinfo {author} {\bibfnamefont {X.}~\bibnamefont {Liu}}, \bibinfo
  {author} {\bibfnamefont {L.}~\bibnamefont {Wang}}, \bibinfo {author}
  {\bibfnamefont {C.~W.}\ \bibnamefont {Tsang}}, \bibinfo {author}
  {\bibfnamefont {Z.}~\bibnamefont {Wang}}, \bibinfo {author} {\bibfnamefont
  {S.~T.}\ \bibnamefont {Lam}}, \bibinfo {author} {\bibfnamefont
  {W.}~\bibnamefont {Wang}}, \bibinfo {author} {\bibfnamefont {J.}~\bibnamefont
  {Xie}}, \bibinfo {author} {\bibfnamefont {X.}~\bibnamefont {Zhou}}, \bibinfo
  {author} {\bibfnamefont {Y.}~\bibnamefont {Zhao}}, \bibinfo {author}
  {\bibfnamefont {S.}~\bibnamefont {Wang}}, \bibinfo {author} {\bibfnamefont
  {J.}~\bibnamefont {Tallon}}, \bibinfo {author} {\bibfnamefont {K.~T.}\
  \bibnamefont {Lai}},\ and\ \bibinfo {author} {\bibfnamefont {S.~K.}\
  \bibnamefont {Goh}},\ }\bibfield  {title} {\bibinfo {title} {Nodeless
  superconductivity in kagome metal {CsV$_3$Sb$_5$} with and without time
  reversal symmetry breaking},\ }\href@noop {} {\bibfield  {journal} {\bibinfo
  {journal} {Nano Lett.}\ }\textbf {\bibinfo {volume} {23}},\ \bibinfo {pages}
  {872} (\bibinfo {year} {2023})}\BibitemShut {NoStop}%
\bibitem [{\citenamefont {Li}\ \emph {et~al.}(2026)\citenamefont {Li},
  \citenamefont {Han}, \citenamefont {Yang}, \citenamefont {He}, \citenamefont
  {Ye}, \citenamefont {Huang}, \citenamefont {Qiao}, \citenamefont {Li},
  \citenamefont {Sun}, \citenamefont {He}, \citenamefont {Han}, \citenamefont
  {Chen}, \citenamefont {Tian}, \citenamefont {Sun}, \citenamefont {Liu},
  \citenamefont {Wu}, \citenamefont {Song}, \citenamefont {Liu}, \citenamefont
  {Ye}, \citenamefont {Huang}, \citenamefont {Ozawa}, \citenamefont {Dai},
  \citenamefont {Tallarida}, \citenamefont {Cui}, \citenamefont {Chen},
  \citenamefont {Jin}, \citenamefont {Zheng}, \citenamefont {Chen},
  \citenamefont {Wang}, \citenamefont {Yu}, \citenamefont {Li},\ and\
  \citenamefont {Yao}}]{Li2026}%
  \BibitemOpen
  \bibfield  {author} {\bibinfo {author} {\bibfnamefont {Y.}~\bibnamefont
  {Li}}, \bibinfo {author} {\bibfnamefont {Y.}~\bibnamefont {Han}}, \bibinfo
  {author} {\bibfnamefont {L.}~\bibnamefont {Yang}}, \bibinfo {author}
  {\bibfnamefont {W.}~\bibnamefont {He}}, \bibinfo {author} {\bibfnamefont
  {P.}~\bibnamefont {Ye}}, \bibinfo {author} {\bibfnamefont {W.}~\bibnamefont
  {Huang}}, \bibinfo {author} {\bibfnamefont {J.}~\bibnamefont {Qiao}},
  \bibinfo {author} {\bibfnamefont {Y.}~\bibnamefont {Li}}, \bibinfo {author}
  {\bibfnamefont {X.}~\bibnamefont {Sun}}, \bibinfo {author} {\bibfnamefont
  {T.}~\bibnamefont {He}}, \bibinfo {author} {\bibfnamefont {J.}~\bibnamefont
  {Han}}, \bibinfo {author} {\bibfnamefont {Y.}~\bibnamefont {Chen}}, \bibinfo
  {author} {\bibfnamefont {R.}~\bibnamefont {Tian}}, \bibinfo {author}
  {\bibfnamefont {H.}~\bibnamefont {Sun}}, \bibinfo {author} {\bibfnamefont
  {Y.}~\bibnamefont {Liu}}, \bibinfo {author} {\bibfnamefont {F.}~\bibnamefont
  {Wu}}, \bibinfo {author} {\bibfnamefont {B.}~\bibnamefont {Song}}, \bibinfo
  {author} {\bibfnamefont {Z.}~\bibnamefont {Liu}}, \bibinfo {author}
  {\bibfnamefont {M.}~\bibnamefont {Ye}}, \bibinfo {author} {\bibfnamefont
  {Y.}~\bibnamefont {Huang}}, \bibinfo {author} {\bibfnamefont
  {K.}~\bibnamefont {Ozawa}}, \bibinfo {author} {\bibfnamefont
  {J.}~\bibnamefont {Dai}}, \bibinfo {author} {\bibfnamefont {M.}~\bibnamefont
  {Tallarida}}, \bibinfo {author} {\bibfnamefont {S.}~\bibnamefont {Cui}},
  \bibinfo {author} {\bibfnamefont {J.}~\bibnamefont {Chen}}, \bibinfo {author}
  {\bibfnamefont {M.}~\bibnamefont {Jin}}, \bibinfo {author} {\bibfnamefont
  {W.}~\bibnamefont {Zheng}}, \bibinfo {author} {\bibfnamefont
  {C.}~\bibnamefont {Chen}}, \bibinfo {author} {\bibfnamefont {Z.}~\bibnamefont
  {Wang}}, \bibinfo {author} {\bibfnamefont {Z.-M.}\ \bibnamefont {Yu}},
  \bibinfo {author} {\bibfnamefont {X.}~\bibnamefont {Li}},\ and\ \bibinfo
  {author} {\bibfnamefont {Y.}~\bibnamefont {Yao}},\ }\href@noop {} {\bibinfo
  {title} {Pressure-induced superconducting-like transition in the $d$-wave
  altermagnet candidate {CsV$_2$Se$_2$O}}} (\bibinfo {year} {2026}),\ \Eprint
  {https://arxiv.org/abs/2604.09457} {arXiv:2604.09457 [cond-mat.supr-con]}
  \BibitemShut {NoStop}%
\bibitem [{\citenamefont {Swain}\ \emph {et~al.}(2022)\citenamefont {Swain},
  \citenamefont {Karmakar},\ and\ \citenamefont {Majumdar}}]{Swain2022}%
  \BibitemOpen
  \bibfield  {author} {\bibinfo {author} {\bibfnamefont {N.}~\bibnamefont
  {Swain}}, \bibinfo {author} {\bibfnamefont {M.}~\bibnamefont {Karmakar}},\
  and\ \bibinfo {author} {\bibfnamefont {P.}~\bibnamefont {Majumdar}},\
  }\bibfield  {title} {\bibinfo {title} {Spin-orbital liquids and
  insulator-metal transitions on the pyrochlore lattice},\ }\href@noop {}
  {\bibfield  {journal} {\bibinfo  {journal} {Phys. Rev. B}\ }\textbf {\bibinfo
  {volume} {106}},\ \bibinfo {pages} {245114} (\bibinfo {year}
  {2022})}\BibitemShut {NoStop}%
\bibitem [{\citenamefont {Karmakar}\ and\ \citenamefont
  {Swain}(2022)}]{Karmakar2022}%
  \BibitemOpen
  \bibfield  {author} {\bibinfo {author} {\bibfnamefont {M.}~\bibnamefont
  {Karmakar}}\ and\ \bibinfo {author} {\bibfnamefont {N.}~\bibnamefont
  {Swain}},\ }\bibfield  {title} {\bibinfo {title} {Transport and spectroscopic
  signatures of a disorder-stabilized metal in two-dimensional frustrated
  {Mott} insulators},\ }\href@noop {} {\bibfield  {journal} {\bibinfo
  {journal} {Phys. Rev. B}\ }\textbf {\bibinfo {volume} {105}},\ \bibinfo
  {pages} {195146} (\bibinfo {year} {2022})}\BibitemShut {NoStop}%
\bibitem [{\citenamefont {Swain}\ and\ \citenamefont
  {Karmakar}(2020)}]{Swain2020}%
  \BibitemOpen
  \bibfield  {author} {\bibinfo {author} {\bibfnamefont {N.}~\bibnamefont
  {Swain}}\ and\ \bibinfo {author} {\bibfnamefont {M.}~\bibnamefont
  {Karmakar}},\ }\bibfield  {title} {\bibinfo {title} {Strain-induced
  superconductor-insulator transition on a {Lieb} lattice},\ }\href
  {https://doi.org/10.1103/PhysRevResearch.2.023136} {\bibfield  {journal}
  {\bibinfo  {journal} {Phys. Rev. Research}\ }\textbf {\bibinfo {volume}
  {2}},\ \bibinfo {pages} {023136} (\bibinfo {year} {2020})}\BibitemShut
  {NoStop}%
\bibitem [{\citenamefont {Kunwar}\ and\ \citenamefont
  {Karmakar}(2024)}]{Kunwar2024}%
  \BibitemOpen
  \bibfield  {author} {\bibinfo {author} {\bibfnamefont {S.~S.}\ \bibnamefont
  {Kunwar}}\ and\ \bibinfo {author} {\bibfnamefont {M.}~\bibnamefont
  {Karmakar}},\ }\href@noop {} {\bibinfo {title} {Kagome {Hubbard} model away
  from the strong coupling limit: Flat band localization and non {Fermi} liquid
  signatures}} (\bibinfo {year} {2024}),\ \Eprint
  {https://arxiv.org/abs/2404.05787} {arXiv:2404.05787 [cond-mat.str-el]}
  \BibitemShut {NoStop}%
\bibitem [{\citenamefont {Kunwar}\ and\ \citenamefont
  {Karmakar}(2026)}]{Kunwar2026}%
  \BibitemOpen
  \bibfield  {author} {\bibinfo {author} {\bibfnamefont {S.~S.}\ \bibnamefont
  {Kunwar}}\ and\ \bibinfo {author} {\bibfnamefont {M.}~\bibnamefont
  {Karmakar}},\ }\bibfield  {title} {\bibinfo {title} {Straintronics across
  {Lieb-kagome} interconversion and variable transport scaling exponents},\
  }\href {https://doi.org/10.1103/PhysRevMaterials.10.L011002} {\bibfield
  {journal} {\bibinfo  {journal} {Phys. Rev. Materials}\ }\textbf {\bibinfo
  {volume} {10}},\ \bibinfo {pages} {L011002} (\bibinfo {year}
  {2026})}\BibitemShut {NoStop}%
\bibitem [{\citenamefont {Kannan}\ \emph {et~al.}(2026)\citenamefont {Kannan},
  \citenamefont {Savla},\ and\ \citenamefont {Karmakar}}]{Kannan2026}%
  \BibitemOpen
  \bibfield  {author} {\bibinfo {author} {\bibfnamefont {S.}~\bibnamefont
  {Kannan}}, \bibinfo {author} {\bibfnamefont {J.}~\bibnamefont {Savla}},\ and\
  \bibinfo {author} {\bibfnamefont {M.}~\bibnamefont {Karmakar}},\ }\bibfield
  {title} {\bibinfo {title} {Correlation-driven metal-insulator transition in a
  $d$-wave altermagnet},\ }\href {https://doi.org/10.1103/38sm-tbwf} {\bibfield
   {journal} {\bibinfo  {journal} {Phys. Rev. B}\ }\textbf {\bibinfo {volume}
  {114}},\ \bibinfo {pages} {L231109} (\bibinfo {year} {2026})}\BibitemShut
  {NoStop}%
\end{thebibliography}%

\clearpage
\onecolumngrid
\smfloats
\renewcommand{\thefigure}{S\arabic{figure}}
\renewcommand{\theequation}{S\arabic{equation}}
\renewcommand{\thetable}{S\arabic{table}}
\renewcommand{\thesection}{S\arabic{section}}
\renewcommand{\theHfigure}{S\arabic{figure}}
\renewcommand{\theHequation}{S\arabic{equation}}
\renewcommand{\theHsection}{S\arabic{section}}
\setcounter{figure}{0}\setcounter{equation}{0}\setcounter{table}{0}\setcounter{section}{0}
\begin{center}
{\normalsize\bfseries Supplemental Material for: Pair density waves in altermagnets: thermal stability and spectroscopic fingerprints}\\[4pt]
{\small Amrutha N Madhusuthanan and Madhuparna Karmakar}\\
{\footnotesize Department of Physics and Nanotechnology, SRM Institute of Science and Technology, Kattankulathur, Chennai 603203, India}
\end{center}
\vspace{4pt}

\section{Microscopic Model, SPA Formalism, and Observables}
\label{sec:S-formalism}

\subsection{Microscopic origin of the altermagnetic hopping $\hat{H}_0$}

In our model, altermagnetism originates from single-particle band-structure reconstruction rather than electronic correlations, so the nearest-neighbor pairing interaction governs only the superconducting correlations while leaving the altermagnetic band splitting intact. Specifically, the noninteracting $d_{x^2-y^2}$ altermagnetic Hamiltonian $\hat{H}_0$ on a two-dimensional square lattice derives from a four-band sublattice-spin model \cite{Smejkal2022beyond,Smejkal2022emerging}. In the sublattice-spin basis $\Psi^{\dagger}_{\mathbf{k}} = \big(c^{\dagger}_{\mathbf{k}A\uparrow}, c^{\dagger}_{\mathbf{k}B\uparrow}, c^{\dagger}_{\mathbf{k}A\downarrow}, c^{\dagger}_{\mathbf{k}B\downarrow}\big)$, the four-band Hamiltonian reads
\begin{equation}
\hat{H}_{4b} = -2t(\cos k_x + \cos k_y)\sigma_0\tau_0 + \frac{t_{\mathrm{am}}}{2}(\cos k_x - \cos k_y)\sigma_0\tau_z - J\,\sigma_z\tau_z - \mu\,\sigma_0\tau_0,
\label{eq:S-H4b}
\end{equation}
where $\sigma_{\alpha}$ and $\tau_{\alpha}$ ($\alpha \in \{0,x,y,z\}$) are Pauli matrices acting in spin and sublattice spaces, respectively, $t_{\mathrm{am}}$ parameterizes the anisotropic electric crystal-field hopping asymmetry between sublattices $A$ and $B$, and $J$ is the momentum-independent staggered spin-sublattice exchange energy. While crystal-field anisotropy alone ($J=0, t_{\mathrm{am}}\neq 0$) splits the bands into spin-degenerate $A$-like and $B$-like pairs and staggered exchange alone ($t_{\mathrm{am}}=0, J\neq 0$) produces opposite spin splittings on the two sublattices while preserving the spin degeneracy of the overall band spectrum, their combination ($t_{\mathrm{am}}\neq 0$ and $J\neq 0$) lifts Kramers degeneracy in a momentum-dependent manner—producing opposite spin splittings near $X=(\pi,0)$ and $Y=(0,\pi)$ while vanishing at $\Gamma$ and along the Brillouin-zone diagonals $|k_x|=|k_y|$.

For a large staggered exchange $J \gg t, t_{\mathrm{am}}$, the two higher-energy bands are pushed well above the Fermi level, locking the low-energy spin and sublattice degrees of freedom into the effective doublet $\Psi^{\dagger}_{2b,\mathbf{k}} = \big(c^{\dagger}_{\mathbf{k}A\uparrow}, c^{\dagger}_{\mathbf{k}B\downarrow}\big)$ \cite{Smejkal2022beyond}. Projecting onto this low-energy subspace yields the effective two-band Hamiltonian
\begin{equation}
\hat{H}_{2b} = -2t(\cos k_x + \cos k_y)s_0 + \frac{t_{\mathrm{am}}}{2}(\cos k_x - \cos k_y)s_z - \mu\,s_0,
\label{eq:S-H2b}
\end{equation}
where $s_{\alpha}$ are Pauli matrices in the joint spin-sublattice basis and the constant energy shift $-J$ of both states of the doublet is absorbed into the chemical potential. Performing an inverse Fourier transform to real space—and including a uniform Zeeman field $h$—gives the noninteracting part of our lattice Hamiltonian (with $t_{ij}=t=1$ and $d$-wave form factor $\eta_{ij}=+1/4$ on $\hat{x}$ bonds and $-1/4$ on $\hat{y}$ bonds):
\begin{equation}
\hat{H}_{0} = \sum_{\langle ij\rangle,\sigma}\left(-t_{ij}+\sigma\, t_{\mathrm{am}}\,\eta_{ij}\right)\left(\hat{c}^{\dagger}_{i\sigma}\hat{c}_{j\sigma}+\mathrm{h.c.}\right) - \sum_{i,\sigma}\left(\mu+\sigma h\right)\hat{n}_{i\sigma}.
\label{eq:S-H0}
\end{equation}
In momentum space this gives the bands $\varepsilon_\sigma(\mathbf{k})=-2t(\cos k_x+\cos k_y)+\sigma\frac{t_{\mathrm{am}}}{2}(\cos k_x-\cos k_y)-\mu-\sigma h$, i.e.\ a spin splitting $\varepsilon_\uparrow-\varepsilon_\downarrow=t_{\mathrm{am}}(\cos k_x-\cos k_y)-2h$. At $h=0$ it vanishes on the zone diagonals and reaches $\pm2t_{\mathrm{am}}$ at $X$ and $Y$; at $h\neq0$ the zero-splitting contour moves to $\cos k_x-\cos k_y=2h/t_{\mathrm{am}}$. The sign of the splitting term in Eqs.~\eqref{eq:S-H4b} and \eqref{eq:S-H2b} is fixed here by the convention $\eta_{ij}=+1/4$ on $\hat{x}$ bonds; the opposite convention simply interchanges the roles of the two spin species.

\subsection{Hubbard-Stratonovich decomposition and SPA Monte Carlo}

Expressing the partition function as a Grassmann functional integral over fermionic fields $\psi_{i\sigma}(\tau)$ and $\bar{\psi}_{i\sigma}(\tau)$ at inverse temperature $\beta=1/T$, we retain the spin-singlet pairing channel of the nearest-neighbor attractive interaction and decouple it via a Hubbard-Stratonovich transformation \cite{Hubbard1959,Schulz1990}:
\begin{equation}
Z = \int \mathcal{D}[\psi,\bar{\psi},\Delta,\Delta^{*}]\, e^{-\int_{0}^{\beta}\!d\tau\,\mathcal{L}(\tau)},
\label{eq:S-Z}
\end{equation}
with $\mathcal{L}(\tau) = \mathcal{L}_{0}(\tau)+\mathcal{L}_{\Delta}(\tau)+\mathcal{L}_{\mathrm{cl}}(\tau)$, where the noninteracting, pairing, and classical stiffness terms read
\begin{align}
\mathcal{L}_{0} &= \sum_{ij,\sigma}\bar{\psi}_{i\sigma}\Big[\left(\partial_{\tau}-\mu-\sigma h\right)\delta_{ij} - t_{ij} + \sigma\, t_{\mathrm{am}}\,\eta_{ij}\Big]\psi_{j\sigma}, \label{eq:S-L0} \\
\mathcal{L}_{\Delta} &= \sum_{\langle ij\rangle}\Big[\Delta_{ij}\left(\bar{\psi}_{i\uparrow}\bar{\psi}_{j\downarrow}+\bar{\psi}_{j\uparrow}\bar{\psi}_{i\downarrow}\right) + \mathrm{h.c.}\Big], \label{eq:S-Ldelta} \\
\mathcal{L}_{\mathrm{cl}} &= 4\sum_{\langle ij\rangle}\frac{|\Delta_{ij}|^{2}}{|U|}, \label{eq:S-Lcl}
\end{align}
with all fields evaluated at imaginary time $\tau$.
Here $\Delta_{ij}(\tau)=|\Delta_{ij}(\tau)|e^{i\phi_{ij}(\tau)}$ is the complex bosonic auxiliary pairing field defined on nearest-neighbor bonds. Integrating out the quadratic fermionic fields generates the effective action solely in terms of the background pairing configurations $\{\Delta_{ij}(\tau)\}$:
\begin{align}
Z &= \int \mathcal{D}[\Delta,\Delta^{*}]\, e^{-S_{\mathrm{eff}}\{\Delta,\Delta^{*}\}}, \label{eq:S-Zeff} \\
S_{\mathrm{eff}} &= -\ln\mathrm{Det}\left[\mathcal{G}^{-1}\{\Delta,\Delta^{*}\}\right] + \int_{0}^{\beta}\!d\tau\,\mathcal{L}_{\mathrm{cl}}(\tau), \label{eq:S-Seff}
\end{align}
where $\mathcal{G}$ is the Nambu-Gor'kov electronic Green's function in a given $\{\Delta,\Delta^{*}\}$ background.

Expanding $\Delta_{ij}(\tau) = \sum_{n}\Delta_{ij,n}e^{-i\Omega_{n}\tau}$ in bosonic Matsubara frequencies $\Omega_{n}=2\pi n T$, the static path approximation (SPA) retains all spatial fluctuations of the pairing field to arbitrary order while keeping only the static ($\Omega_{n}=0$) thermal mode, $\Delta_{ij}(\tau)\rightarrow\Delta_{ij}$ \cite{Mayr2005,Tarat2015,Karmakar2016}. The problem then maps onto fermions moving in a classical, spatially correlated background $\{\Delta_{ij}\}$ governed by the effective Hamiltonian
\begin{equation}
\begin{split}
H_{\mathrm{eff}} ={}& \sum_{\langle ij\rangle,\sigma}\left(-t_{ij}+\sigma\, t_{\mathrm{am}}\,\eta_{ij}\right)\left(c^{\dagger}_{i\sigma}c_{j\sigma}+\mathrm{h.c.}\right) - \sum_{i,\sigma}\left(\mu+\sigma h\right)n_{i\sigma} \\
&+ \sum_{\langle ij\rangle}\Big[\Delta_{ij}\left(c^{\dagger}_{i\uparrow}c^{\dagger}_{j\downarrow}+c^{\dagger}_{j\uparrow}c^{\dagger}_{i\downarrow}\right)+\mathrm{h.c.}\Big] + 4\sum_{\langle ij\rangle}\frac{|\Delta_{ij}|^{2}}{|U|}.
\end{split}
\label{eq:S-Heff}
\end{equation}
Equilibrium configurations obey the Boltzmann weight
\begin{equation}
P\{\Delta_{ij}\} \propto \mathrm{Tr}_{c,c^{\dagger}}\,e^{-\beta H_{\mathrm{eff}}},
\label{eq:S-boltz}
\end{equation}
which is sampled via classical Metropolis Monte Carlo. To reduce the $\mathcal{O}(N^3)$ cost of exact diagonalization at every local update, we employ the traveling cluster approximation (TCA) \cite{Kumar2006,Karmakar2024}, updating bonds within a moving cluster embedded in the full $L\times L$ lattice. While SPA neglects quantum fluctuations and reduces to mean-field theory as $T\rightarrow 0$, it captures nonperturbative thermal amplitude and phase fluctuations across the superconducting, pseudogap, and normal regimes—a capability benchmarked extensively across unconventional superconductors \cite{Mayr2005,Karmakar2020,Karmakar2024}, frustrated Mott systems \cite{Swain2022,Karmakar2022}, flat-band lattices \cite{Swain2020,Kunwar2026}, and altermagnetic metals \cite{Kannan2026}.

\subsection{Definitions of thermodynamic and spectroscopic observables}

All observables are averaged over equilibrated Monte Carlo configurations ($\langle\cdots\rangle$):
\begin{itemize}
\item \textit{Pairing-amplitude structure factors:}
\begin{equation}
S_{\alpha}(\mathbf{q}) = \Bigg\langle\frac{\big|\sum_{i}\Delta^{\alpha}_{i}\,e^{i\mathbf{q}\cdot\mathbf{r}_{i}}\big|^{2}}{N\sum_{i}|\Delta^{\alpha}_{i}|^{2}}\Bigg\rangle,
\label{eq:S-Sq}
\end{equation}
with $\alpha=x,y$, where $\Delta^{\alpha}_{i}=|\Delta_{i}|\,e^{i\theta^{\alpha}_{i}}$ is the complex pairing field (amplitude and phase) on the nearest-neighbor bond along $\hat{\alpha}$ emerging from site $i$, and $N=L^{2}$. The normalization by $N\sum_{i}|\Delta^{\alpha}_{i}|^{2}$ makes $S_{\alpha}(\mathbf{q})=1$ for a uniform pairing field at $\mathbf{q}=(0,0)$. At $h=0$ the model is invariant under a $\pi/2$ rotation combined with a spin flip, so $S_y(\mathbf{q})$ is expected to show the ordering peaks of $S_x(\mathbf{q})$ rotated by $\pi/2$, up to sampling noise; a Zeeman field breaks this connection, and we therefore show both channels separately.
\item \textit{Local and average spin polarization:}
\begin{equation}
m_{i} = \langle n_{i\uparrow} - n_{i\downarrow}\rangle, \qquad m = \frac{1}{N}\sum_{i} m_{i}.
\label{eq:S-mag}
\end{equation}
\item \textit{Single-particle density of states (DOS):}
\begin{equation}
N(\omega) = \Big\langle\frac{1}{N}\sum_{i,n}\Big[|u^{i}_{n}|^{2}\,\delta(\omega-E_{n}) + |v^{i}_{n}|^{2}\,\delta(\omega+E_{n})\Big]\Big\rangle,
\label{eq:S-dos}
\end{equation}
where $E_{n}$ and $(u^{i}_{n},v^{i}_{n})$ are the Bogoliubov-de Gennes eigenvalues and eigenvectors of $H_{\mathrm{eff}}$.
\item \textit{Spin-resolved spectral function and Fermi-surface weight:}
\begin{equation}
A_{\sigma}(\mathbf{k},\omega) = -\frac{1}{\pi}\,\mathrm{Im}\,G_{\sigma}(\mathbf{k},\omega+i0^{+}),
\label{eq:S-Akw}
\end{equation}
where $G_{\sigma}(\mathbf{k},\omega) = \lim_{\delta\rightarrow 0} G_{\sigma}(\mathbf{k},i\omega_{n})|_{i\omega_{n}\rightarrow\omega+i\delta}$ is the real-frequency continuation of the Matsubara Green's function $G_{\sigma}(\mathbf{k},\tau)=-\langle c_{\mathbf{k}\sigma}(\tau)c^{\dagger}_{\mathbf{k}\sigma}(0)\rangle$. Because the pairing field $\{\Delta_{ij}\}$ is static in SPA, $G_{\sigma}(\mathbf{k},\omega)$ is obtained directly on the real frequency axis without numerical analytic continuation. We track both the spin-summed low-energy weight $A(\mathbf{k},0)=\sum_{\sigma}A_{\sigma}(\mathbf{k},0)$ (mapping the Fermi surface) and the spin difference $\delta A(\mathbf{k},\omega) = A_{\uparrow}(\mathbf{k},\omega) - A_{\downarrow}(\mathbf{k},\omega)$.
\end{itemize}

\section{Fermi-Surface and Spectral Function Evolution at \texorpdfstring{$h=0$}{h=0}}
\label{sec:S-h0}

\begin{figure}[!htbp]
  \centering
  \subc{0.92}{a}{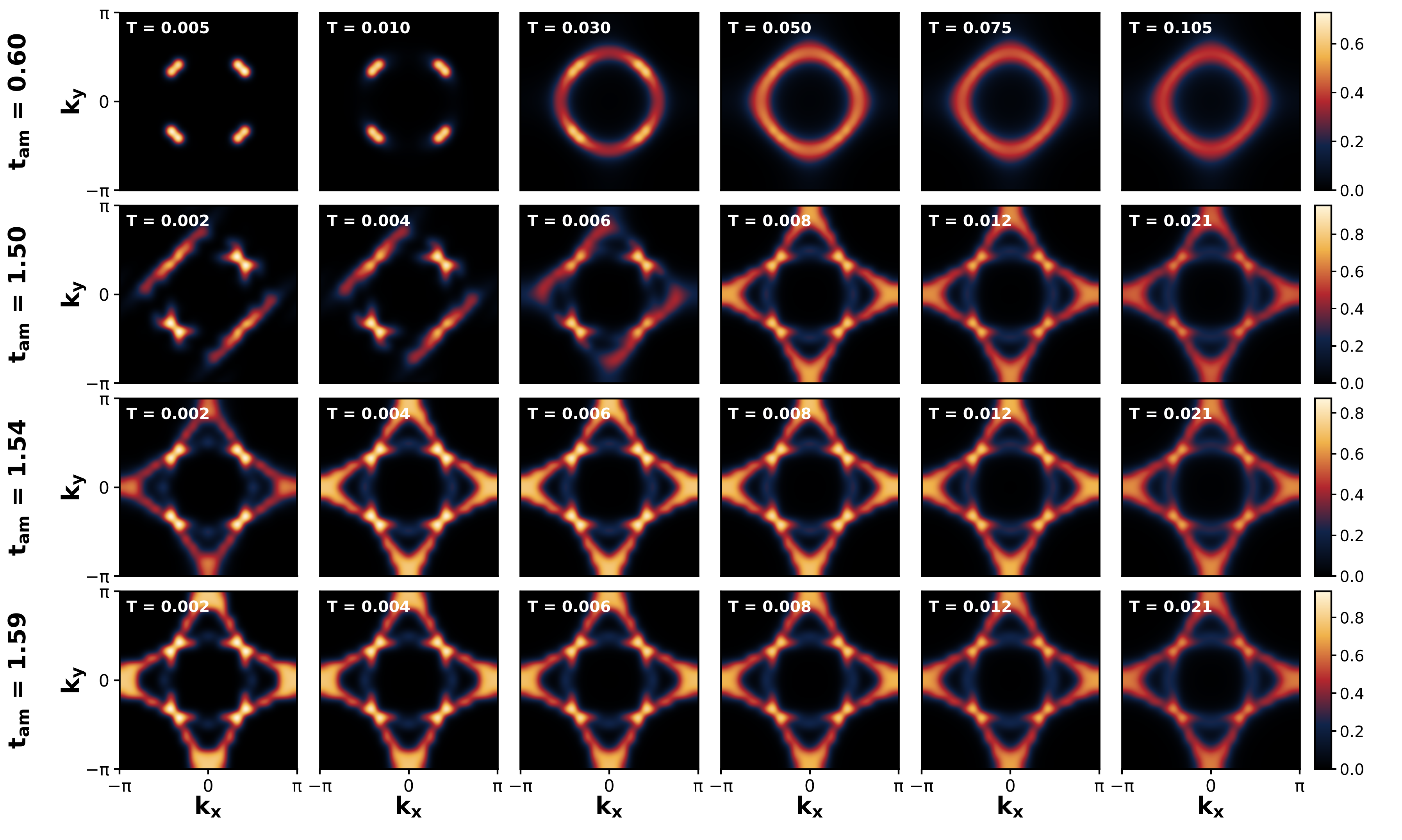}
  \subc{0.92}{b}{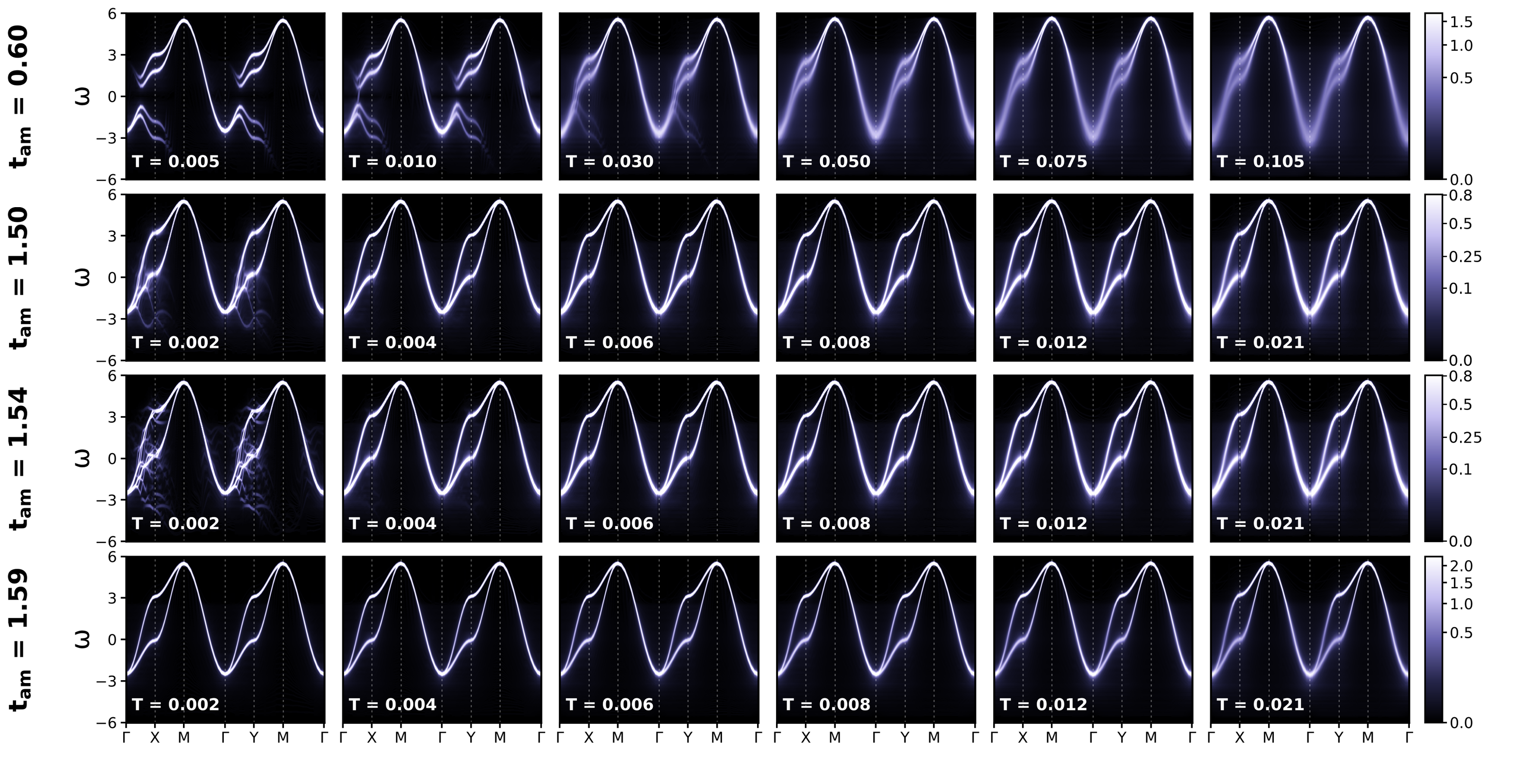}
  \caption{(a) Temperature evolution of the zero-energy spectral weight $A(\mathbf{k},\omega=0)$ at $h=0$ on an $L=24$ lattice for four representative couplings: $t_{\mathrm{am}}=0.60t$ ($d$-BCS, top row), $t_{\mathrm{am}}=1.50t$ ($\text{PDW 1D}$, second row), $t_{\mathrm{am}}=1.54t$ ($\text{PDW 2D}$, third row), and $t_{\mathrm{am}}=1.59t$ (normal altermagnetic metal, bottom row). The temperature is printed in each panel (in units of $t$); the three rows with $t_{\mathrm{am}}\ge1.50t$ share the same temperature grid. (b) Spin-summed spectral function $A(\mathbf{k},\omega)$ along the path $\Gamma\text{--}X\text{--}M\text{--}\Gamma\text{--}Y\text{--}M\text{--}\Gamma$ at $h=0$ for the same regimes and temperature grids.}
  \label{fig:S1}
\end{figure}

Figures~\ref{fig:S1}(a) and \ref{fig:S1}(b) show how the Fermi-surface map $A(\mathbf{k},0)$ and the spin-summed spectral function $A(\mathbf{k},\omega)$ evolve with temperature in the four zero-field regimes. In the uniform $d$-BCS state at $t_{\mathrm{am}}=0.60t$ (top row), $A(\mathbf{k},0)$ at $T=0.005t$ consists of four point nodes on the Brillouin-zone diagonals, where the $d_{x^2-y^2}$ gap vanishes, and Fig.~\ref{fig:S1}(b) shows a clean Bogoliubov gap along $\Gamma\text{--}X\text{--}M$ and $\Gamma\text{--}Y\text{--}M$. At $T=0.010t$ the nodal spots begin to stretch into arcs, and by $T=0.030t$, close to the gap-closing scale $T_{\mathrm{gap}}\approx0.031t$, the contour is almost closed. Above $T_c\approx0.037t$ the spectral peaks broaden and the gap in $A(\mathbf{k},\omega)$ fills in. The broadening continues through the pseudogap window ($T=0.050t$, below $T_{\mathrm{pg}}\approx0.053t$ at this $t_{\mathrm{am}}$) and into the normal state ($T=0.075t$ and $0.105t$).

In the $\text{PDW 1D}$ regime ($t_{\mathrm{am}}=1.50t$) the zero-energy map looks very different from the normal metal at low temperature. At $T=0.002t$ and $0.004t$ the spin-split, star-shaped contour breaks up into short diagonal arcs and two diagonal streaks, which is what one expects when pairs with a finite center-of-mass momentum $\mathbf{Q}$ gap only part of the Fermi surface. These features weaken at $T=0.006t$ and are essentially gone by $T=0.008t$, close to $T_c\approx0.0086t$ obtained from $S_x(\mathbf{Q})$ (Fig.~\ref{fig:S4_h0}). In the $\text{PDW 2D}$ regime ($t_{\mathrm{am}}=1.54t$) the contour keeps its star shape and the reconstruction appears as faint extra arcs inside the contour at the lowest temperatures, which are absent in the normal metal at $t_{\mathrm{am}}=1.59t$. The imprint of the pairing is clearer in $A(\mathbf{k},\omega)$ [Fig.~\ref{fig:S1}(b)]: at $T=0.002t$ there are replica branches and avoided crossings close to $\omega=0$ along $\Gamma\text{--}X$ and $\Gamma\text{--}Y$ (again more visible in the $t_{\mathrm{am}}=1.54t$ row), and they fade by $T=0.004t$. The spectral reconstruction is therefore strongest at the lowest temperatures, while the order itself, measured through $S_x(\mathbf{Q})$, persists up to $T_c$. At $t_{\mathrm{am}}=1.59t$ both $A(\mathbf{k},0)$ and $A(\mathbf{k},\omega)$ are those of an altermagnetic normal metal.

\section{Finite Zeeman Field (\texorpdfstring{$h \neq 0$}{h nonzero}) Thermal Evolution}
\label{sec:S-h}

\begin{figure}[!htbp]
  \centering
  \subc{0.92}{a}{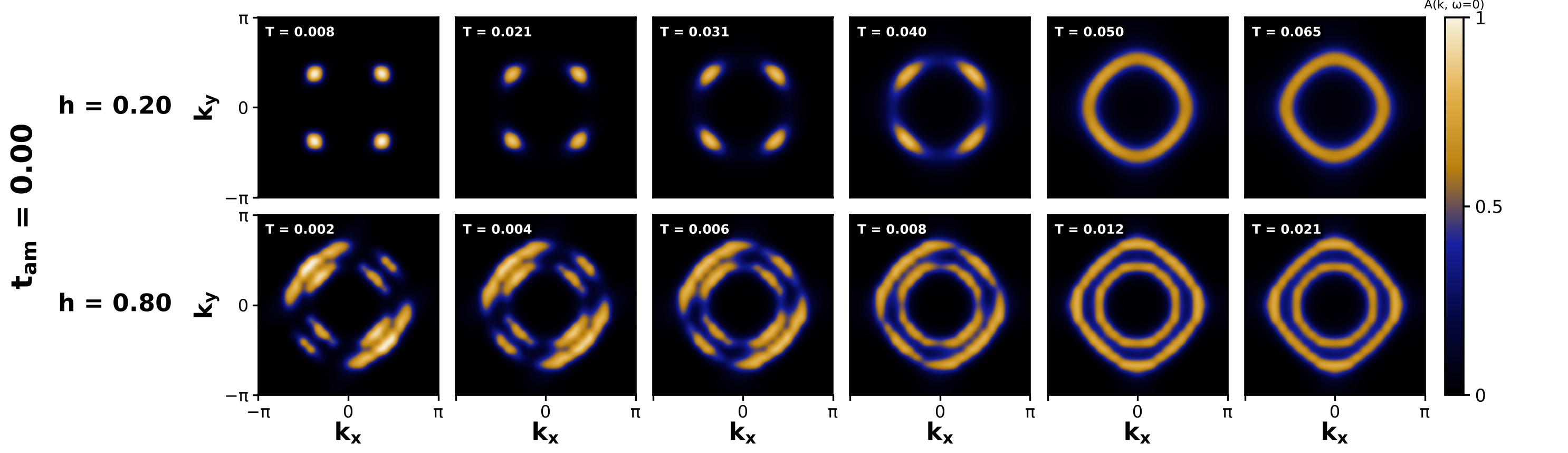}
  \subc{0.92}{b}{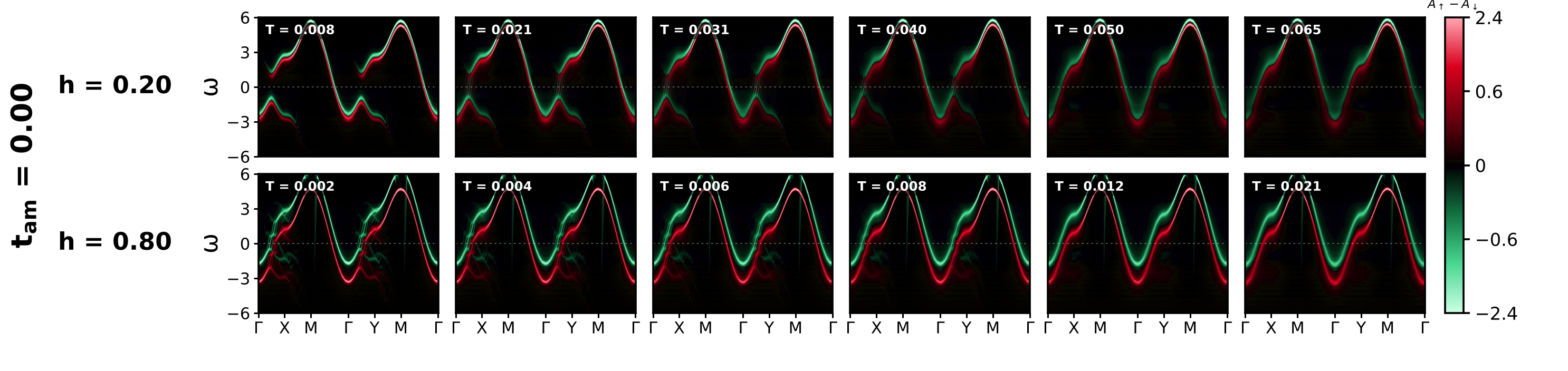}
  \subc{0.55}{c}{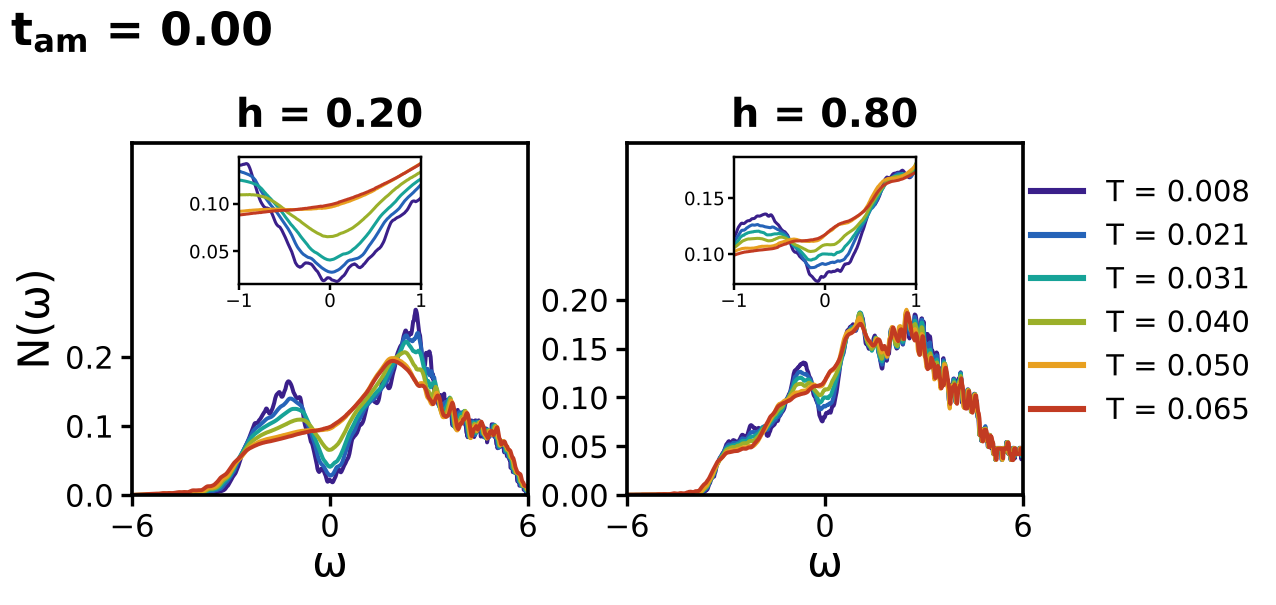}
  \caption{Spectroscopic evolution at $t_{\mathrm{am}}=0.00t$ (conventional Pauli-limited $d$-wave limit) for $h=0.20t$ (QBP, $T=0.008$--$0.065t$) and $h=0.80t$ (diagonal FFLO phase, $T=0.002$--$0.021t$); the temperatures are given in each panel. (a) zero-energy spectral weight $A(\mathbf{k},0)$. (b) spin-difference spectral function $\delta A(\mathbf{k},\omega) = A_{\uparrow}(\mathbf{k},\omega)-A_{\downarrow}(\mathbf{k},\omega)$ along $\Gamma\text{--}X\text{--}M\text{--}\Gamma\text{--}Y\text{--}M\text{--}\Gamma$. (c) single-particle density of states $N(\omega)$ with low-energy insets.}
  \label{fig:S3a_spec}
\end{figure}

\begin{figure}[!htbp]
  \centering
  \subc{0.92}{a}{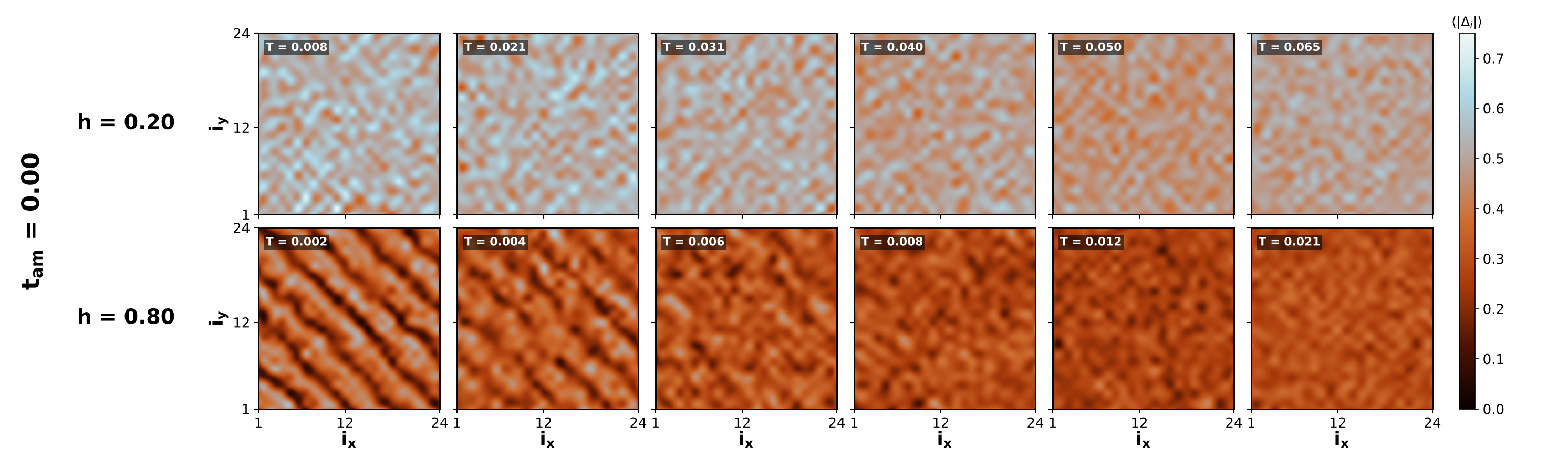}
  \subc{0.92}{b}{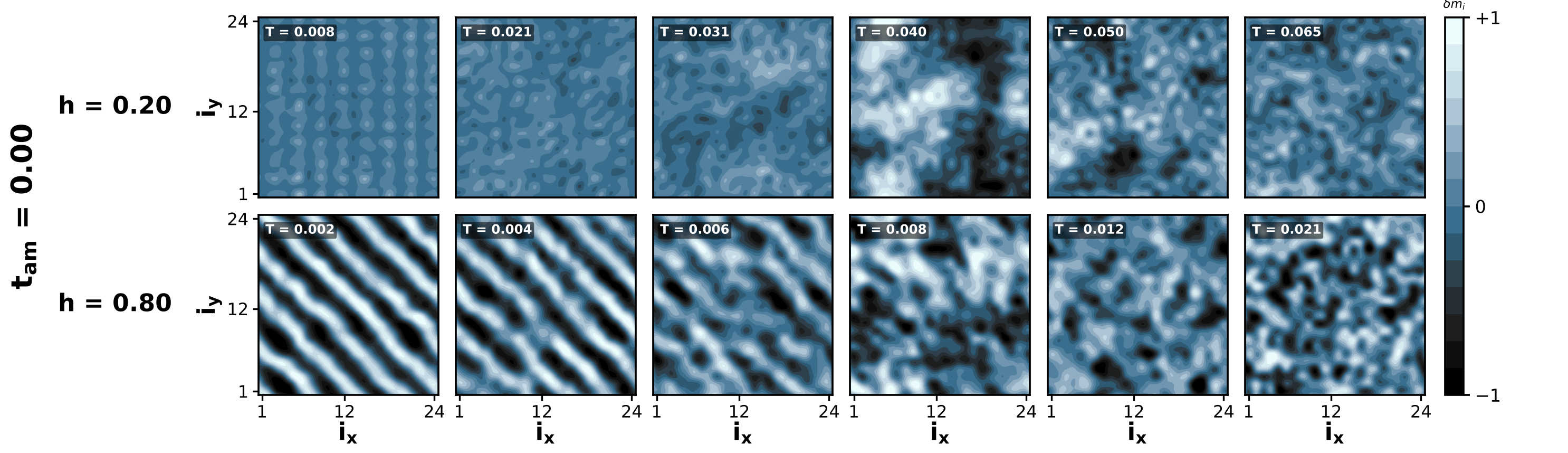}
  \caption{Real-space thermal evolution at $t_{\mathrm{am}}=0.00t$ for $h=0.20t$ and $h=0.80t$. (a) thermally averaged pairing amplitude $\langle|\Delta_i|\rangle$. (b) local spin-polarization modulation $\delta m_i = m_i - m$, highlighting diagonal FFLO stripes at $h=0.80t$ below $T_c \approx 0.0115t$.}
  \label{fig:S3a_real}
\end{figure}

\begin{figure}[!htbp]
  \centering
  \subc{0.80}{a}{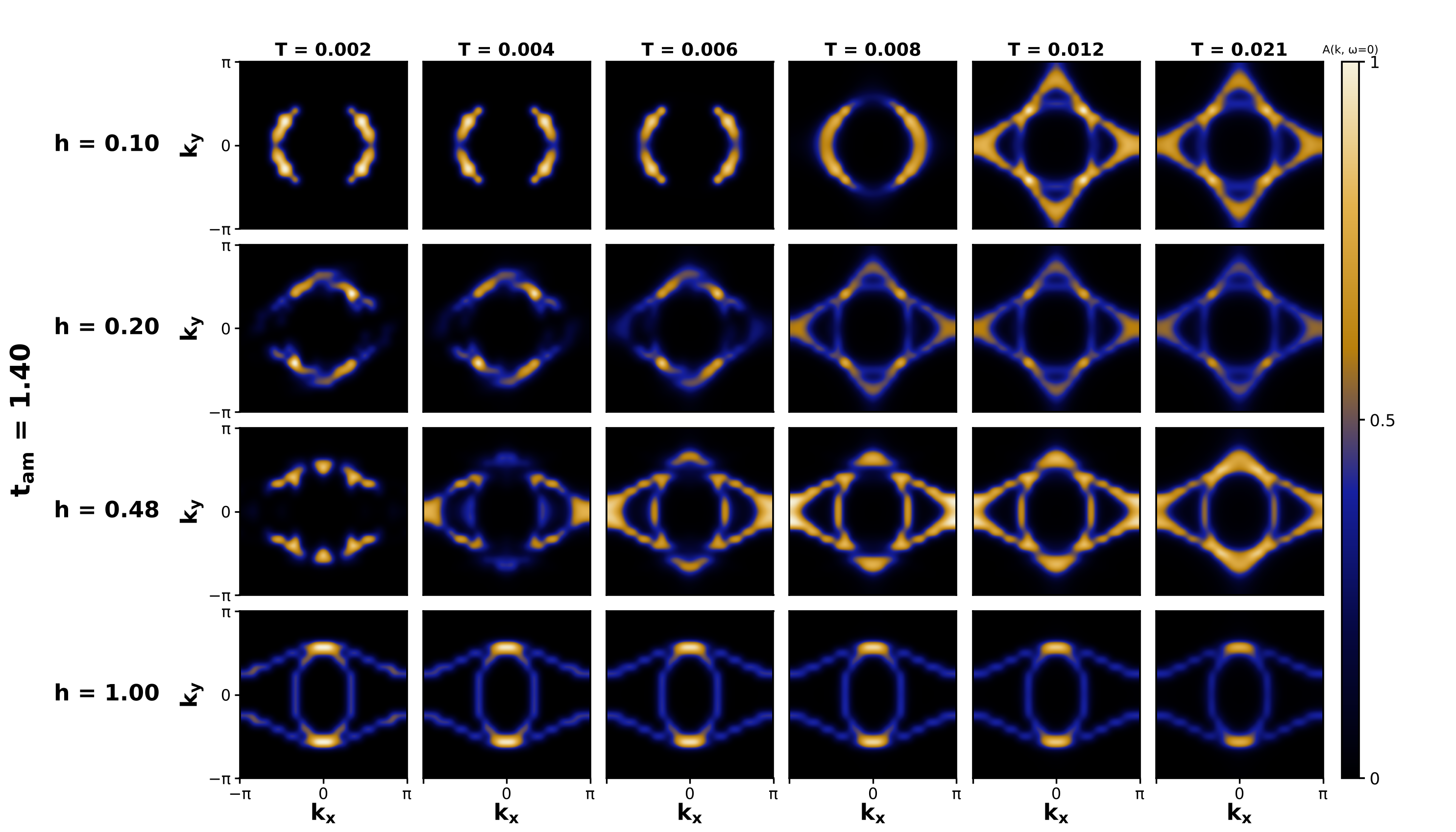}
  \subc{0.80}{b}{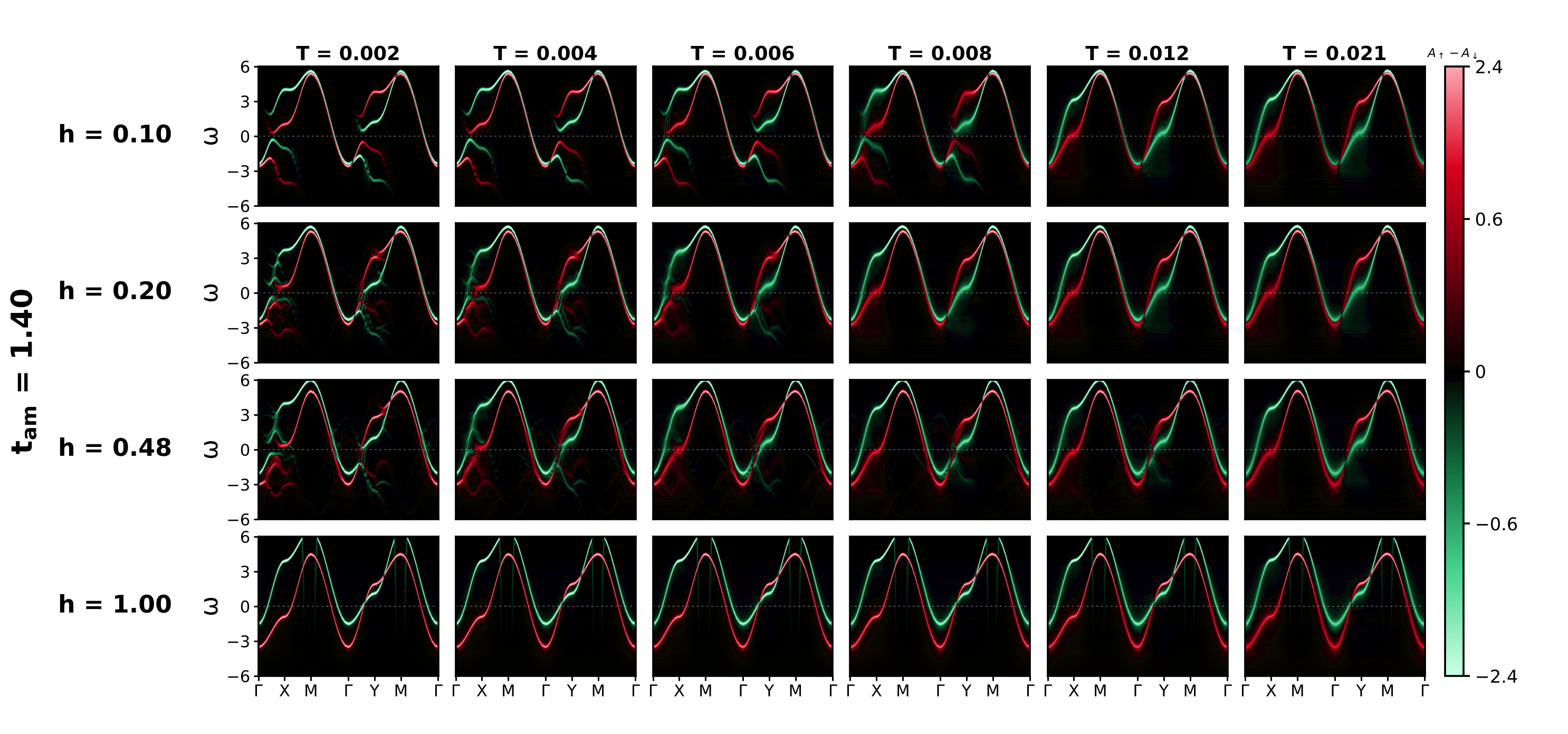}
  \subc{0.80}{c}{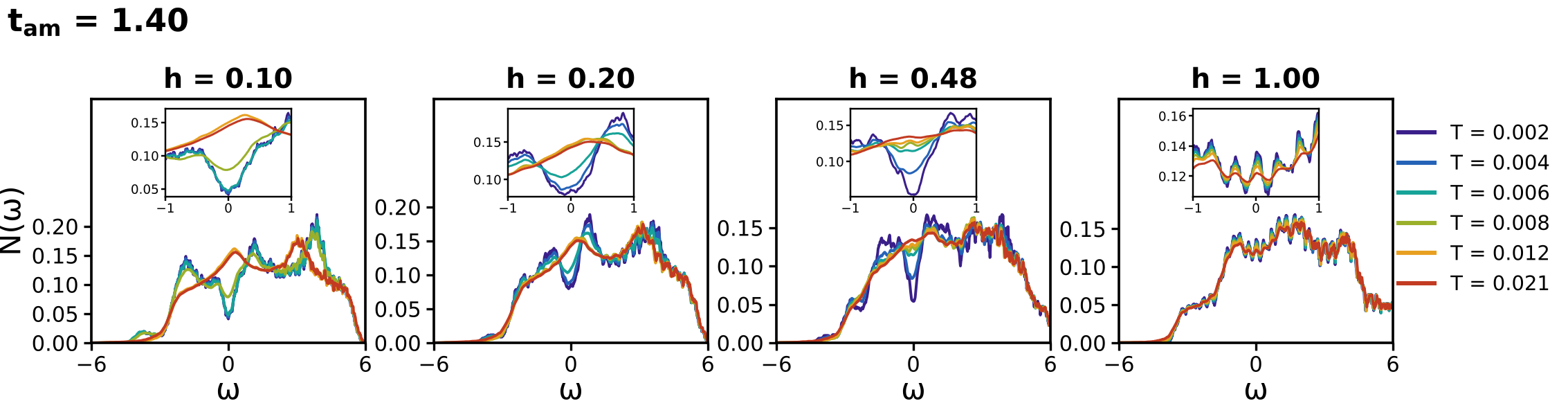}
  \caption{Spectroscopic evolution in the strong altermagnetic regime $t_{\mathrm{am}}=1.40t$ across $h=0.10t$ (QBP, $\mathbf{Q}=(0,0)$), $h=0.20t$ (non-collinear LO), $h=0.48t$ (collinear Type-I LO), and $h=1.00t$ (PFL). (a) Fermi-surface weight $A(\mathbf{k},0)$. (b) spin-difference spectral function $\delta A(\mathbf{k},\omega)$. (c) density of states $N(\omega)$.}
  \label{fig:S3b_spec}
\end{figure}

\begin{figure}[!htbp]
  \centering
  \subc{0.78}{a}{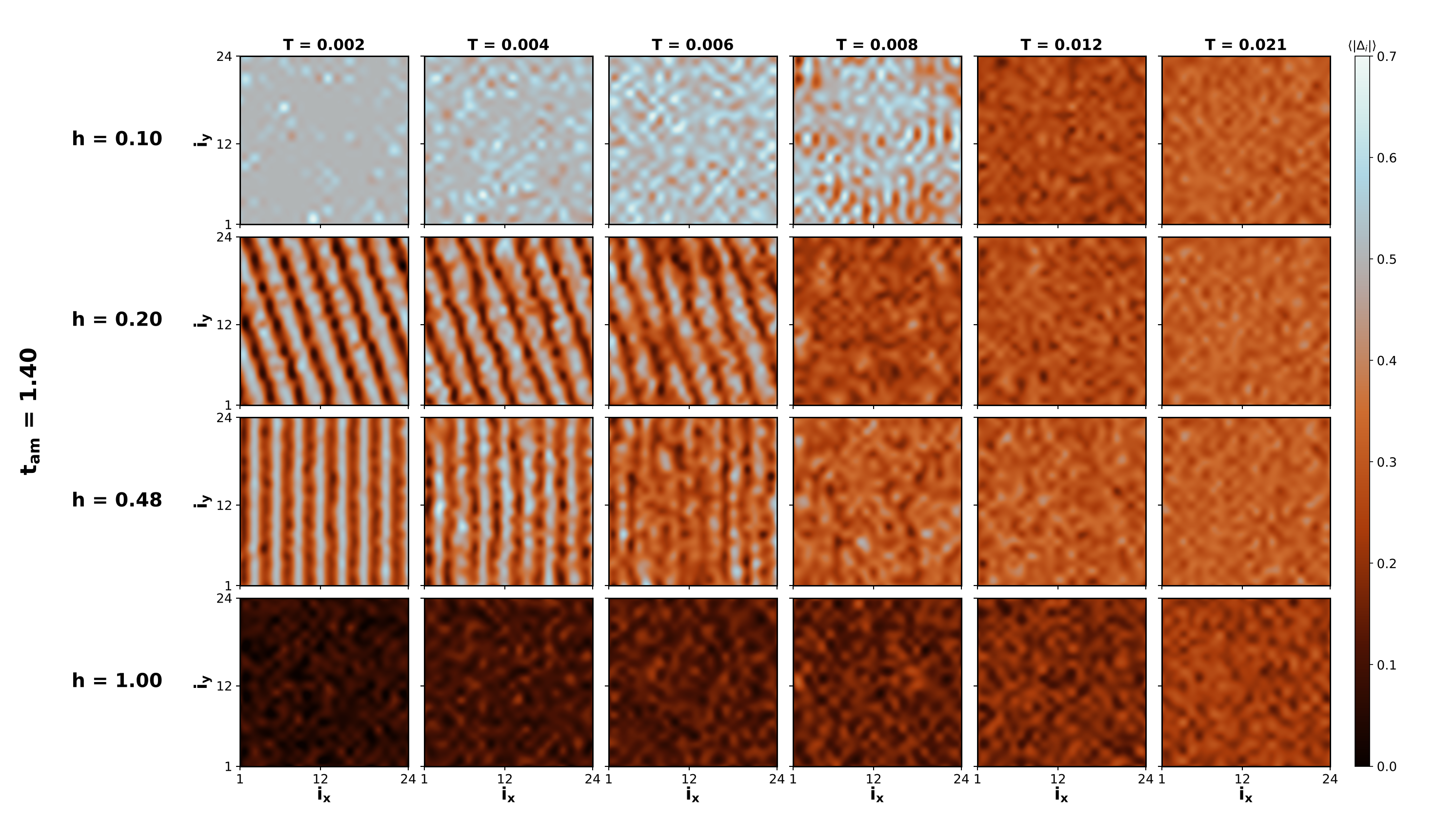}
  \subc{0.78}{b}{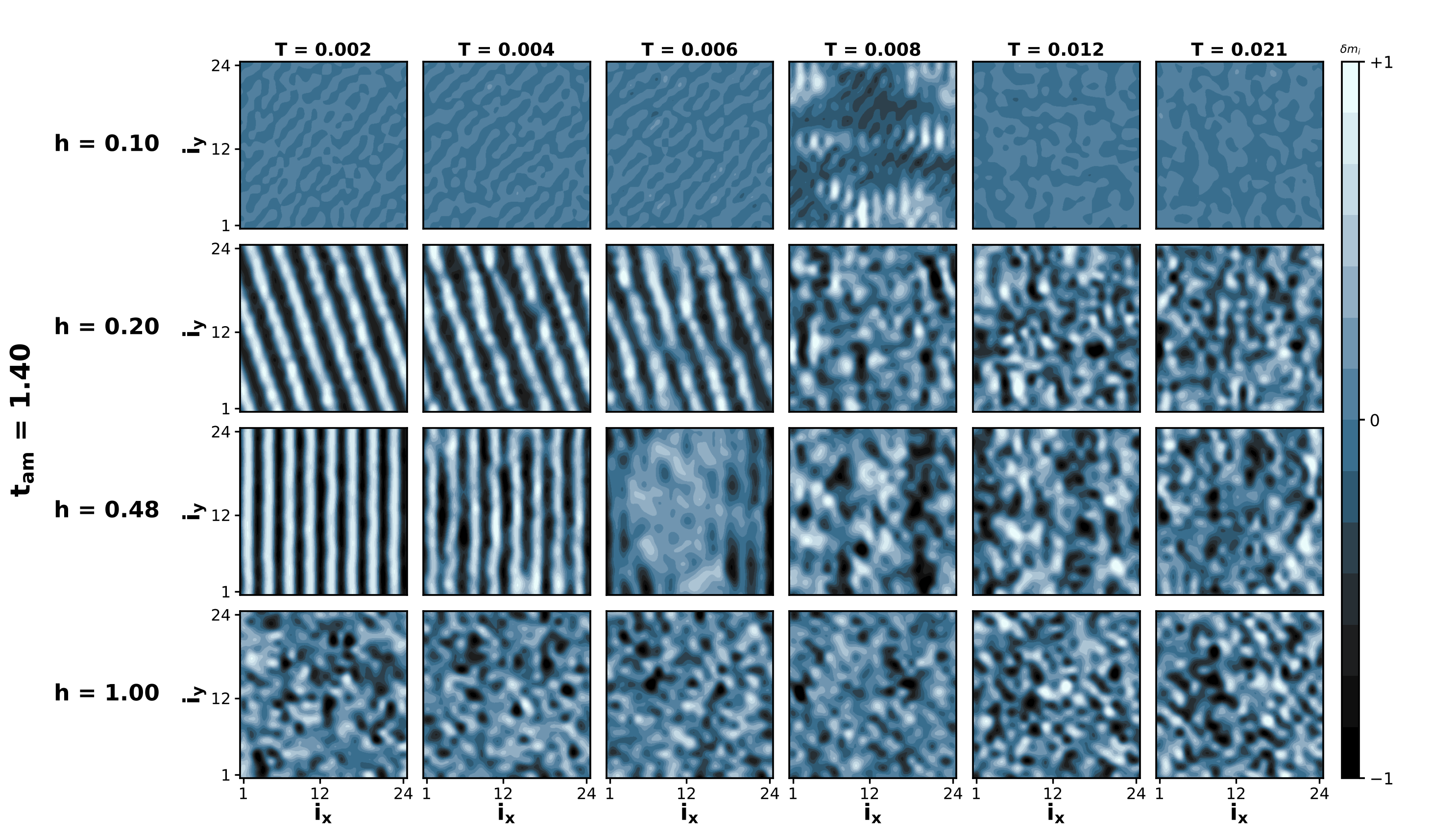}
  \caption{Real-space maps of $\langle|\Delta_i|\rangle$ (a) and $\delta m_i$ (b) at $t_{\mathrm{am}}=1.40t$ across $h=0.10t$, $0.20t$, $0.48t$, and $1.00t$, showing the field-driven rotation from the $\mathbf{Q}=(0,0)$ QBP state ($h=0.10t$) to diagonal stripes ($h=0.20t$) and vertical collinear stripes ($h=0.48t$).}
  \label{fig:S3b_real}
\end{figure}

\begin{figure}[!htbp]
  \centering
  \subc{0.80}{a}{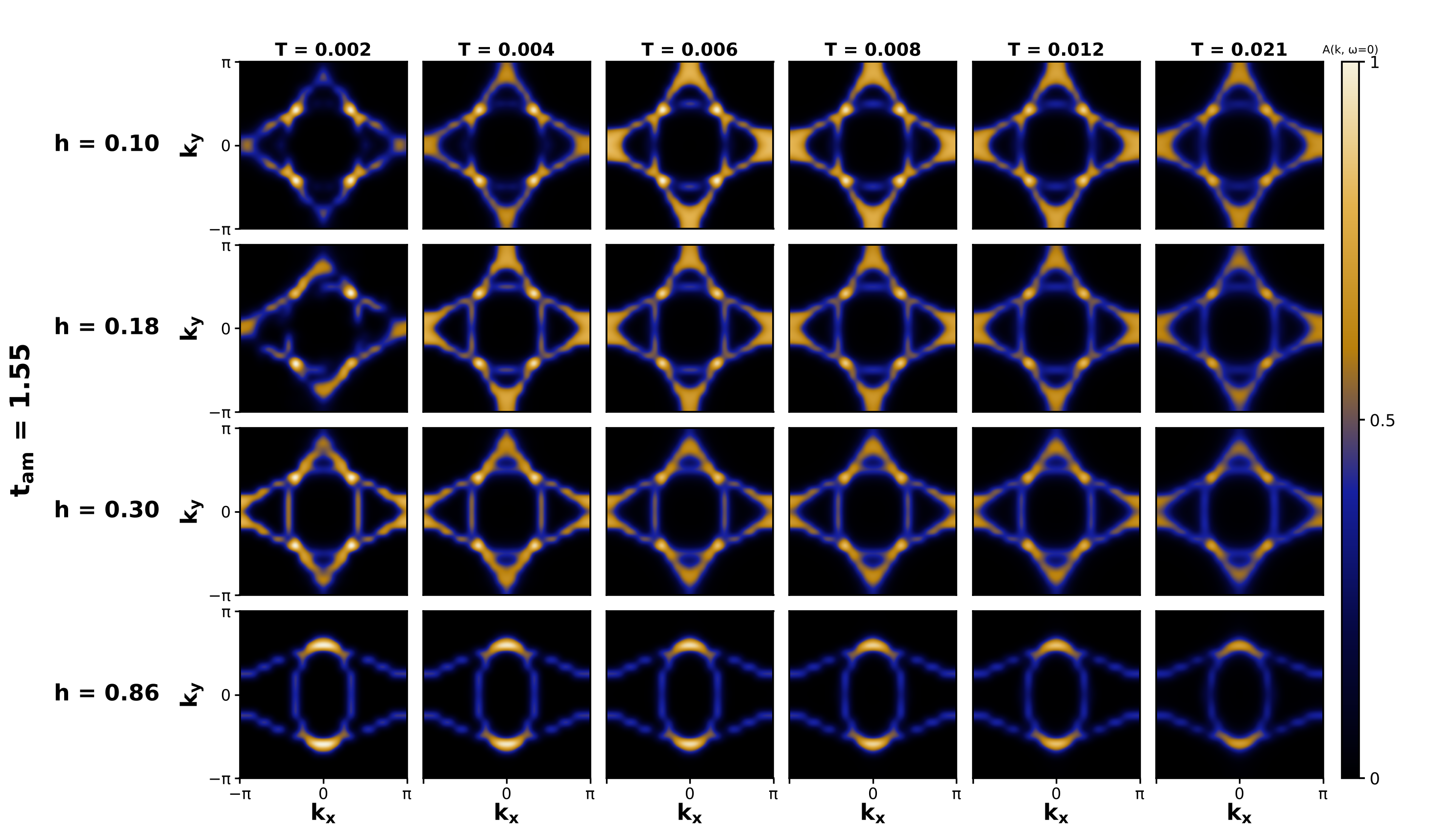}
  \subc{0.80}{b}{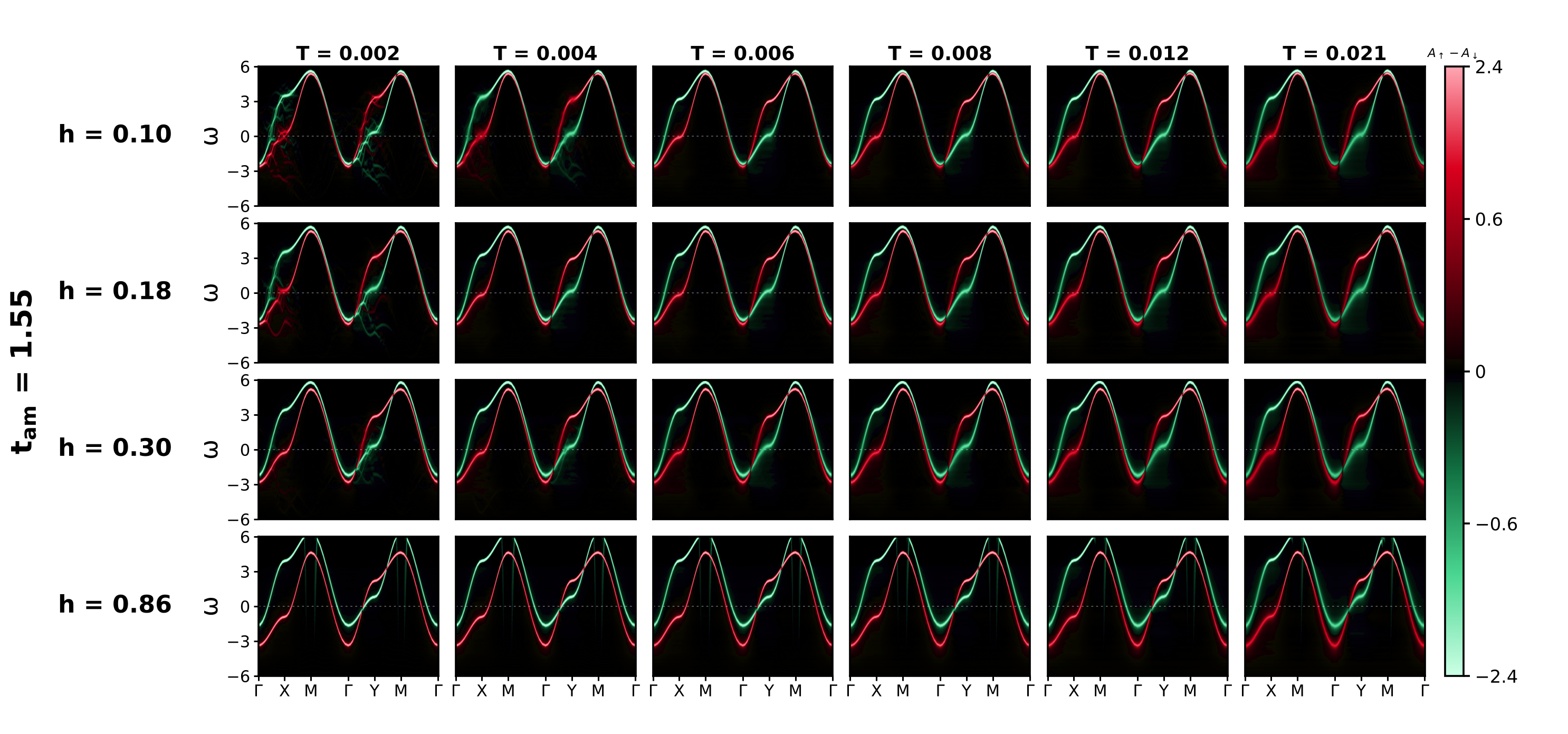}
  \subc{0.80}{c}{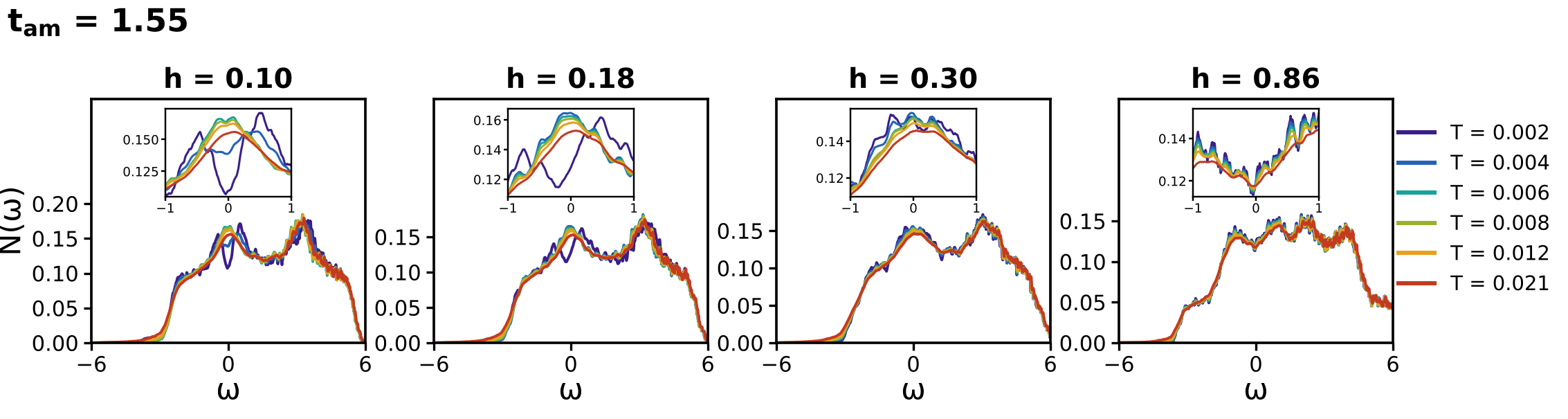}
  \caption{Spectroscopic evolution in the bidirectional PDW regime $t_{\mathrm{am}}=1.55t$ across $h=0.10t$ ($\text{2D-LO}$), $h=0.18t$ (non-collinear $\text{1D-LO}$), $h=0.30t$ (collinear Type-I LO), and $h=0.86t$ (PFL). (a) $A(\mathbf{k},0)$. (b) $\delta A(\mathbf{k},\omega)$. (c) $N(\omega)$.}
  \label{fig:S3c_spec}
\end{figure}

\begin{figure}[!htbp]
  \centering
  \subc{0.78}{a}{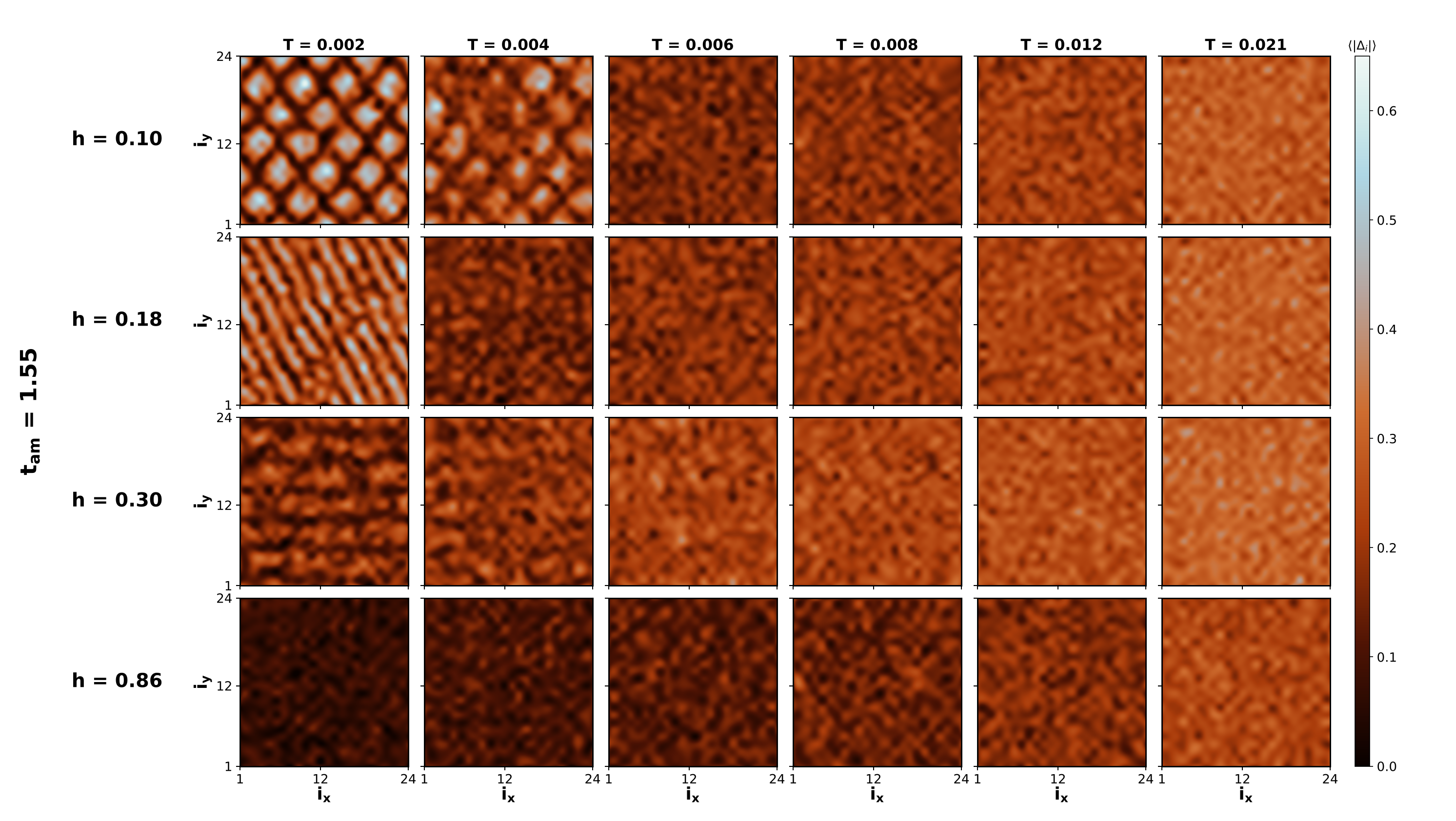}
  \subc{0.78}{b}{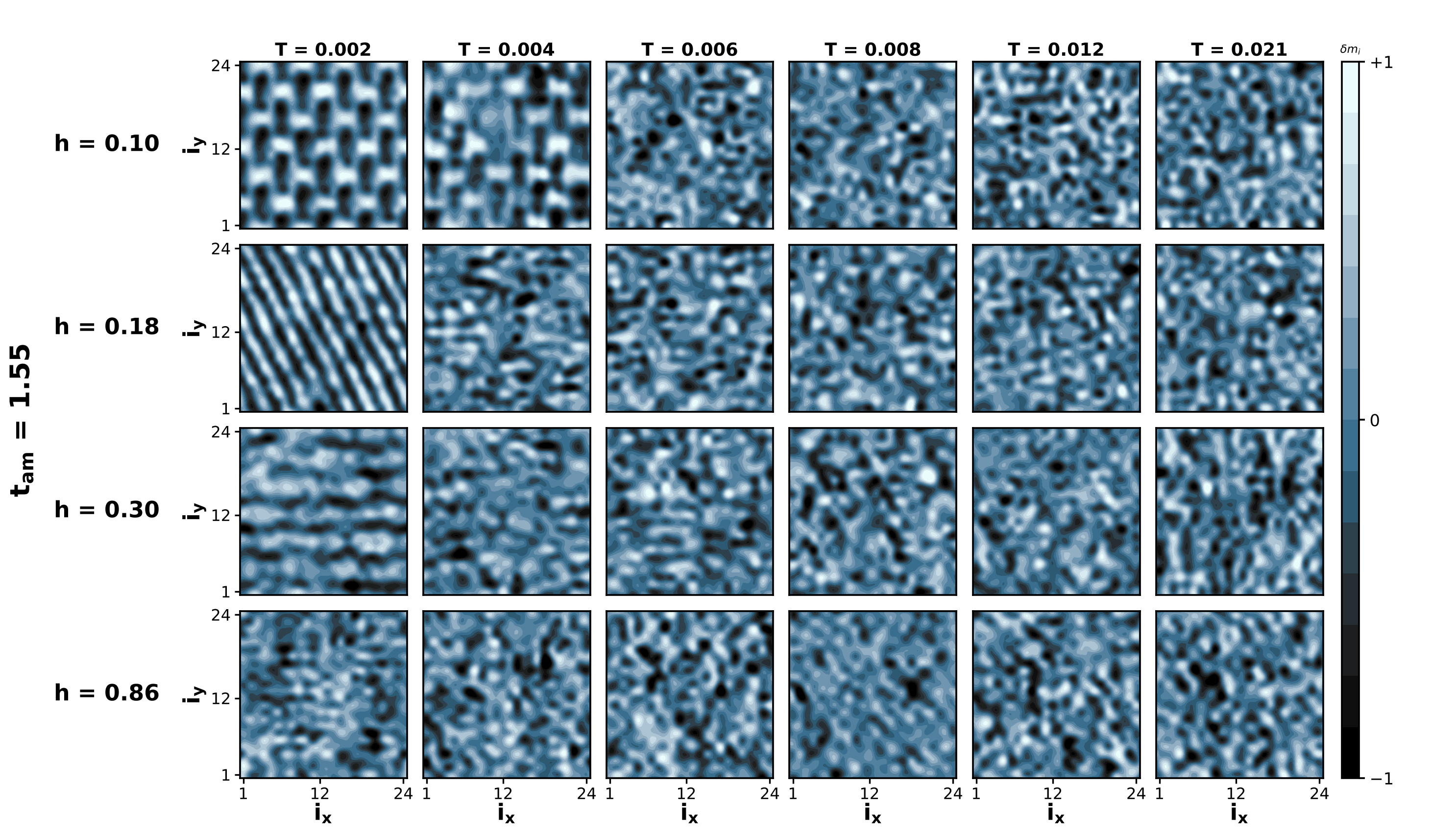}
  \caption{Real-space maps of $\langle|\Delta_i|\rangle$ (a) and $\delta m_i$ (b) at $t_{\mathrm{am}}=1.55t$ across $h=0.10t$, $0.18t$, $0.30t$, and $0.86t$, illustrating the progression from a 2D checkerboard ($h=0.10t$) to diagonal stripes ($h=0.18t$) and horizontal stripes ($h=0.30t$) before melting into the PFL ($h=0.86t$).}
  \label{fig:S3c_real}
\end{figure}

Figures~\ref{fig:S3a_spec}--\ref{fig:S3c_real} map out the combined effect of thermal fluctuations and an external Zeeman field across three representative regimes: the conventional Pauli-limited $d$-wave superconductor ($t_{\mathrm{am}}=0.00t$), the pre-PDW altermagnetic regime ($t_{\mathrm{am}}=1.40t$), and the bidirectional PDW regime ($t_{\mathrm{am}}=1.55t$).

At $t_{\mathrm{am}}=0.00t$ (Figs.~\ref{fig:S3a_spec} and \ref{fig:S3a_real}), a weak Zeeman field ($h=0.20t$) stabilizes the quantum breached pair (QBP) state with $\mathbf{Q}=(0,0)$ and nonzero magnetization, which keeps the $d_{x^2-y^2}$ nodal structure, with four nodal spots in $A(\mathbf{k},0)$ and a V-shaped gap in $N(\omega)$ at low $T$, with $T_c\approx0.0475t$ from the structure-factor analysis in Fig.~\ref{fig:S4_h}. Increasing the field to $h=0.80t$ drives a Pauli-limited transition into a conventional diagonal FFLO state ($\mathbf{Q}=(-2,-2)\cdot 2\pi/24$, $T_c \approx 0.0115t$), characterized by two concentric Zeeman-split Fermi sheets in $A(\mathbf{k},0)$, a nearly uniform energy offset between the spin-$\uparrow$ and spin-$\downarrow$ bands in $\delta A(\mathbf{k},\omega)$, and diagonal stripes in $\langle|\Delta_i|\rangle$ and $\delta m_i = m_i - m$.

Introducing strong altermagnetic hopping at $t_{\mathrm{am}}=1.40t$ (Figs.~\ref{fig:S3b_spec} and \ref{fig:S3b_real}) fundamentally reshapes this field evolution. Whereas $h=0.10t$ remains in the QBP state with $\mathbf{Q}=(0,0)$ below $T_c \approx 0.0114t$ with a pronounced low-energy gap in $N(\omega)$, a moderate field of $h=0.20t$ already induces a non-collinear LO state ($\mathbf{Q}=(3,1)\cdot 2\pi/24$, $T_c \approx 0.0090t$) with tilted real-space stripes. At $h=0.48t$ the modulation locks onto the $x$ axis, giving vertical collinear Type-I LO stripes ($\mathbf{Q}=(4,0)\cdot 2\pi/24$, $T_c \approx 0.0086t$ from $S_x$ and $0.0073t$ from $S_y$). Pair breaking destroys superconductivity at $h=1.00t$ (PFL).

Finally, at $t_{\mathrm{am}}=1.55t$ (Figs.~\ref{fig:S3c_spec} and \ref{fig:S3c_real}), where zero-field pairing already forms a bidirectional checkerboard, a small field $h=0.10t$ stabilizes a two-dimensional $\text{2D-LO}$ lattice of pairing islands surrounded by a spin-polarized nodal web ($T_c \approx 0.0071t$). Increasing $h$ to $0.18t$ and $0.30t$ breaks the fourfold checkerboard symmetry in favor of unidirectional diagonal ($\mathbf{Q}=(-4,-2)\cdot 2\pi/24$) and horizontal ($\mathbf{Q}=(0,3)\cdot 2\pi/24$) stripes, respectively, with folded Bogoliubov shadow replicas in $\delta A(\mathbf{k},\omega)$ that vanish upon entering the PFL at $h=0.86t$.

\subsection{Thermal evolution of the $\text{2D-LO}$ phase at $t_{\mathrm{am}}=1.55t$ and $h=0.10t$}
\label{sec:S-2dlo}

\begin{figure}[!htbp]
  \centering
  \includegraphics[width=0.62\linewidth]{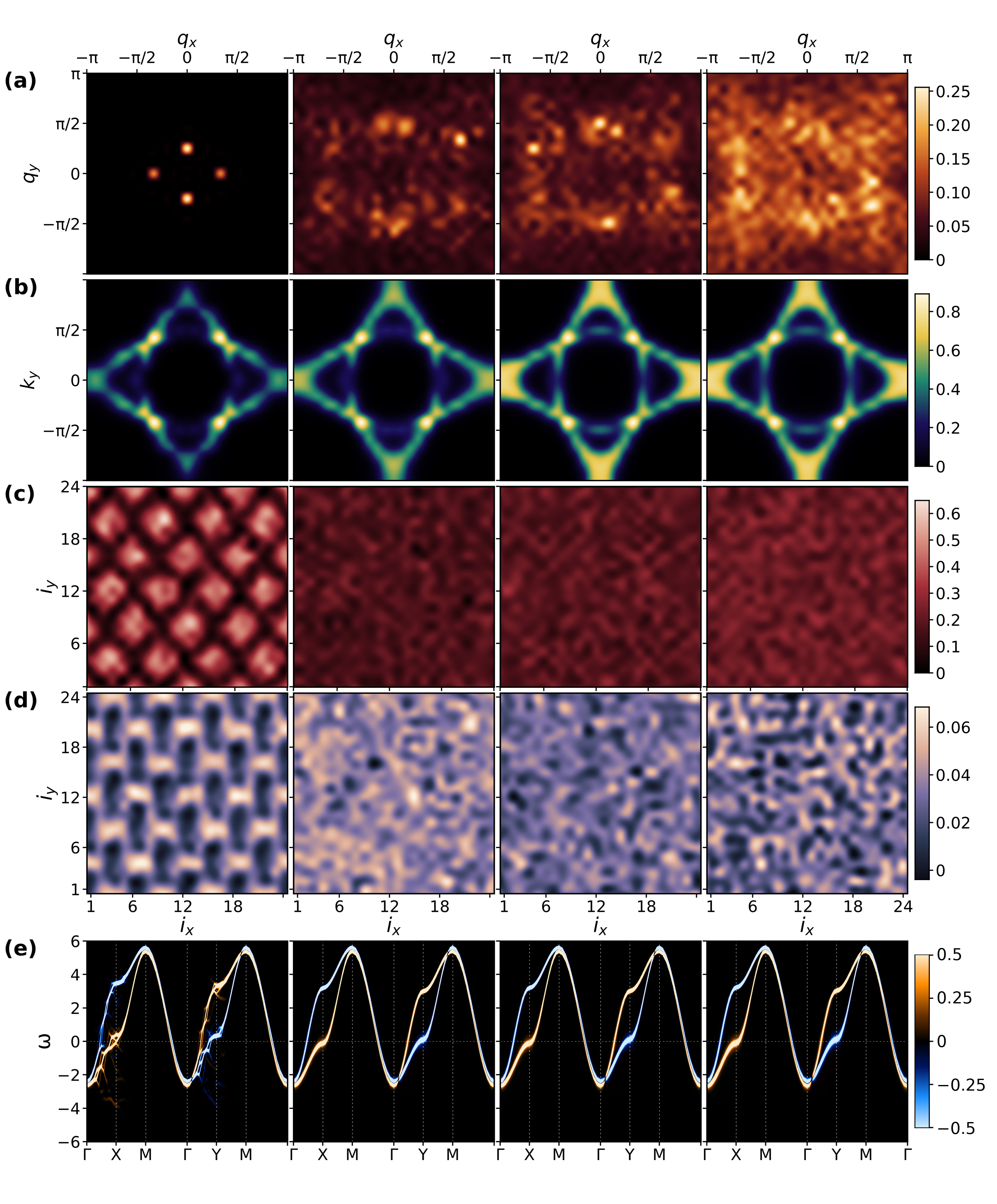}
  \caption{Thermal evolution of the bidirectional two-dimensional Larkin-Ovchinnikov ($\text{2D-LO}$) phase at $t_{\mathrm{am}}=1.55t$ and $h=0.10t$ for $T=0.002t$, $0.006t$, $0.008t$, and $0.012t$ (from left to right). (a) Structure factor $S_x(\mathbf{q})$, displaying four Bragg peaks at $\mathbf{Q}=(\pm q,\pm q)$ below $T_c$ that melt into a diffuse ring at higher $T$. (b) Zero-energy spectral weight $A(\mathbf{k},0)$, showing antinodal gapping along both crystal axes and diagonal nodal pockets at $T=0.002t$ that evolve into the normal-state Fermi surface upon heating. Single-snapshot real-space maps of (c) the pairing amplitude $|\Delta_i|$ and (d) the local spin polarization $m_i$, revealing an interlocking checkerboard in which spin polarization concentrates along the nodal web of $|\Delta_i|$. (e) Spin-difference spectral function $A_\uparrow(\mathbf{k},\omega)-A_\downarrow(\mathbf{k},\omega)$ along $\Gamma\text{--}X\text{--}M\text{--}\Gamma\text{--}Y\text{--}M\text{--}\Gamma$ (orange and blue denote spin-$\uparrow$ and spin-$\downarrow$ excess, respectively).}
  \label{fig:S3d_2dlo}
\end{figure}

At stronger altermagnetic hopping ($t_{\mathrm{am}}=1.55t$), where the zero-field ground state already favors a bidirectional checkerboard condensate ($\text{PDW 2D}$), applying a weak Zeeman field ($h=0.10t$) gives a thermally fragile $\text{2D-LO}$ phase, as shown in Fig.~\ref{fig:S3d_2dlo} across $T=0.002t$, $0.006t$, $0.008t$, and $0.012t$. Deep in the ordered state at $T=0.002t$, the structure factor $S_x(\mathbf{q})$ in Fig.~\ref{fig:S3d_2dlo}(a) exhibits four sharp Bragg peaks of comparable intensity at $\mathbf{Q}=(\pm q,\pm q)$ with peak intensity $\approx 0.25$. In contrast to the two-peak signature of unidirectional stripes, this four-peak pattern confirms simultaneous spatial modulation along two orthogonal directions. The peak height is about half that of the unidirectional $\text{PDW 1D}$ state, consistent with the weight being shared between two perpendicular wavevectors. The order also melts quickly on heating: by $T=0.006t$ the four peaks have dissolved into a diffuse ring with $S_x(\mathbf{q}) < 0.007$.

In real space at $T=0.002t$, the pairing amplitude $|\Delta_i|$ [Fig.~\ref{fig:S3d_2dlo}(c)] forms a periodic two-dimensional checkerboard of bright superconducting islands ($|\Delta_i|\approx 0.65$) bounded by an intersecting grid of dark nodal lines ($|\Delta_i|\rightarrow 0$). The corresponding spin-polarization map $m_i$ in Fig.~\ref{fig:S3d_2dlo}(d) displays an interlocking complementary network: $m_i$ peaks sharply (up to $\approx 0.07$) at the nodal intersections where the pairing amplitude vanishes, while remaining near zero inside the superconducting islands. As in the unidirectional LO phase, the excess spin polarization stays out of the paired regions and sits in the nodal web of the bidirectional condensate. At $T=0.006t$--$0.012t$ both the pairing checkerboard and the spin-polarization grid have melted into disordered, short-range puddles.

The bidirectional modulation also imprints an approximately fourfold-symmetric reconstruction on the electronic spectra. At $T=0.002t$, the zero-energy spectral weight $A(\mathbf{k},0)$ in Fig.~\ref{fig:S3d_2dlo}(b) shows gapped antinodal regions near both $(\pm\pi,0)$ and $(0,\pm\pi)$, leaving four prominent nodal pockets along the Brillouin-zone diagonals flanked by faint folded shadow arcs. Simultaneously, $A_\uparrow(\mathbf{k},\omega)-A_\downarrow(\mathbf{k},\omega)$ in Fig.~\ref{fig:S3d_2dlo}(e) exhibits complex multi-branch Bogoliubov band folding and avoided crossings near $\omega=0$ along both $\Gamma\text{--}X\text{--}M$ and $\Gamma\text{--}Y\text{--}M$. Once the bidirectional coherence is lost at $T \ge 0.006t$, these shadow replicas disappear, antinodal spectral weight is restored, and the excitation spectrum reverts to two slightly Zeeman-offset, altermagnetically spin-split normal-state bands.

\section{Pairing Structure Factor versus Temperature}
\label{sec:S-sq}

\begin{figure}[!htbp]
  \centering
  \includegraphics[width=\linewidth]{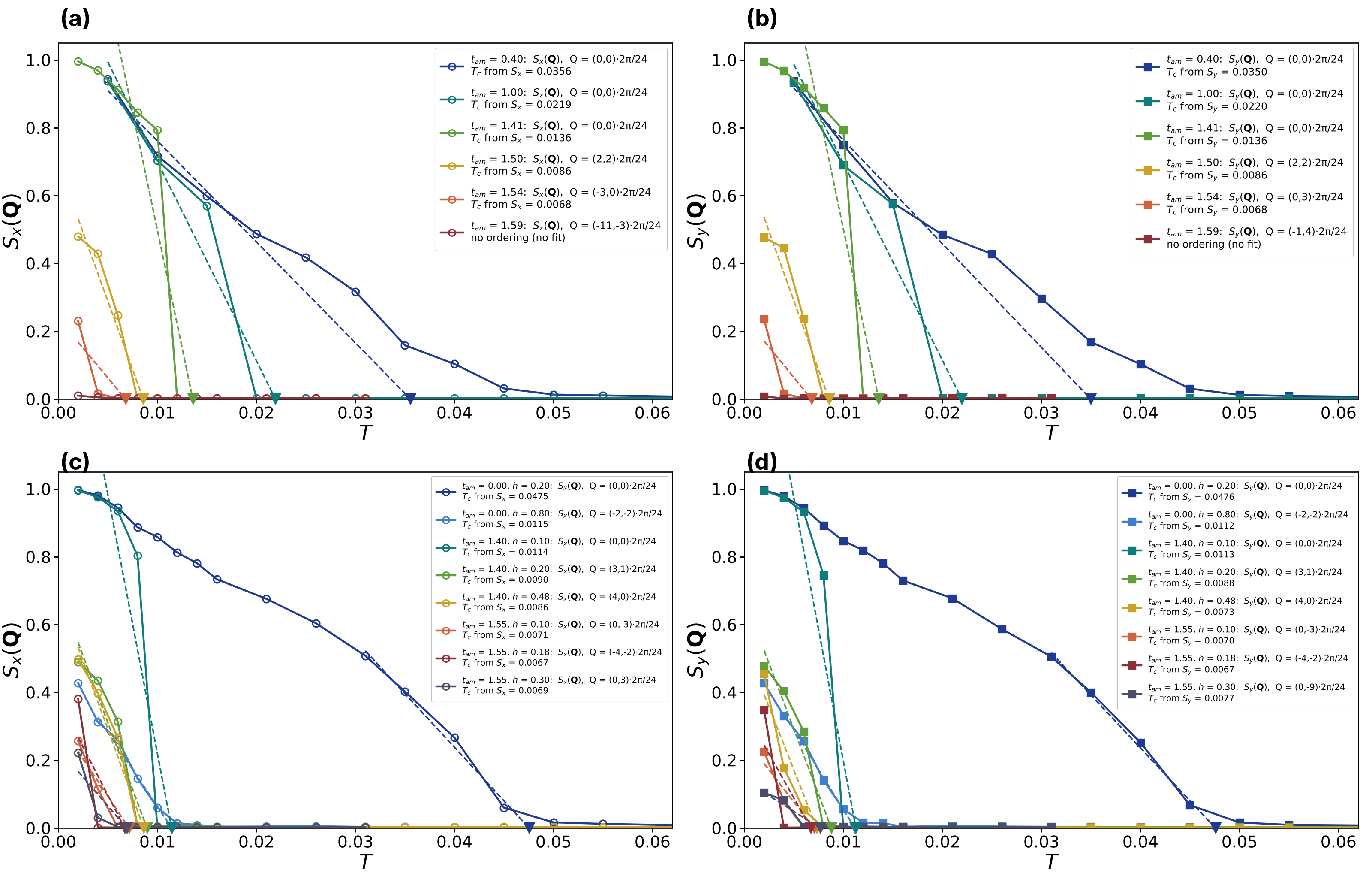}
  \caption{Temperature dependence of the ordering-wavevector structure-factor peaks $S_x(\mathbf{Q})$ and $S_y(\mathbf{Q})$ on an $L=24$ lattice. (a),(b) $S_x(\mathbf{Q})$ and $S_y(\mathbf{Q})$ at $h=0$ for representative $t_{\mathrm{am}}$ values; the $\pi/2$ rotation between the dominant $x$- and $y$-bond peaks is visible in the $\text{PDW 2D}$ regime ($t_{\mathrm{am}}=1.54t$). (c),(d) $S_x(\mathbf{Q})$ and $S_y(\mathbf{Q})$ at finite Zeeman fields $h\neq0$ for selected $t_{\mathrm{am}}$ values. The wavevector $\mathbf{Q}$ of each peak is listed in the legends in units of $2\pi/24$. Dashed lines show the linear fits used to estimate the ordering temperature $T_c$ (triangles).}
  \label{fig:S4}
  \label{fig:S4_h0}
  \label{fig:S4_h}
\end{figure}

To determine the superconducting critical temperature $T_c$ consistently across both uniform and spatially modulated phases, Figs.~\ref{fig:S4_h0} and \ref{fig:S4_h} show the thermal decay of the dominant structure-factor peaks $S_x(\mathbf{Q})$ and $S_y(\mathbf{Q})$ at $h=0$ and $h \neq 0$, respectively. In the uniform $d$-BCS regime ($t_{\mathrm{am}} \le 1.41t$ at $h=0$), the peak sits at $\mathbf{Q}=(0,0)$ and approaches unity as $T\rightarrow 0$, vanishing smoothly at $T_c$. Entering the PDW and LO phases ($t_{\mathrm{am}}=1.50t, 1.54t$ at $h=0$, or finite $h$), spectral weight transfers to finite ordering wavevectors $\mathbf{Q}\neq (0,0)$, accompanied by a reduced low-$T$ peak amplitude ($S_{\alpha}(\mathbf{Q}) \approx 0.2$--$0.5$). With the normalization of Eq.~\eqref{eq:S-Sq}, the Fourier weight of a nodal (modulated) pairing field is shared between $\pm\mathbf{Q}$, so an ideal single-$\mathbf{Q}$ stripe gives $1/2$ per peak, and in the bidirectional states it is shared further among the wavevectors along both axes.

Comparing $S_x(\mathbf{Q})$ and $S_y(\mathbf{Q})$ shows that both bond orientations give nearly the same transition temperature: the two values agree to better than $\approx5\%$ in most cases ($t_{\mathrm{am}}=1.54t$, $h=0$: $0.0068t$ in both; $t_{\mathrm{am}}=1.40t$, $h=0.20t$: $0.0090t$ and $0.0088t$; $t_{\mathrm{am}}=1.55t$, $h=0.10t$: $0.0071t$ and $0.0070t$), and to roughly $10$--$15\%$ in the worst cases ($t_{\mathrm{am}}=1.40t$, $h=0.48t$: $0.0086t$ and $0.0073t$; $t_{\mathrm{am}}=1.55t$, $h=0.30t$: $0.0069t$ and $0.0077t$). For instance, at $t_{\mathrm{am}}=1.54t$ and $h=0$ [Fig.~\ref{fig:S4_h0}], the dominant peak at $\mathbf{Q}=(-3,0)\cdot 2\pi/24$ in $S_x(\mathbf{Q})$ appears rotated by $\pi/2$ at $\mathbf{Q}=(0,3)\cdot 2\pi/24$ in $S_y(\mathbf{Q})$, with $T_c = 0.0068t$ in both channels. At $t_{\mathrm{am}}=1.59t$ ($h=0$), $S_{\alpha}(\mathbf{Q})$ remains negligible across all temperatures (of order $1/N$ for uncorrelated fields), indicating the absence of appreciable spatially coherent superconducting order.

\section{Variational Mean-Field Theory Benchmarks on \texorpdfstring{$L=80$}{L=80}}
\label{sec:S-mft}

\begin{figure}[!htbp]
  \centering
  \subnt{0.54}{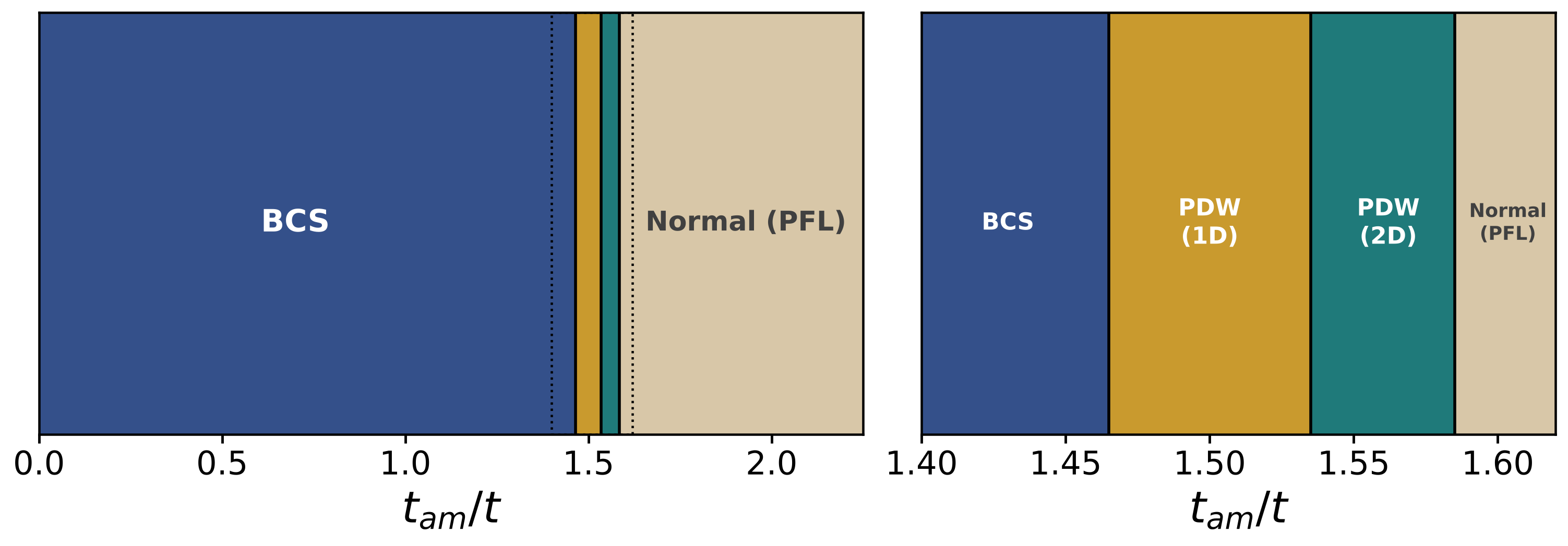}\\[4pt]
  \subnt{0.54}{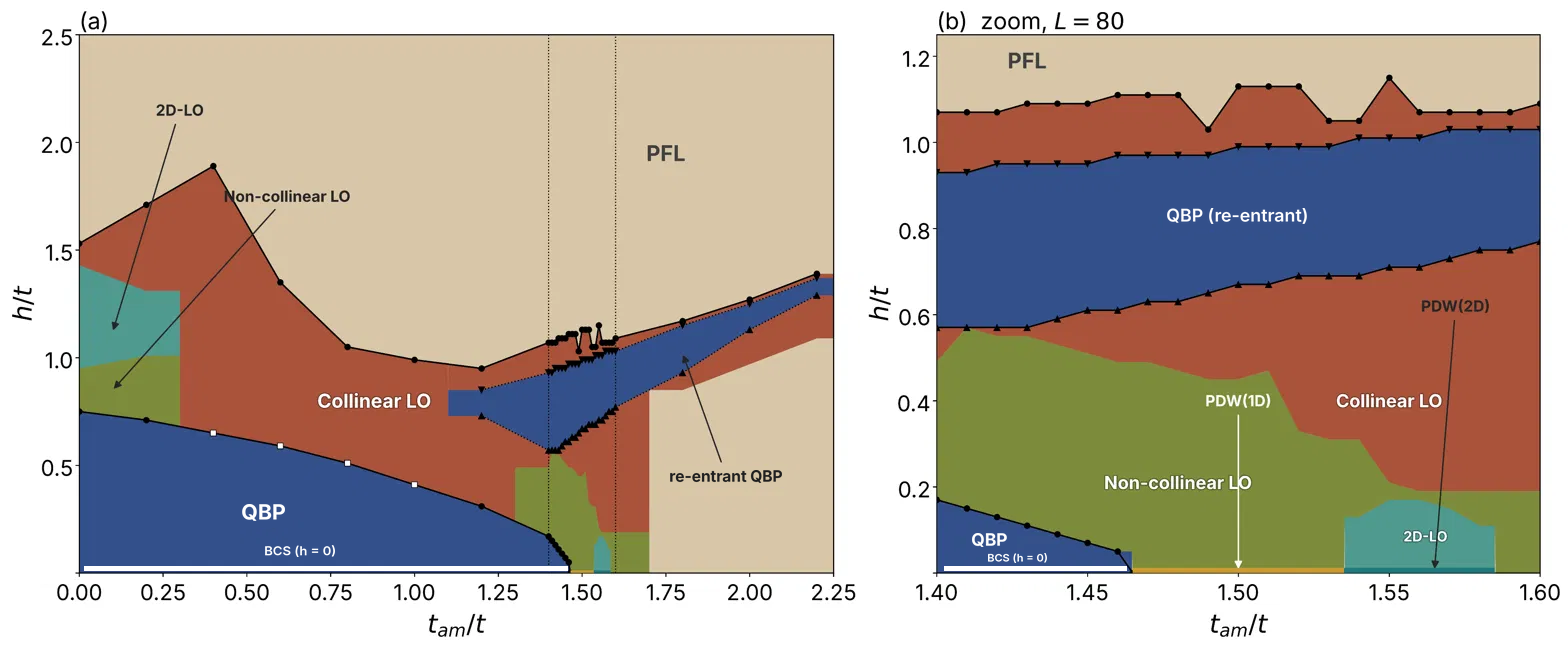}
  \caption{Variational mean-field ground-state ($T=0.005t$) phase diagrams on an $L=80$ lattice at $|U|=4t$ and $\mu=-1.5t$. Top: zero-field ($h=0$) phase progression from uniform $d$-wave BCS ($t_{\mathrm{am}} \lesssim 1.465t$) to unidirectional $\text{PDW 1D}$ ($1.465t \lesssim t_{\mathrm{am}} \lesssim 1.535t$), bidirectional $\text{PDW 2D}$ ($1.535t \lesssim t_{\mathrm{am}} \lesssim 1.585t$), and the normal metal ($t_{\mathrm{am}} \gtrsim 1.585t$), with the right panel magnifying the PDW window. Bottom: $t_{\mathrm{am}}$--$h$ ground-state phase diagram showing the $d$-wave BCS state at $h=0$ (marked by the white line), the QBP phase at finite field, non-collinear LO, collinear LO, $\text{2D-LO}$, re-entrant QBP, and polarized Fermi liquid (PFL) regimes, with the right panel zooming in on $1.40 \le t_{\mathrm{am}}/t \le 1.60$.}
  \label{fig:S5_gs}
\end{figure}

\begin{figure}[!htbp]
  \centering
  \subnt{0.52}{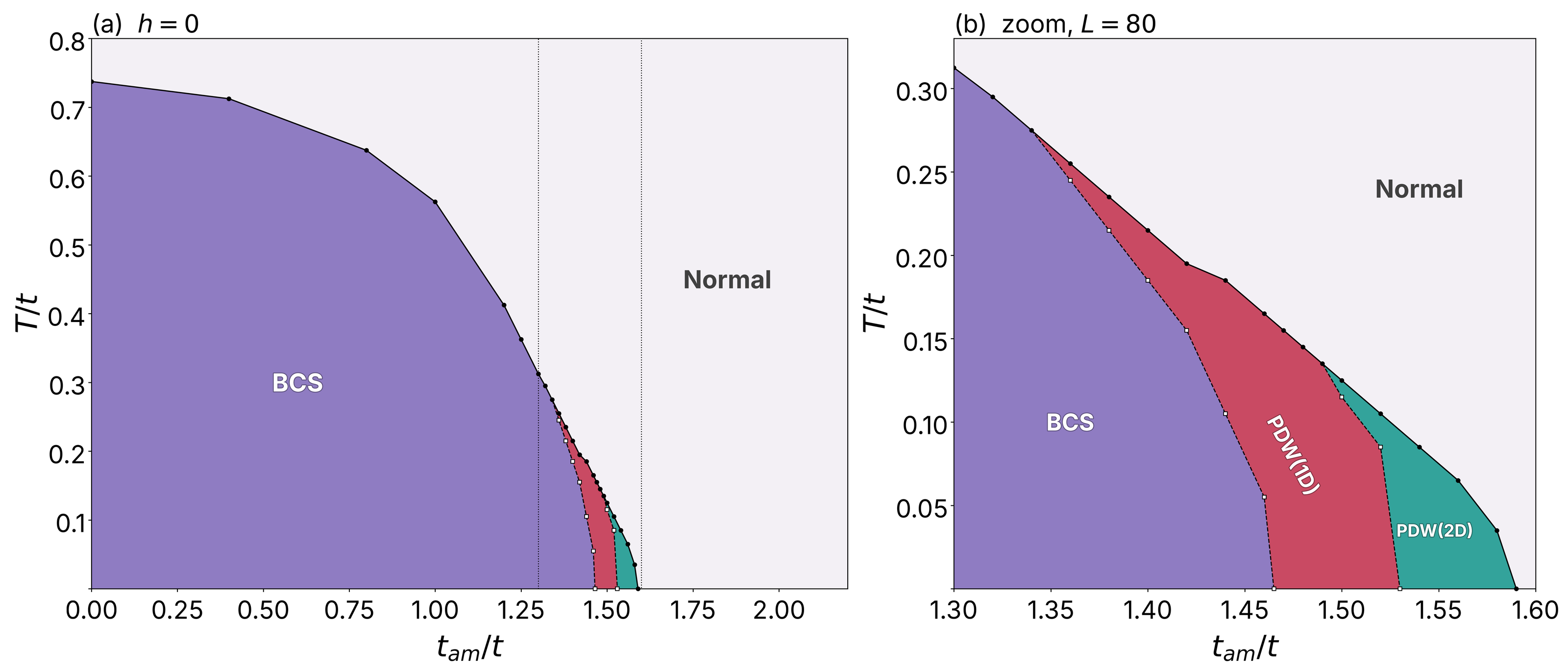}\\[4pt]
  \subnt{0.44}{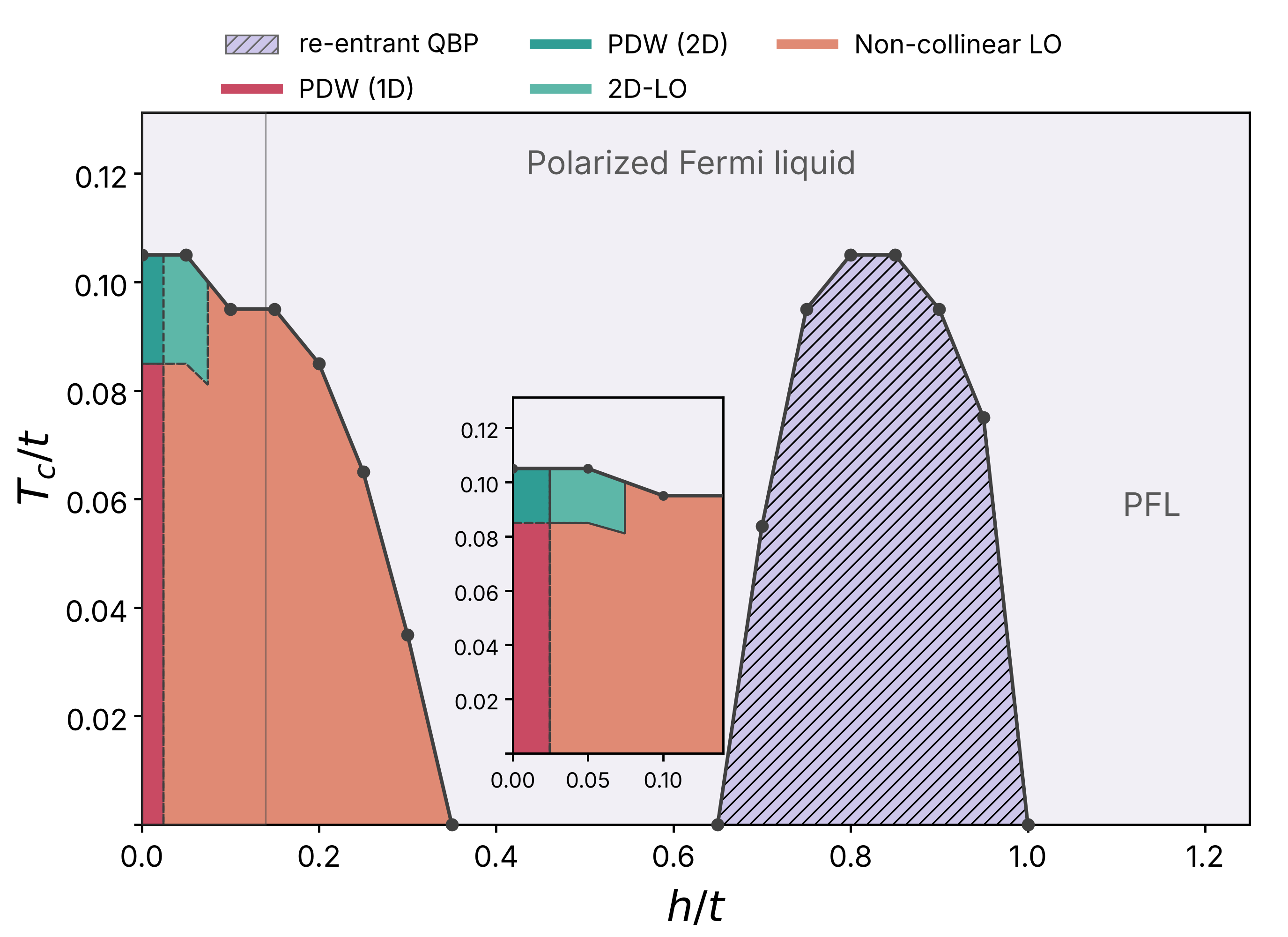}
  \caption{Finite-temperature variational mean-field phase diagrams on $L=80$. Top: $t_{\mathrm{am}}$--$T$ plane at $h=0$ (with the right panel magnifying $1.30 \le t_{\mathrm{am}}/t \le 1.60$). Bottom: $h$--$T$ plane at $t_{\mathrm{am}}=1.52t$ showing the $\text{PDW 1D}$, $\text{PDW 2D}$, $\text{2D-LO}$, non-collinear LO, and re-entrant QBP domes. While MFT captures the qualitative sequence of phases, neglecting thermal phase fluctuations overestimates $T_c$ by a factor of roughly $6$--$20$ compared to SPA Monte Carlo (see text).}
  \label{fig:S5_thermal}
\end{figure}

\begin{figure}[!htbp]
  \centering
  \subnt{0.66}{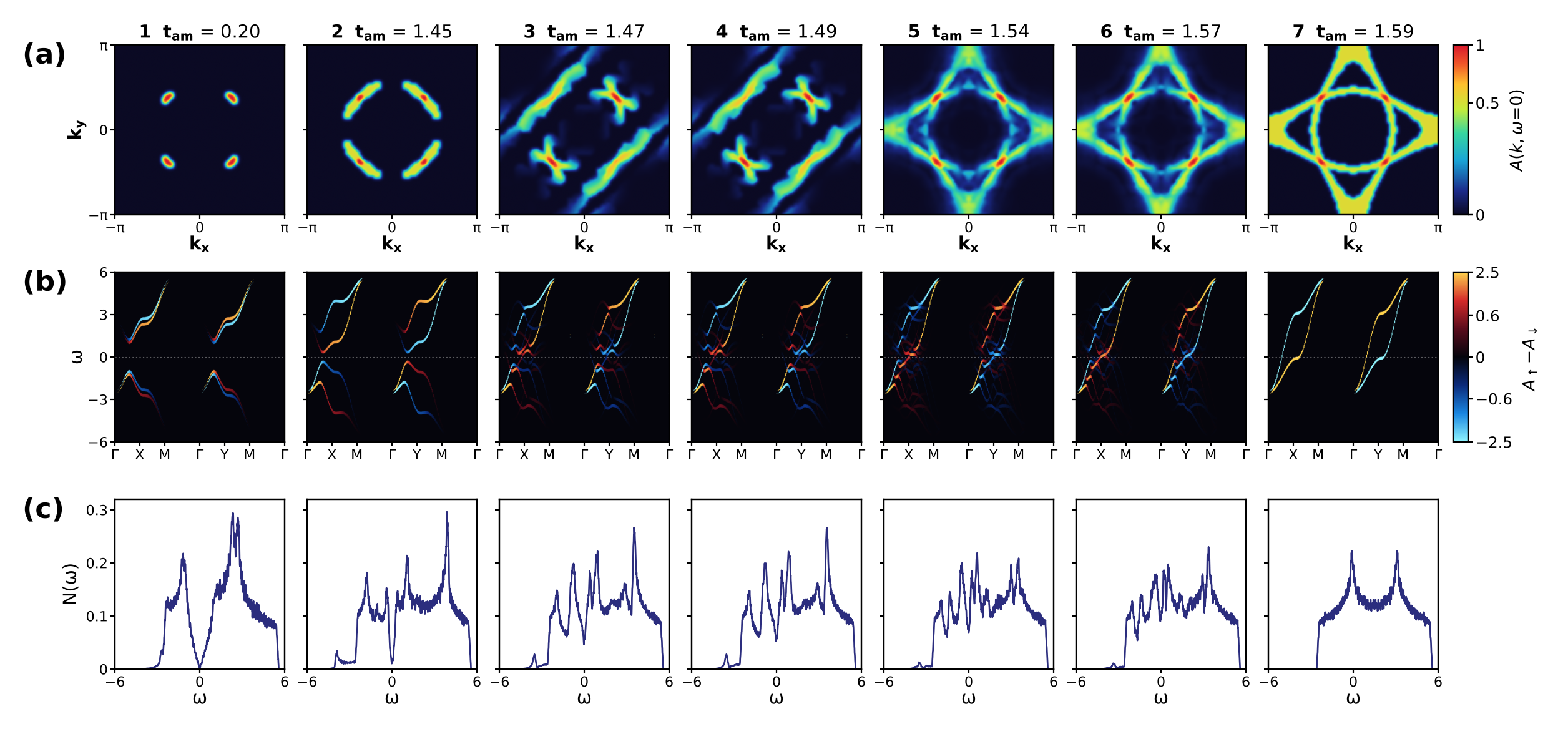}\\[4pt]
  \subnt{0.66}{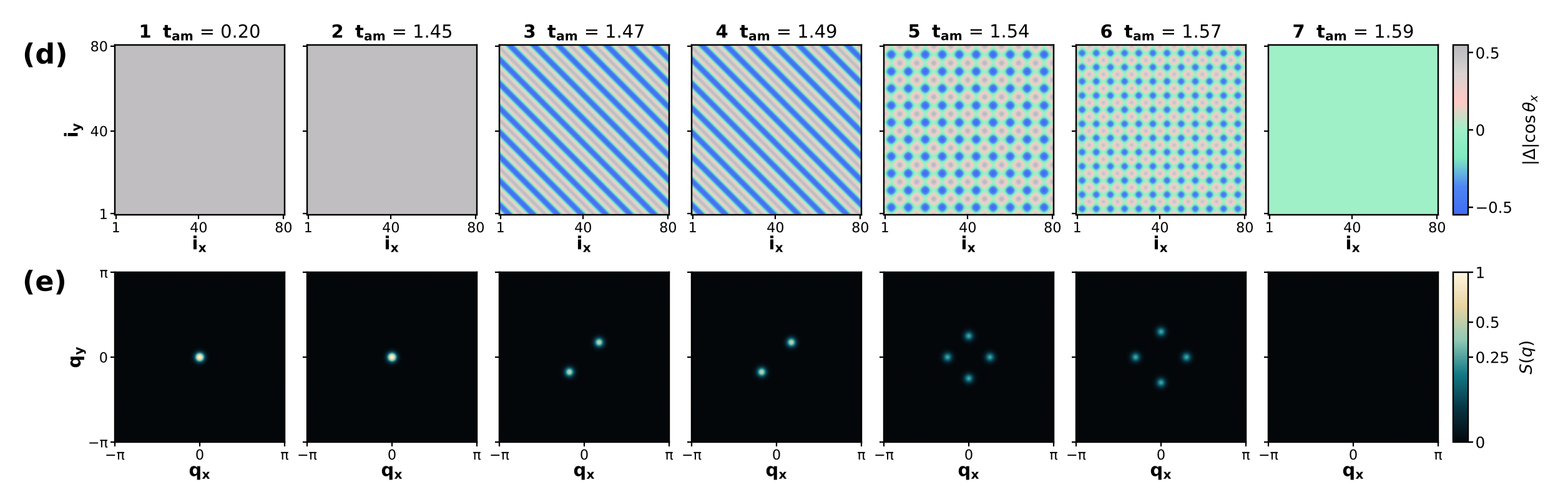}
  \caption{Variational mean-field evolution across representative zero-field couplings $t_{\mathrm{am}} = 0.20t, 1.45t, 1.47t, 1.49t, 1.54t, 1.57t,$ and $1.59t$ (columns 1--7) on $L=80$ at $T=0.005t$. From top to bottom: (a) low-energy spectral weight $A(\mathbf{k},0)$, (b) spin-difference spectral function $\delta A(\mathbf{k},\omega)$ along $\Gamma\text{--}X\text{--}M\text{--}\Gamma\text{--}Y\text{--}M\text{--}\Gamma$, (c) single-particle density of states $N(\omega)$, (d) real-space order-parameter modulation $|\Delta|\cos\theta_x$, and (e) pairing structure factor $S(\mathbf{q})$.}
  \label{fig:S5_evolution}
\end{figure}

\begin{figure}[!htbp]
  \centering
  \includegraphics[width=0.45\linewidth]{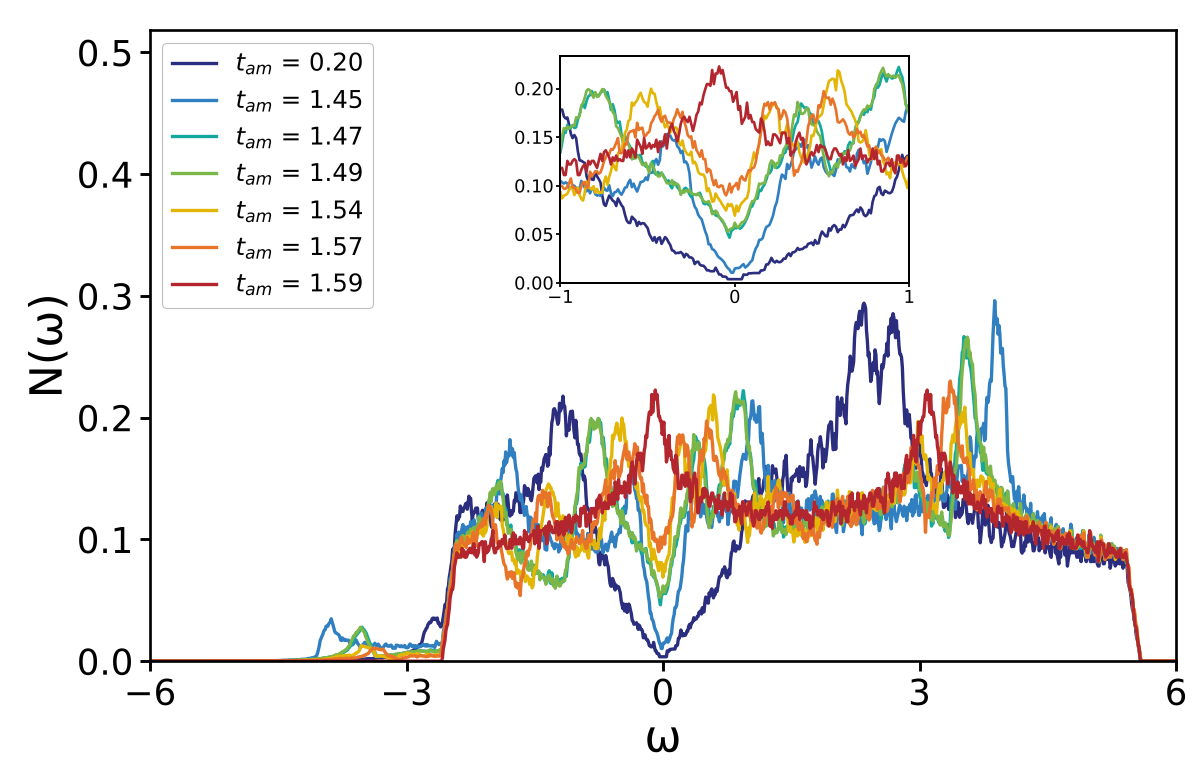}
  \caption{Overlay of the variational mean-field density of states $N(\omega)$ on $L=80$ across the seven representative $t_{\mathrm{am}}$ values shown in Fig.~\ref{fig:S5_evolution}, with the inset magnifying the low-energy region $|\omega|\le t$ around the Fermi level.}
  \label{fig:S5_dos}
\end{figure}

To isolate the role of thermal fluctuations and benchmark our finite-lattice SPA simulations against a large system size ($L=80$), we perform variational mean-field theory (MFT) calculations at $|U|=4t$ and $\mu=-1.5t$. Following standard variational MFT for modulated superconductors \cite{Karmakar2020,Karmakar2024}, we take a real, spatially modulated pairing amplitude with fixed $d$-wave relative bond phase $\phi_{ij}^{\mathrm{rel}}=\pi$ and optimize the free energy over uniform $\Delta_{ij}=|\Delta|$, unidirectional $\Delta_{ij}\sim|\Delta|\cos(q_x x_i + q_y y_i)$, and bidirectional $\Delta_{ij}\sim|\Delta|\left[\cos(q_x x_i)+\cos(q_y y_i)\right]$ trial states.

As shown in Fig.~\ref{fig:S5_gs} (top), the $L=80$ ground-state ($T=0.005t$) phase progression at $h=0$ is close to our $L=24$ SPA results (the boundaries differ by $\lesssim0.03t$): uniform $d$-wave BCS order persists up to $t_{\mathrm{am}}\approx 1.465t$, followed by a unidirectional $\text{PDW 1D}$ window ($1.465t \lesssim t_{\mathrm{am}} \lesssim 1.535t$), a bidirectional $\text{PDW 2D}$ window ($1.535t \lesssim t_{\mathrm{am}} \lesssim 1.585t$), and a normal altermagnetic metal beyond $t_{\mathrm{am}}\approx 1.585t$. In the $t_{\mathrm{am}}$--$h$ plane [Fig.~\ref{fig:S5_gs}, bottom], the weak-$t_{\mathrm{am}}$ ground state is the $d$-wave BCS superconductor only on the line $h=0$; any finite Zeeman field turns it into the QBP phase ($\mathbf{Q}=(0,0)$ with a finite magnetization), which persists up to the lower critical field. Beyond it, MFT gives the same sequence of field-driven transitions from the zero-field PDW into non-collinear LO, collinear LO, and $\text{2D-LO}$ phases, and in addition a re-entrant QBP region at $0.65\lesssim h/t\lesssim1.0$, where the Zeeman field partly compensates the altermagnetic splitting. 

The spectroscopic evolution across this zero-field cascade is resolved on $L=80$ in Figs.~\ref{fig:S5_evolution} and \ref{fig:S5_dos}. Deep in the $d$-BCS regime ($t_{\mathrm{am}}=0.20t$), $A(\mathbf{k},0)$ exhibits four sharp diagonal point nodes and $N(\omega)$ displays a clean V-shaped gap ($N(0)\rightarrow 0$). Approaching the PDW threshold ($t_{\mathrm{am}}=1.45t$) distorts the Fermi surface and narrows the gap, while entering $\text{PDW 1D}$ ($t_{\mathrm{am}}=1.47t, 1.49t$) and $\text{PDW 2D}$ ($t_{\mathrm{am}}=1.54t, 1.57t$) produces extended Bogoliubov Fermi arcs, folded shadow bands in $\delta A(\mathbf{k},\omega)$, and multiple subgap van Hove peaks in $N(\omega)$—accompanied by crisp diagonal stripe (two-peak $S(\mathbf{q})$) and checkerboard (four-peak $S(\mathbf{q})$) modulations in real and momentum space. Once pairing vanishes at $t_{\mathrm{am}}=1.59t$, $N(\omega)$ collapses onto a normal-state van Hove peak at $\omega=0$.

Extending variational MFT to finite temperature (Fig.~\ref{fig:S5_thermal}) shows why the SPA sampling is needed in two dimensions. Because MFT ignores the phase fluctuations that destroy long-range order in two dimensions, it overestimates all transition temperatures: $T_c^{\mathrm{MFT}} \approx 0.74t$ at $t_{\mathrm{am}}=0$ (about twenty times larger than $T_c^{\mathrm{SPA}}\approx 0.036t$), and $T_c^{\mathrm{MFT}}\approx 0.04t$--$0.15t$ inside the PDW window $1.47\lesssim t_{\mathrm{am}}/t\lesssim1.58$ and about $0.1t$ in the field-driven phases at $t_{\mathrm{am}}=1.52t$ (compared to $T_c^{\mathrm{SPA}} \approx 0.007t$--$0.017t$).

\section{Finite-Size Comparison (\texorpdfstring{$L=24$}{L=24} vs.\ \texorpdfstring{$L=40$}{L=40})}
\label{sec:S-fss}

\begin{figure}[!htbp]
  \centering
  \sub{0.47}{a}{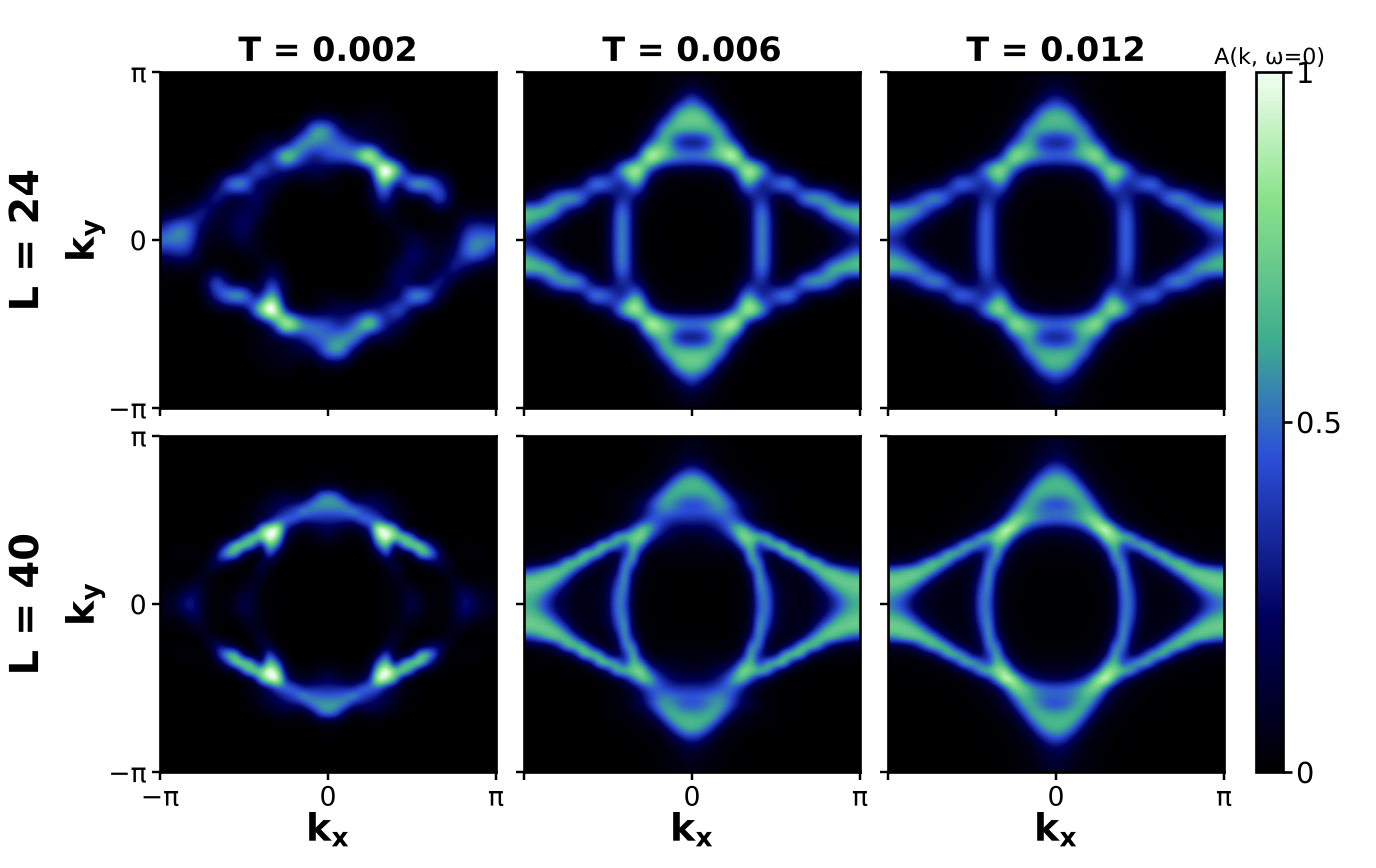}\hfill
  \sub{0.47}{b}{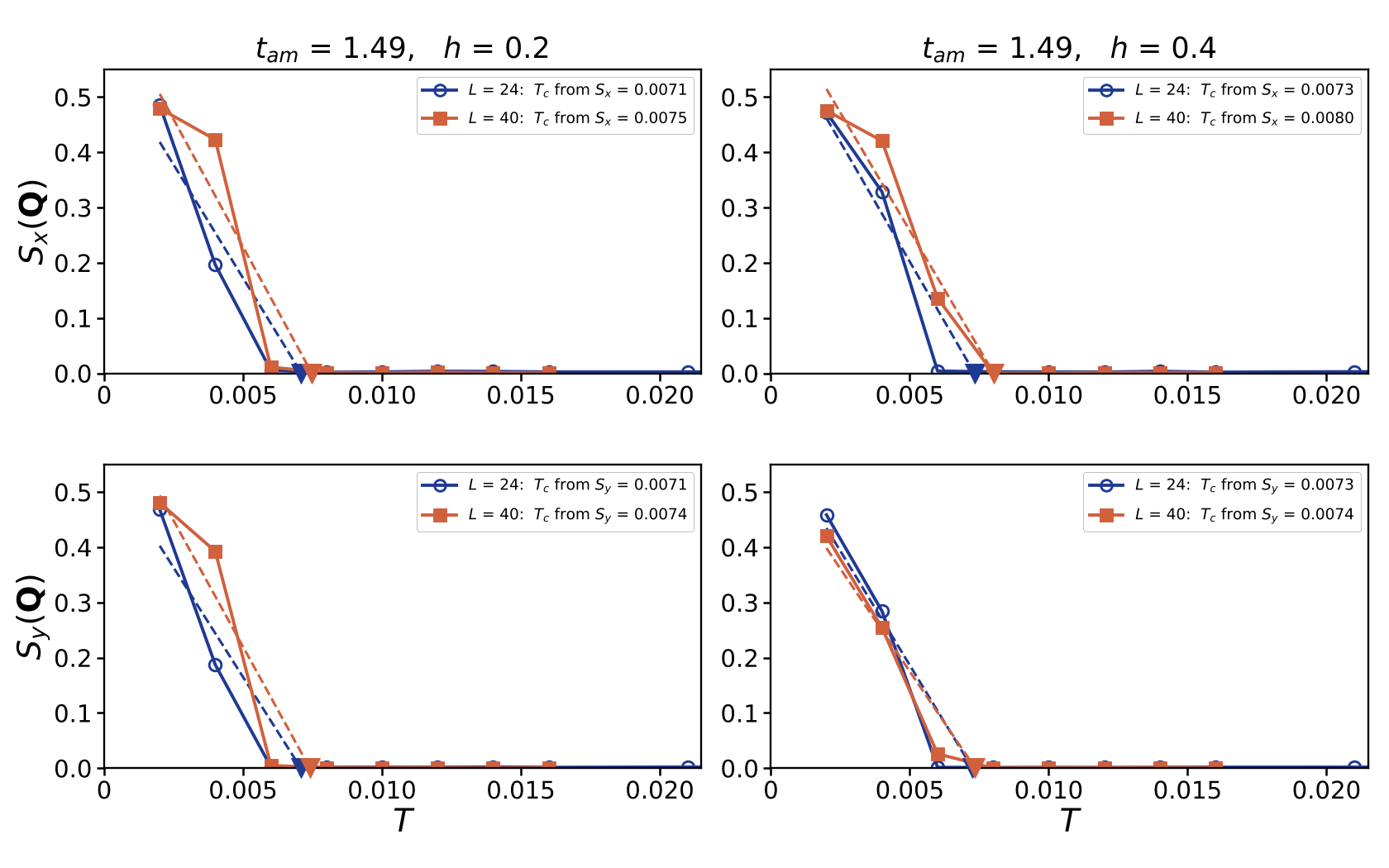}
  \caption{(a) Comparison of the zero-energy spectral weight $A(\mathbf{k},\omega=0)$ between $L=24$ (top row) and $L=40$ (bottom row) SPA Monte Carlo simulations at $t_{\mathrm{am}}=1.49t$ across $T=0.002t$, $0.006t$, and $0.012t$, showing close agreement of the Fermi-surface reconstruction between the two sizes. (b) Finite-size comparison of the structure-factor peaks $S_x(\mathbf{Q})$ (top row) and $S_y(\mathbf{Q})$ (bottom row) versus temperature on $L=24$ (blue circles) and $L=40$ (orange squares) lattices at $t_{\mathrm{am}}=1.49t$ for $h=0.20t$ (left column) and $h=0.40t$ (right column). Dashed lines and triangles indicate the linear fits used to estimate $T_c$.}
  \label{fig:S6}
\end{figure}

Finally, to check that our finite-temperature phase boundaries and spectral reconstructions do not depend sensitively on the lattice size, Figs.~\ref{fig:S6}(a) and \ref{fig:S6}(b) compare SPA Monte Carlo results on $L=24$ and $L=40$ lattices at $t_{\mathrm{am}}=1.49t$. In Fig.~\ref{fig:S6}(a), the momentum-resolved Fermi-surface weight $A(\mathbf{k},0)$ shows virtually identical Bogoliubov arc segmentation at $T=0.002t$ and antinodal refilling across $T=0.006t$ and $0.012t$ on both lattice sizes, with $L=40$ simply providing finer momentum resolution. Likewise, the thermal decay of both $S_x(\mathbf{Q})$ and $S_y(\mathbf{Q})$ in Fig.~\ref{fig:S6}(b) gives close critical temperatures on $L=24$ and $L=40$: from $S_x$, $T_c = 0.0071t$ ($L=24$) versus $0.0075t$ ($L=40$) at $h=0.20t$, and $T_c = 0.0073t$ ($L=24$) versus $0.0080t$ ($L=40$) at $h=0.40t$; from $S_y$ the corresponding values are $0.0071t$ versus $0.0074t$ and $0.0073t$ versus $0.0074t$. The larger lattice therefore gives a $T_c$ that is higher by about $1$--$10\%$.

\end{document}